\documentclass[aps,prb,twocolumn,superscriptaddress,longbibliography]{revtex4-2}

\usepackage{xcolor}	
\usepackage[utf8]{inputenc}

\usepackage{siunitx}
\usepackage{amsmath}
\usepackage{mathrsfs}
\usepackage{graphicx}
\usepackage{dcolumn}
\usepackage{bm}
\usepackage[normalem]{ulem}
\usepackage{float}
\usepackage{braket}
\usepackage{amssymb}
\usepackage{soul,xcolor}

\newcommand{\beq}{\begin{eqnarray}}
\newcommand{\eeq}{\end{eqnarray}}

\usepackage{booktabs}

\usepackage{comment}
\usepackage{xr-hyper}
\usepackage{hyperref}
\makeatletter

\hypersetup{
colorlinks=true,
citecolor=blue,
linkcolor=blue,
urlcolor=blue,
}

\begin{document}

\title{Vibrationally-mediated Dzyaloshinskii-Moriya interaction \\ as the origin of Chirality-Induced Spin Selectivity in donor-acceptor molecules}

\author{Alessandro Chiesa}
\email{alessandro.chiesa@unipr.it}
\affiliation{Dipartimento di Scienze Matematiche, Fisiche e Informatiche, Universit\`a  di Parma, Parco Area delle Scienze, 53/A, I-43124 Parma, Italy.}
\affiliation{Gruppo Collegato di Parma, INFN-Sezione Milano-Bicocca, I-43124 Parma, Italy.}
\affiliation{UdR Parma, INSTM, I-43124 Parma, Italy.}
\author{D. K. Andrea Phan Huu}
\thanks{These authors contributed equally to this work.}
\affiliation{Dipartimento di Scienze Matematiche, Fisiche e Informatiche, Universit\`a  di Parma, Parco Area delle Scienze, 53/A, I-43124 Parma, Italy.}
\author{Arianna Cantarella}
\thanks{These authors contributed equally to this work.}
\affiliation{Dipartimento di Scienze Matematiche, Fisiche e Informatiche, Universit\`a  di Parma, Parco Area delle Scienze, 53/A, I-43124 Parma, Italy.}
\affiliation{Gruppo Collegato di Parma, INFN-Sezione Milano-Bicocca, I-43124 Parma, Italy.}
\author{Leonardo Celada}
\affiliation{Dipartimento di Scienze Matematiche, Fisiche e Informatiche, Universit\`a  di Parma, Parco Area delle Scienze, 53/A, I-43124 Parma, Italy.}
\affiliation{Gruppo Collegato di Parma, INFN-Sezione Milano-Bicocca, I-43124 Parma, Italy.}
\author{Michael R. Wasielewski}
\affiliation{Department of Chemistry, Institute for Quantum Information Research
and Engineering, and Center for Molecular Quantum Transduction, Northwestern
University, Evanston, IL 60208-3113}
\author{Paolo Santini}
\affiliation{Dipartimento di Scienze 
Matematiche, Fisiche e Informatiche, Universit\`a  di Parma, Parco Area delle Scienze, 53/A, I-43124 Parma, Italy.}
\affiliation{Gruppo Collegato di Parma, INFN-Sezione Milano-Bicocca, I-43124 Parma, Italy.}
\affiliation{UdR Parma, INSTM, I-43124 Parma, Italy.}
\author{Stefano Carretta}
\email{stefano.carretta@unipr.it}
\affiliation{Dipartimento di Scienze Matematiche, Fisiche e Informatiche, Universit\`a  di Parma, Parco Area delle Scienze, 53/A, I-43124 Parma, Italy.}
\affiliation{Gruppo Collegato di Parma, INFN-Sezione Milano-Bicocca, I-43124 Parma, Italy.}
\affiliation{UdR Parma, INSTM, I-43124 Parma, Italy.}

\begin{abstract} 
\noindent
Chirality-induced spin selectivity (CISS) was recently observed in photo-excited donor-chiral bridge-acceptor molecules, but a predictive theory able to explain available experiments is still lacking. Here we show that low-energy torsional modes modulating hopping and spin-orbit coupling give rise to a Dzyaloshinskii-Moriya interaction between the transferred electron and the one sitting on the donor, producing high spin polarization for perfectly realistic parameters. 
Our model introduces a low energy scale in the spin dynamics which explains the magnetic field dependence observed in EPR measurements and predicts a non-trivial temperature dependence, as demonstrated by numerical simulations.
The present theory lays the foundations for future test-bed experiments and for the design of applications in spintronics and quantum technologies.
\end{abstract}

\maketitle

After many different observations in photo-emission, transport and polarization-on-surface experiments \cite{Naaman2019,Naaman2020b,Bloom2024,Kumar2026,Waldeck2026} Chirality-Induced Spin Selectivity (CISS) was recently evidenced also at the molecular level, in donor-acceptor molecules involving a chiral bridge ($D-\chi-A$) dispersed in solution \cite{Eckvahl2023,Eckvahl2024,Latawiec2025}.  
In these systems, fast electron transfer (ET) is induced by photo-excitation of either the electron donor or acceptor subunit. A sizable triplet component in the radical-pair state is observed by time-resolved electron paramagnetic resonance (EPR), completely absent in the achiral analogues. 

The important simplification of the experimental setup in ET, with the removal of complex interfaces and substrates, could be the key for a substantial step forward in our comprehension of the phenomenon \cite{Chiesa2021,Chiesa2025}. The fundamental dilemma for any theory of CISS  is how to reconcile the small spin-orbit coupling of organic chiral molecules (triggering the effect) with the large electronic energy gaps between the states involved in the electron motion \cite{evers2022}. 
This large energy mismatch prevents any single-electron model to yield a significant spin polarization \cite{Alwan2024,Fransson2025b}, unless very strong spin relaxation is assumed \cite{PhysRevB.111.205417,Matityahu2016PRB}. The transferred electron must therefore be involved in some form of interaction \cite{Fransson2025}.   
For instance, the interaction of the transferred electron with the one remaining on the donor was shown to give a spin polarization when combined with SOC \cite{Fay2021,Fay2021b}. A significant polarization was also demonstrated by including electron-electron correlations within the chiral bridge \cite{fransson_chirality-induced_2019,chiesa_many-body_2024} or low-energy vibrations coupled to electronic degrees of freedom \cite{Fransson2020,Fransson2021,das_temperature-dependent_2022,Subotnik2022,Cuniberti2023,Rudge2025,NostroJPCL,Fransson2026}. In both these latter cases, the effect of spin-orbit (SOC) is amplified, yielding a large spin polarization in a many-electron picture of the bridge. 
However, molecular systems undergoing photo-induced ET typically show large energy gaps between completely filled (HOMO) and empty (LUMO) bridge orbitals. An example is provided by PXX-NMI$_2$-NDI molecule \cite{Eckvahl2023}, on which we recently performed an extensive ab-initio study  \cite{phan_huu_ab_2025}. 
This makes electron-electron interactions practically ineffective and reduces the ET to a single-electron hopping process through the empty LUMOs. Nonetheless, experimental evidence of a strong magnetic-field dependence of CISS \cite{Latawiec2025} imply the existence of a low-energy scale involved in the effect.

We focus on low-energy Peierls vibrations, such as torsional modes typical of chiral molecules \cite{phan_huu_ab_2025,NitzanandSubo2025}, modulating both hopping and SOC \cite{Bissesar2022,dhali2021understanding,Poh2024,savi2025organic}. 
Here we show that they can give rise to an effective spin-spin interaction between the moving electron and the one sitting on the donor, much larger than that arising from a purely electronic model as in \cite{Fay2021b}. 
In particular, the interplay between modulation of hopping and SOC introduces a Dzyaloshinskii-Moriya interaction (DMI) of the same order of the isotropic exchange. By mixing singlet and triplet within the low-energy spin subspace, this DMI produces a large spin polarization and a triplet component in the charge-separated radical pair perfectly compatible with experiments.  



We demonstrate the key role of the vibrationally-mediated DMI for CISS in ET by numerically solving the system dynamics in a Redfield framework. We study the dependence of the resulting spin polarization on model parameters, finding large values in a realistic range of couplings. 
Our model explains the observed magnetic field dependence of the triplet component of the radical pair state probed by EPR experiments.
In spite of the low-energy scale introduced by spin-spin interactions, the effect is shown to be robust in temperature. Spin-spin interactions are enhanced by temperature 
and spin polarization has a non trivial temperature dependence which could be tested in future experiments. \\
The sizable spin polarization we predict (not limited to 50 \% for non-trivial models) will be the starting point to design applications to exploit CISS for quantum technologies. \\

\begin{figure}[t!]
    \centering
    \includegraphics[width=\linewidth]{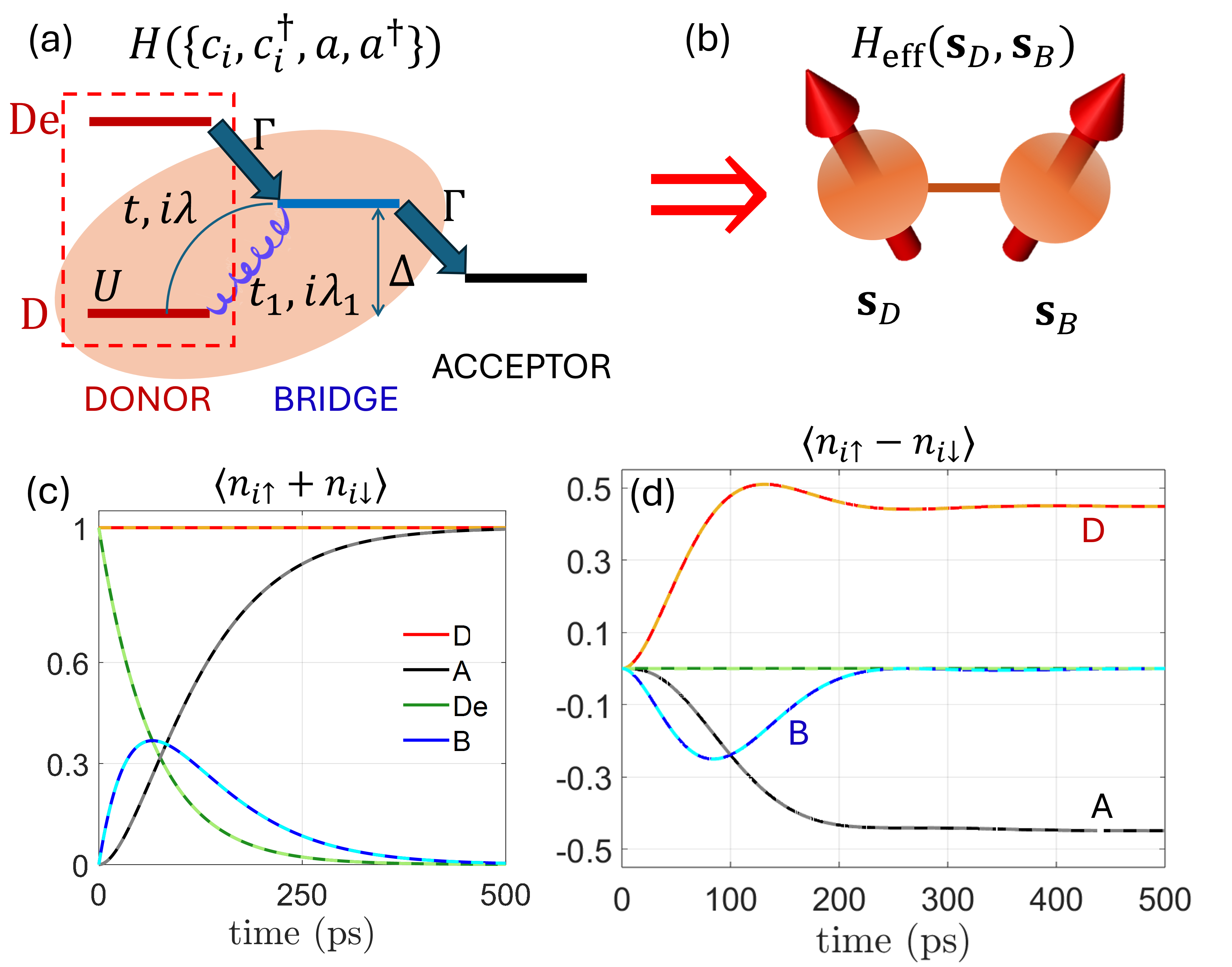}
    \caption{(a) Scheme of the minimal electron-transfer model with ground and excited orbitals on the donor (D, De), an intermediate orbital on the bridge (B) and one on the acceptor (A). The dynamics is ruled by incoherent spin-independent jumps from De to B and from B to A at rates $\Gamma$ and by a Hamiltonian $H$ including fermionic and bosonic degrees of freedom. (b) The Hamiltonian $H$ with a single electron localized on $D$ and one on $B$ is mapped onto an effective spin Hamiltonian involving spin operators ${\bf s}_D$ and ${\bf s}_B$. (c,d) Simulated time evolution of the charge $\langle n_{i\uparrow}+n_{i\downarrow}\rangle$ (c) and of the local spin polarization $2\langle s_{zi} \rangle=\langle n_{i\uparrow}-n_{i\downarrow}\rangle$ (d) on different orbitals. Solid lines: simulation with the full Hamiltonian $H$ and up to 9 bosons. Dashed: simulation with the effective Hamiltonian $H_{\rm eff}$. Parameters: $t=1$ meV, $\lambda=0.1$ meV, $U=3.5$ eV, $\Delta=5$ eV, $t_1=1$ meV, $\lambda_1=1$ meV, $\hbar\omega=2$ meV, $J_{CE} = -10^{-3}$ meV, $\Gamma=5\times 10^{-3}$ meV and we initialized the system with a fixed number of bosons $n=3$.}
    \label{fig:scheme}
\end{figure}

\noindent
{\bf LOW-ENERGY HAMILTONIAN}\\
Photo-induced ET in molecules displaying CISS \cite{Eckvahl2023,Eckvahl2024,Latawiec2025} can be described by a sequential incoherent hopping mechanism from the excited donor orbital (De) to the acceptor (A), via an intermediate orbital (B) on the bridge [as sketched in Fig. \ref{fig:scheme}-(a)]. During the process, the transferred electron interacts with the one sitting on the ground orbital of the donor (D). Such interaction arises both from a delocalization of the electronic wavefunction through hopping and SOC and from the corresponding modulations induced by low-energy vibrations.
The related dynamics just after photo-excitation can be described by incoherent spin-independent transfer rates $\Gamma$ from the De to B and from B to A, combined with a coherent evolution ruled by the Hamiltonian $H=H_0+H_1$, with
\begin{eqnarray} \nonumber
    H_0 &=& \Delta \sum_{\sigma=\uparrow,\downarrow} c^\dagger_{\rm B\sigma}c_{\rm B\sigma} + U \sum_{j= \rm D,B}n_{j\uparrow} n_{j\downarrow} \\ &+& \sum_\nu \hbar \omega_\nu a^\dagger_\nu a_\nu  \\ \nonumber
    H_1 &=& (t+i\lambda) c^\dagger_{\rm B\uparrow} c_{\rm D\uparrow} + (t-i\lambda) c^\dagger_{\rm B\downarrow} c_{\rm D\downarrow}  \\ 
    &+& \sum_\nu (a_\nu+a^\dagger_\nu)[(t_{1\nu}+i\lambda_{1\nu}) c^\dagger_{\rm B\uparrow} c_{\rm D\uparrow} \\ \nonumber
    && \;\;\;\;\,\,\, + (t_{1\nu}-i\lambda_{1\nu}) c^\dagger_{\rm B\downarrow} c_{\rm D\downarrow} ] + {\rm h.c.},  
    \label{eq:2orb}
\end{eqnarray}
where $c_{j\sigma}^\dagger$ ($c_{j\sigma}$) are fermionic creation (annihilation) operators of an electron with spin $\sigma$ either on the ground donor or on the bridge orbital ($j=\rm D,B$) and $n_{j\sigma} = c^\dagger_{j\sigma}c_{j\sigma}$; 
$a_\nu^\dagger$ ($a_\nu$) is a bosonic creation (annihilation) operator of a mode of frequency $\hbar\omega_\nu$. 
In the system Hamiltonian we have separated the leading terms diagonal in the occupation number basis (i.e. the energy gap $\Delta$, the on-site Coulomb repulsion $U$ and the vibration energy) from the much smaller spin independent hopping (of strength $t$) and SOC (parameterized by $\lambda$ and assumed axial for simplicity). In the perturbation term $H_1$ we have also included the coupling of the mode $\nu$ with hopping and SOC, of strength  $t_{1\nu}$ and $\lambda_{1\nu}$, respectively \footnote{The electronic system could also be coupled to Holstein modes modulating on-site energies, but these cannot mediate a spin-spin interaction and hence are not considered here}. \\
Since $t,\lambda, t_1, \lambda_1 \ll  U,\Delta$, charge is practically localized on D and B in the intermediate step of the ET between the two incoherent jumps and hence we can consider $H_1$ acting as a second-order perturbation on {\it spin-states} 
\begin{equation}
    \ket{\sigma\sigma'}_{\{n_\nu\}} \equiv 
    c^\dagger_{\rm D\sigma} c^\dagger_{\rm B\sigma'} \prod_{\nu} \frac{(a^\dagger)^{n_\nu}}{\sqrt{n_\nu!}} \, \ket{\emptyset}, \;\;\;\;\;\;\; \sigma,\sigma'=\uparrow,\downarrow  
    \label{eq:states}
\end{equation}
where $n_\nu$ is the number of bosons in mode $\nu$ and the bosons state is factorized from the electronic one.  
This treatment results in an effective low-energy spin Hamiltonian of the form
\begin{eqnarray} \nonumber
    H_{\rm eff} &=& J {\bf s}_1 \cdot {\bf s}_2 + J_D (2 s_{z1} s_{z2} -s_{x1} s_{x2} - s_{y1}s_{y2}) \\ 
    &+& D_z (s_{x1}s_{y2}-s_{y1}s_{x2})
    \label{eq:Hspin}
\end{eqnarray}
where the three contributions account for an isotropic, axial anisotropic and anti-symmetric (Dzyaloshinskii-Moriya) exchange terms (see Supporting Information). The values of the couplings are
\begin{subequations}
\label{eq:couplings}
    \begin{align} \label{eq:coupligs}
    J &= J_{CE} + \frac{2t^2-2\lambda^2/3}{\Delta'} + 2 \sum_\nu \left(t_{1\nu}^2-\frac{\lambda_{1\nu}^2}{3} \right) f(n_\nu) \\ 
    J_D &= \frac{4\lambda^2}{3\Delta'} + \sum_\nu \frac{4\lambda^2_{1\nu}}{3} f(n_\nu)  \\
    D_z &= -\frac{4\lambda t}{\Delta'} -\sum_\nu 4\lambda_{1\nu} t_{1\nu}f(n_\nu) .
    \end{align}
\end{subequations}
with $1/\Delta' = 1/(U-\Delta) + 1/(U+\Delta)$ and 
\begin{eqnarray} 
    f(n_\nu) &=& \frac{n_\nu+1}{U-\Delta+\hbar\omega_\nu} + \frac{n_\nu+1}{U+\Delta+\hbar\omega_\nu} \\ \nonumber &+& \frac{n_\nu}{U-\Delta-\hbar\omega_\nu} + \frac{n_\nu}{U+\Delta-\hbar\omega_\nu} \approx \frac{2n_\nu+1}{\Delta'} .
\end{eqnarray}
The last approximation holds because we are considering low-energy vibrations ($\hbar\omega \sim$ few meV), while $U,\Delta$ are typically several eV (see below).
Besides the second-order contributions discussed above, Eq. \eqref{eq:coupligs} includes a ferromagnetic direct exchange $J_{CE}$. \\
\begin{figure*}[ht!]
    \centering
    \includegraphics[width=1\linewidth]{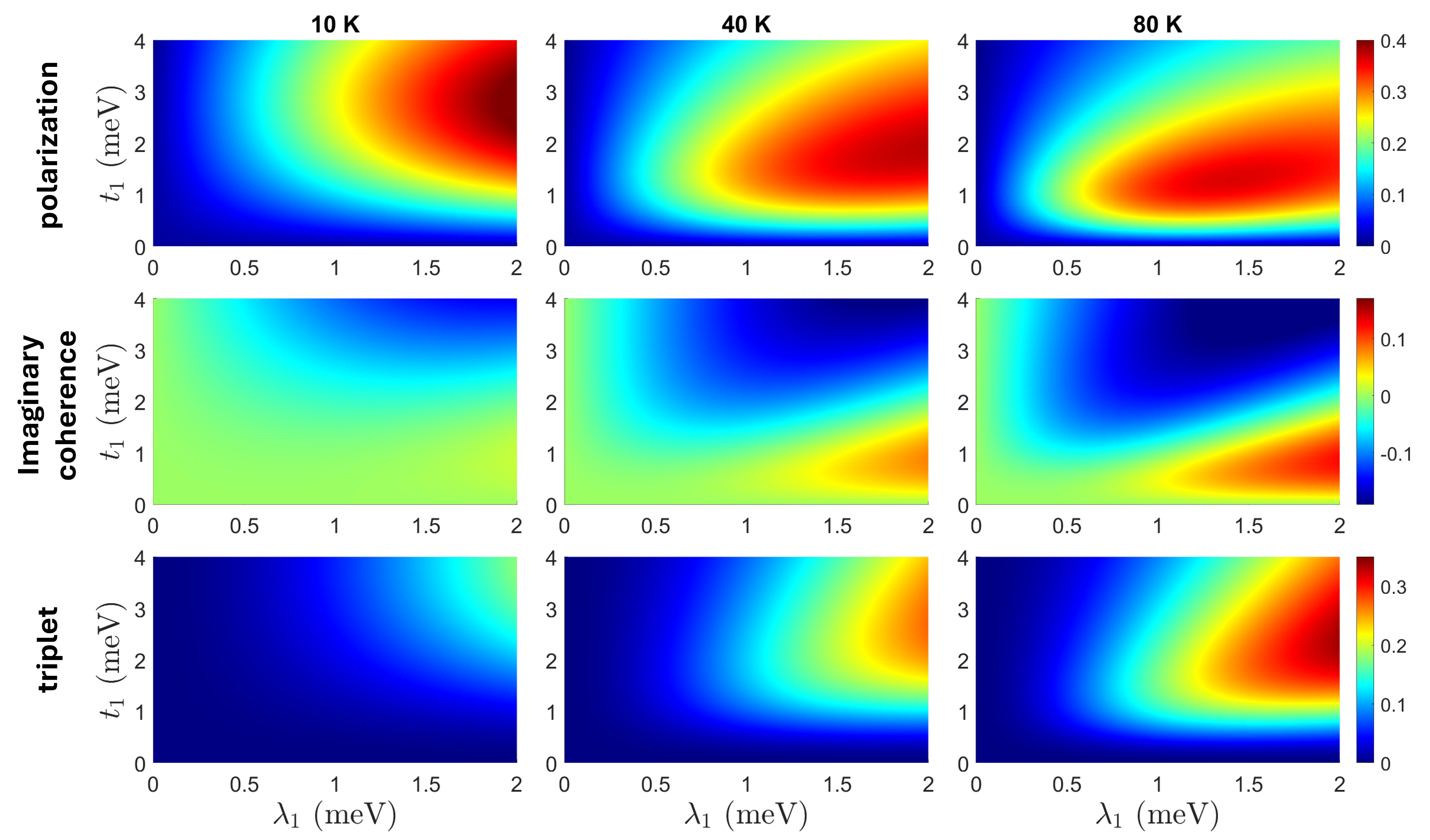}
    \caption{Top panels: spin polarization, corresponding to twice the real part of the singlet-triplet coherence. Middle panels: spin rotation about chiral axis, i.e. imaginary component of the singlet-triplet coherence. Bottom: triplet population. All the values are at the end of the ET, as a function of $\lambda_1$ and $t_1$. Other parameters of the simulation: $t=1$ meV, $\Gamma = 5\times10^{-3}$ meV (corresponding to ET time in the few hundreds of ps range $\hbar/\Gamma \approx 100$ ps), direct exchange contribution $J_{CE}=-10^{-3}$ meV, $\lambda=0.1$ meV, $U=3.5$ eV, $\Delta=5$ eV, $\hbar \omega_0 = 2$ meV. 
    Simulations with larger $t_1$ and $\Gamma$ are reported in the Supporting Information.}
    \label{fig:polmaps}
\end{figure*}
Some remarks are now in order. 
First, the leading contribution to the spin-spin interactions come from vibrationally-modulated terms. Indeed, (i) different modes provide additional contributions; (ii) the factor $f(n_\nu)$ gives an enhancement which becomes increasingly important with temperature and (iii) we can expect $t_1$ and $\lambda_1$ of the same order but often larger than $t$ and $\lambda$ \cite{Marian2018,phan_huu_ab_2025,manian2025vibronic}. Indeed, Peierls coupling in systems showing torsional flexibility can yield a strong modulation of hopping and SOC with $\lambda_1$ even significantly larger than $\lambda$ \cite{dhali2021understanding,phan2022thermally,Poh2024}. \\
Second, since $U<\Delta$ (as required to get a stable state before photo-excitation) and $t>\lambda$, the static contribution to $J$ is ferromagnetic (negative), exactly  as $J_{CE}$. \\
Third, only axial components appear in Hamiltonian \eqref{eq:Hspin}, because we started for sake of simplicity from an axially symmetric Hamiltonian \eqref{eq:2orb}. A more complex form of the SOC would lead to other components of the DMI in $H_{\rm eff}$, but this would not qualitatively alter our conclusions \footnote{In principle, any system where inversion symmetry is broken could display the DMI considered in the present model.}. \\
Finally, it is worth noting that second-order perturbation theory provides here a very good approximation of the exact dynamics, even for strongly coupled modes ($t_1, \lambda_1 > \hbar \omega$), as far as $t_1, \lambda_1 \ll U-\Delta$, i.e. the exact eigenstates of $H$ are very close to the factorized states in Eq. \eqref{eq:states} (see Supplementary Information for details).  
Hence, we can trace out the vibrational degrees of freedom and consider the dynamics ruled by $H_{\rm eff}$.  \\

\noindent
{\bf LARGE SPIN POLARIZATION}\label{largespinpol}\\
We simulate the ET dynamics by numerically integrating the Redfield equation with Hamiltonian $H_0+H_{\rm eff}$ and spin-independent jump operators $\sqrt{\Gamma} \sum_\sigma c^\dagger_{B\sigma}c_{De\sigma}$ and $\sqrt{\Gamma} \sum_\sigma c^\dagger_{A\sigma}c_{B\sigma}$. We have checked that the inclusion of a spin-dependent term (due to modulation of the SOC) in the master equation does not practically affect our results (see Supplementary Information). For computational reasons we include a single effective boson mode, keeping in mind that in real systems the couplings will be enhanced by the sum on several modes in Eq. \eqref{eq:couplings}. 
An example of the computed time evolution of charge and spin polarization along the chiral axis is reported in Fig. \ref{fig:scheme}-(c,d) for a specific number of vibrations in the mode, $n=3$. 
Panel (d) also illustrates the agreement between the perturbative and exact dynamics (dashed vs solid lines). \\

The origin of the spin polarization is the vibrationally-mediated DMI, which mixes singlet (S) and triplet (T). Starting from a singlet state just after the spin independent transfer from De, this in general yields both a real and an imaginary coherence in the S-T basis, the former of which corresponds to a polarization of the transferred spin. 
Remarkably, for realistic choice of the parameters the spin polarization accumulates on A, because it undergoes coherent oscillations with angular frequency $\sqrt{(J+J_D)^2+D_z^2}$ comparable to the ET rates. The restriction of the relevant dynamics to the low-energy spin subspace (in which $J$ and $D_z$ are comparable) is the key to achieve a high spin polarization. 

To provide a realistic description of typical molecules displaying CISS (without restricting to a specific one), we derive the parameters $U, \Delta, t,\lambda$
from ab-initio calculations \cite{FMO,orca,qchem} on PXX-NMI$_2$-NDI and we perform numerical simulations as a function of $t_1$ and $\lambda_1$ 
(see Supplementary Section \ref{ab-initio} for further details) \footnote{Since the orientation of the spin-orbit coupling vector depends on the details of the molecular structure, we consider for $\lambda$ in our minimal axial model the magnitude of the spin-orbit coupling vector, as shown in the Supporting Information.}. In particular, in the following we use $U=3.5$ eV, $\Delta = 5$ eV, $t=1$ meV, $\lambda=0.1$ meV.  
We stress that these are typical values for $D-\chi-A$ systems and that their precise value does not significantly affect our results.  
We set $\Gamma=5\times 10^{-3}$ meV, which implies a time constant for the ET $\hbar/\Gamma\approx 100$ ps and an ET complete in a few hundreds of ps. We consider a thermal equilibrium state of the vibrational mode at different temperatures.\\
Computed results for the accumulated spin polarization, imaginary coherence and triplet component as a function of $t_1$ and $\lambda_1$ are reported in Fig. \ref{fig:polmaps}.
The first line of Fig. \ref{fig:polmaps} shows the spin polarization on the radical pair $P_z=\langle s_{zD}-s_{zA}\rangle$ after the end of the ET, at different temperatures. In the S-T basis and for the present axial model, $P_z$ corresponds to twice the real-part of the coherence between singlet ($\ket{S}$) and $\ket{T_0}$ states, i.e. $P_z = \langle\ket{S}\bra{T_0}+\ket{T_0}\bra{S}\rangle$. Here $\ket{T_0}$ is the $M=0$ component of the triplet, where the $z$ axis is set by the chirality. 
By raising the temperature more vibrational quanta are introduced and hence the values of $J, J_D$ and $D_z$ are increased. This leads to a corresponding growth in the oscillation frequency and hence, for fixed ET time, changes both the maximum value of $P_z$ and its position in $\{\lambda_1,t_1\}$. In particular, while at lower temperatures the ET time matches the maximum of the first oscillation in $P_z$ for $\lambda_1\approx t_1\approx 2$ meV, at higher temperatures the value of $P_z$ is already decreasing before the end of the ET. By reducing the ET time, the maximum of $P_z$ is reached at higher temperatures (see simulation with doubled $\Gamma$ in the  Supplementary Information).
It is worth noting that, in spite of the small energy scale of these coherent oscillations, $P_z$ is robust with temperature, because the initial photo-excited state is out-of-equilibrium and the vibrational state remains at thermal equilibrium for the whole dynamics. \\
In the second line of panels we show the imaginary component of the $\ket{S}-\ket{T_0}$ {\it coherence}, i.e. the expectation value $C$ of the two-spin operator $s_{xD} s_{yA} - s_{yD} s_{xA}$ on the radical pair state. We note that the maximum of $|C|$ increases with temperature and $C$ changes sign in correspondence of the change of sign of $J$, i.e. when $t_1$ becomes leading compared to $\lambda_1$. \\
The last line reports the triplet component $P_T$ in the final spin pair, i.e. the projection of the final state on $3/4 + {\bf s}_D \cdot {\bf s}_A$. In the present axial model, this corresponds to a projection on $\ket{T_0}\bra{T_0}$. We note that $P_T$ increases with $\lambda_1$ and has a maximum in $t_1$, because the mixing with the triplet component decreases at larger $t_1$. \\
Note that assuming a different form of the SOC (thus introducing other components of the DMI) does not alter the overall picture, but only the precise numerical values of the quantities shown in Fig. \ref{fig:polmaps}. \\


\begin{figure}[ht!]
    \centering
    \includegraphics[width=\linewidth]{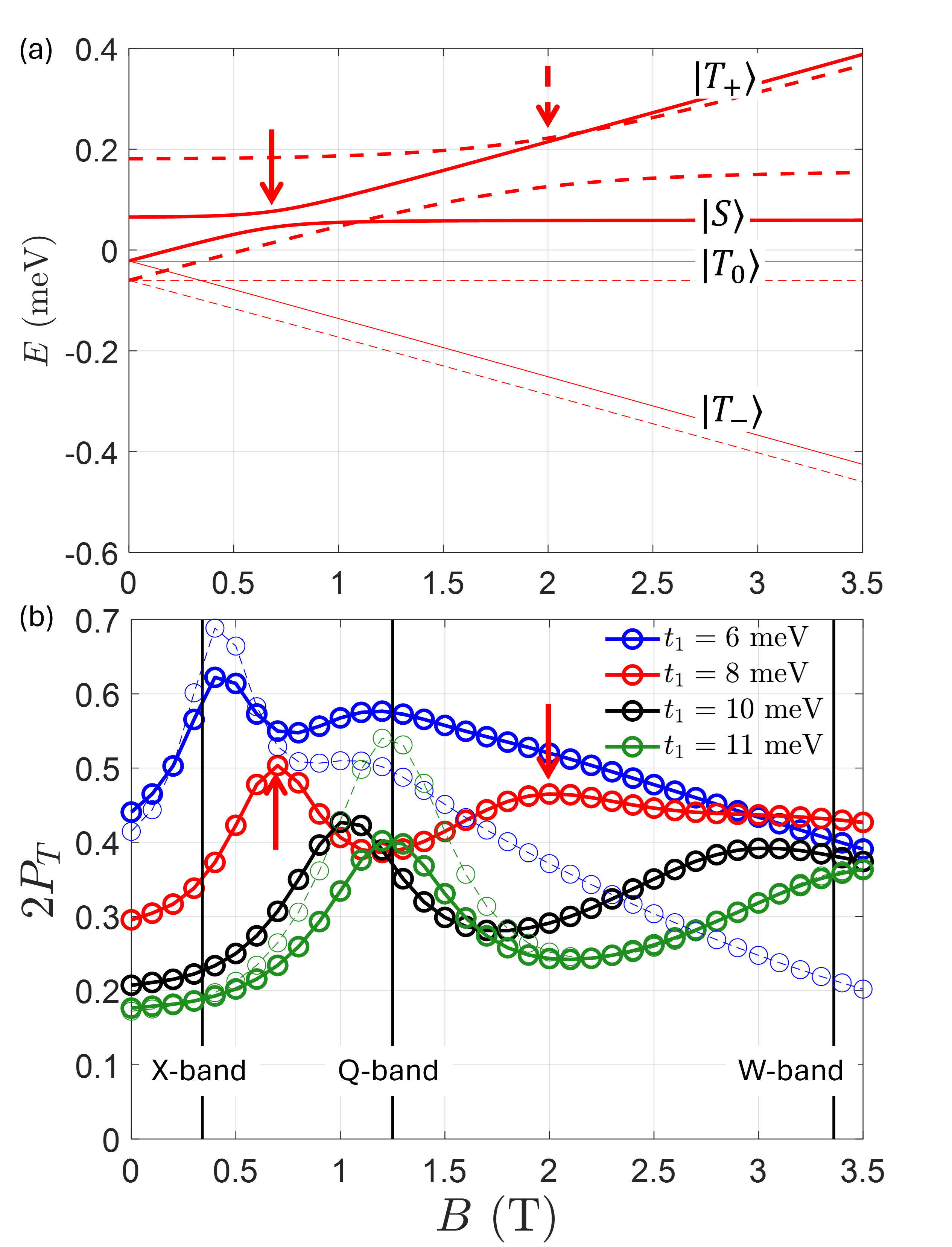}
    \caption{(a) Energy level diagram as a function of the magnetic field applied at $\theta=90^\circ$ with respect to the chiral anisotropy axis, using parameters $t_1=8$ meV, $\lambda_1=2.5$ meV. Solid (dashed) curves refer to  $n=0(1)$ boson, leading to avoided level crossing at different fields and of different width. (b) CISS efficiency at 80 K as usually obtained by fitting EPR spectra, corresponding to twice the triplet component $P_T$. Different solid lines refer to different $t_1$ as indicated in the legend, while dashed line of the same color are obtained by considering doubling the boson energy while keeping the temperature fixed ($\hbar \omega=2\rightarrow4$ meV).
    Dashed lines indicate the fields probed by EPR at X-, Q- and W-band.
    The other parameters are kept fixed to  $J_{CE}=-1\times10^{-2}$ meV, $t=1$ meV, $\lambda=0.1$ meV, $U=3.5$ eV, $\Delta=5$ eV.}
    \label{fig:field}
\end{figure}

\noindent
{\bf COMPARISON WITH EXPERIMENTS} \\
The triplet component of the radical pair is the quantity accessed by time-resolved electron paramagnetic resonance (TREPR) experiments, the experimental technique used so far to probe CISS in photo-induced ET \cite{Luo2021,Chiesa2021,Ren2023,Chiesa2025}.
By focusing on the low-energy spin dynamics, the present theory can reproduce experimental observations and in particular the non-trivial magnetic field dependence. 

We consider hereafter a situation with ${\bf B}$ perpendicular to the chiral anisotropy axis, because this is the orientation which is most sensitive to CISS effect in experiments performed either on isotropic solutions or on molecules dispersed in liquid crystals (as done so far). Liquid crystals orient the direction of chiral molecules but leave the two orientations equally likely. Therefore, we perform simulations on ensembles including both orientations of the molecules, but fixed direction.  \\ 
To gain insight into the magnetic field dependence it is useful to plot the energy level diagram of the two-spin system as a function of the applied field $B$. This is shown in Fig. \ref{fig:field}-(a) for a specific choice of parameters and fixed number of bosons $n=0$ (continuous lines) or $n=1$ (dashed). At zero field the three levels of the triplet remain practically degenerate ($J_D$ is very small) and are split by the isotropic exchange $J$ from the higher energy singlet. Then, for both solid and dashed lines, we note an avoided level crossing (AC) around 0.7 and 2.1 T, respectively, between $\ket{S}$ and the $M=1$ component of the triplet ($\ket{T_+}$) along the direction of the field. The width of the AC is determined by $D_z$ and hence increases with $n$ \footnote{In the present minimal axial model we obtain an AC for any $\theta \neq 0$. For non-axial SOC and hence DMI we will get ACs also for $\theta=0$}. \\
The corresponding CISS efficiency (represented by $2P_T$ according to the usual definition in TREPR) is reported in Fig. \ref{fig:field}-(b) (red line and circles). Peaks are visible at each AC of panel (a), due to the increased mixing between $\ket{S}$ and $\ket{T_+}$ when levels come close. $P_T$ also includes population of $\ket{T_-}$, which decreases due to the increasing Zeeman gap, while $\ket{T_0}$ is never populated in the present axial model at $\theta=90^\circ$ \footnote{In the liquid crystals alignment we are considering both $\ket{T_+}$ and $\ket{T_-}$ undergo an avoided level crossing for oppositely oriented sets of molecules in the solution.}.

Other curves in Fig. \ref{fig:field}-(b)  refer to different choices of the parameter $t_1$, leading to ACs and corresponding peaks in $2P_T$ at different magnetic fields and of different width.
By analyzing Fig. \ref{fig:field}-(b) we immediately note that CISS efficiency comparable with experiments (in the 30\%- 60\% range \cite{Eckvahl2023,Eckvahl2024,Latawiec2025}) can be achieved in a realistic parameter range. Here the values of $t_1$ are slightly higher than those employed in Fig. \ref{fig:polmaps}, but are still realistic since they account for the sum on several contributing modes (simulations for other parameter sets are reported in the Supporting Information). Moreover, depending on the parameters we can obtain different trends with magnetic fields. In particular, black and green curves show an efficiency doubled in going from X to Q band (vertical lines) and only slightly reduced from Q to W band, as observed in DNA hairpins reported in \cite{Latawiec2025}. Conversely, similar efficiency at the three probed bands are found in the red dataset, in substantial agreement with Ref. \cite{Eckvahl2024}. 
Besides changing $t_1$ and $\lambda_1$, a different ratio between CISS efficiencies can be obtained also by varying the frequency of the vibrational mode, while keeping the temperature fixed. A few examples are represented by the thin blue and green lines in Fig. \ref{fig:field}-(b), where the frequency of the mode is doubled from 2 to 4 meV, thus making the two maxima sharper and yielding a more pronounced decrease with the field for the blue curve. \\
In general, we expect a non trivial magnetic field dependence for most set of parameters which yield a sizable CISS effect at zero or low field. 
Indeed, for small $t_1$ and $J$, the AC will occur at very low field and thus the efficiency will decrease with $B$. For larger $t_1$ (leading to larger $J$ but still smaller than the Zeeman splitting in W band), at least one AC will occur at a higher magnetic field.   \\

\noindent
{\bf DISCUSSION and FUTURE EXPERIMENTS} \\
We have introduced a vibrationally-assisted mechanism explaining the observed triplet component in spin-correlated radical pairs generated by photo-induced electron transfer through a chiral bridge. Peierls vibrations give rise to an effective Dzyaloshinskii-Moriya interaction acting on the spin pair during the electron transfer and explaining several seemingly conflicting observations: besides the large triplet component and spin polarization (i), the coexistence of large electronic energy gaps, giving rise to a single-electron picture of the electron-transfer process (ii) and of a low-energy dynamics yielding 
the measured magnetic-field dependence of the CISS efficiency (iii). Finally, the predicted polarization is robust with temperature (iv) as observed in several experiments \cite{Bloom2024} and shows a non-trivial temperature dependence which could be tested in future experiments. Here, the only effect of temperature is to provide bosons to amplify the spin Hamiltonian couplings. 
In principle, one could also incorporate an explicit temperature dependence of the electron-transfer rates, for example within a Marcus-type formalism (as in Ref.~\cite{Fay2021b}). However, in the present work we deliberately separate these effects in order to isolate the role of vibrationally-induced spin interactions, and to avoid introducing additional assumptions on the bath spectral density.

\begin{figure}[ht!]
    \centering
    \includegraphics[width=1.\linewidth]{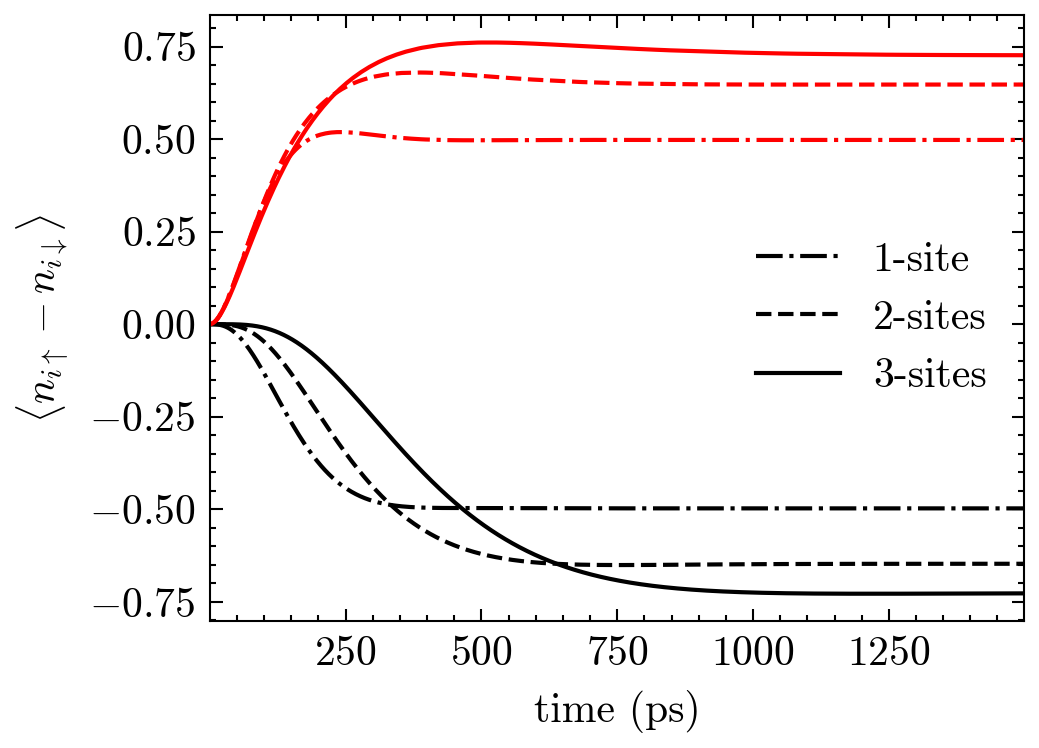}
    \caption{Evolution of the spin polarization on A (black) and D (red) for systems with one, two, and three sites on the bridge. Parameters are reported in Table~\ref{tab:dinamiche_over50incoherent} of the Supporting Information. Full population and spin polarization evolution is displayed in Fig.~\ref{fig:dinamiche_over50incoherent}
    }
    \label{fig:over50-main2}
\end{figure}

The interplay of energy scales between spin-spin coupling and Zeeman splitting leads to avoided level crossings in the spectrum which could be probed by tuning their position and width via the orientation of the molecules with respect to the magnetic field. Pulse EPR experiments could be employed to access the sensitivity to magnetic field fluctuations close to the avoided level crossings, which should yield maxima in the spin coherence times $T_2$ \cite{Hill2016}. \\

As a benchmark of the present theory, one could design experiments on molecules characterized either by very large {\it static} (i.e. not vibrationally assisted) $J$, where any mixing of the singlet with the triplet should be suppressed, or on molecules in which the donor is a radical, thus reducing the transfer to a single-electron model. In this case we only expect a very small singlet-triplet imaginary coherence \cite{Fay2021} and a triplet component of the order of $\lambda^2/t^2$. 
Molecules with chirality confined to the donor or to the acceptor could also provide an interesting benchmark of the theory: only the former should give a significant CISS effect. 
Finally, to distinguish polarization and singlet-triplet imaginary coherence from the triplet population and discriminate the role of the two enantiomers we would need instead preparation of sample with absolute orientation of the chiral molecules, as already discussed in \cite{Chiesa2025}. 

For applications in Quantum Technologies (such as high-temperature initialization of a spin qubit or quantum sensing \cite{Aiello2022,Chiesa2023,Chiesa2025}), an important point to be discussed is the maximum achievable spin polarization. It can be easily demonstrated (see Supplementary Section \ref{50perclimitandbeyond}) that the maximum spin polarization accumulated on A is limited to 50\% if we consider a monochromatic oscillating function on B transferred to A by an exponential decay rate. Note that this limit is dictated by the simple form of the sinusoidal oscillation on the bridge and hence it is not specific of the present model (see, e.g. \cite{Fay2021b}). 
In fact, several possibilities to overcome the 50\% polarization can be conceived, based on introducing additional harmonics in the coherent oscillations of the spin polarization on the bridge. The simplest option consists in increasing the length of the bridge by additional orbitals. 
However, it must be noted that non-vanishing hopping amplitudes lead to delocalized states across the bridge, which can be mapped onto the previous single-site description (see Supplementary Section \ref{50perclimitandbeyond}) \footnote{Indeed, only for very small intra-site hopping (on the same order of magnitude of the effective spin-spin couplings) the 50\% limit on spin polarization is surpassed}.
A more realistic scenario is to consider a multistep incoherent hopping mechanism between multiple bridge sites coupled to D by exchange and DMI. In this case, large spin polarization can be achieved, and can be systematically enhanced by adding more sites, as shown in Fig.~\ref{fig:over50-main2} and detailed in the Supporting Information.

In summary, the mechanism introduced here provides a physically transparent and robust route to spin polarization in photo-induced electron transfer through chiral systems. It explains the triplet component observed in EPR experiments and its magnetic filed dependence, it offers several experimentally testable predictions and suggests synthetic strategies towards efficient CISS for applications in Quantum Technologies.


\section*{Acknowledgments}
\noindent
We warmly thank M. Mezzadri for developing a tool used in numerical simulations. 
The work was funded by the Horizon Europe Programme within the ERC-Synergy project CASTLE (proj. n. 101071533). Views and
opinions expressed are however those of the author(s) only
and do not necessarily reflect those of the European Union or
the European Commission. Neither the European Union nor
the granting authority can be held responsible for them. 

%

\clearpage
\onecolumngrid

\begin{center}
    {\Large \textbf{Supplementary Material}}\\[1em]
    {\Large \textbf{Vibrationally-mediated Dzyaloshinskii-Moriya interaction \\ as the origin of Chirality-Induced Spin Selectivity in donor-acceptor molecules}}
\end{center}

\setcounter{section}{0}
\renewcommand{\theequation}{S\arabic{equation}}
\renewcommand{\thesection}{S\arabic{section}}
\renewcommand{\thefigure}{S\arabic{figure}}
\renewcommand{\thetable}{S\arabic{table}}
\setcounter{figure}{0}
\setcounter{table}{0}
\setcounter{equation}{0}

\section{Derivation of the spin-spin interactions}
Hereafter we derive the effective Hamiltonian $H_{\rm eff}$ of the main text. For sake of clarity, we start by considering only the electronic contribution and then we add the coupling to vibrations.

\subsection{Electronic super-exchange}
The {\it electronic} Hamiltonian describing a pair of electrons on the two orbitals D and $j$ can be separated into a leading term $H_0$ and a perturbation $H_1$ given by 
\begin{subequations}
\label{eq:HamEl}
\begin{align}
    H &= H_0 + H_1 \\
    H_0 &= U_D n_{D\uparrow} n_{D\downarrow} + U_j n_{j\uparrow} n_{j\downarrow} + \Delta (n_{j\uparrow} + n_{j \downarrow}) \\
    H_1 &= -t \sum_\sigma (c^\dagger_{j\sigma} c_{D\sigma} + {\rm h.c.}) + i \lambda (c^\dagger_{j\uparrow} c_{D\uparrow}-c^\dagger_{j\downarrow} c_{D\downarrow} - {\rm h.c.}) \\ \nonumber  &= (-t+i\lambda) c^\dagger_{j\uparrow} c_{D\uparrow} + (-t-i\lambda) c^\dagger_{j\downarrow} c_{D\downarrow} + {\rm h.c.} \\ \nonumber &= 
    \Lambda \left[ e^{-i\varphi} c^\dagger_{j\uparrow} c_{D\uparrow} + e^{i\varphi} c^\dagger_{j\downarrow} c_{D\downarrow}  +  e^{i\varphi} c^\dagger_{D\uparrow} c_{j\uparrow} + e^{-i\varphi} c^\dagger_{D\downarrow} c_{j\downarrow} \right]
\end{align}
\end{subequations}
where $\Lambda = \sqrt{t^2 + \lambda^2}$, $\varphi = \arccos t/\Lambda = \arcsin \lambda/\Lambda = \arctan \lambda/t$.  We have included for simplicity only the $z$ component of the spin orbit coupling, but this assumption does not qualitatively affects our conclusions. To be more general, we have considering a generic site $j$ on the bridge, coupled to the donor ground orbital. Although in principle the Coulomb interaction can vary from site to site, we simplify the model by taking $U_j=U_D=U$. \\

To proceed, we can either eliminate $H_1$ to first order by a proper Schrieffer-Wolf transformation $H_{\rm eff} = e^S H e^{-S}$ or apply second-order perturbation theory for $t,\lambda \ll U_D, \Delta$. The low energy manifold consists of the four localized spin states $c^\dagger_{D\sigma} c^\dagger_{j \sigma'} \ket{\emptyset} \equiv \ket{\sigma \sigma'}$, $\{{\sigma,\sigma'=\uparrow,\downarrow}\}$. The effective Hamiltonian in this low-energy subspace becomes
\begin{equation}
\bra{\sigma \sigma'} H_{\rm eff} \ket{\sigma'' \sigma'''} = -\sum_k \frac{\bra{\sigma \sigma'} H_1 \ket{k} \bra{k} H_1 \ket{\sigma'' \sigma'''}}{E_k},
    \label{eq:Heff}
\end{equation}
where we have shifted to zero the energy of the low energy manifold and the excited states are $c^\dagger_{D\uparrow} c^\dagger_{D \downarrow}\ket{\emptyset}$ and $c^\dagger_{j\uparrow} c^\dagger_{j \downarrow}\ket{\emptyset}$, at energies $U-\Delta$ and $U+\Delta$. 
We find the following matrix form for $H_{\rm eff}$
\begin{equation}
    \begin{pmatrix}
        0 & 0 & 0 & 0 \\
        0 & -\frac{t^2+\lambda^2}{\Delta'} & \frac{t^2-\lambda^2-2i\lambda t}{\Delta'} & 0 \\
        0 & \frac{t^2-\lambda^2+2i\lambda t}{\Delta'} & -\frac{t^2+\lambda^2}{\Delta'} & 0 \\
        0 & 0 & 0 & 0
    \end{pmatrix}
\end{equation}
with $1/\Delta' = 1/(U-\Delta) + 1/(U+\Delta)$. This effective spin Hamiltonian can be recast in the form
\begin{equation}
    H_{\rm spin} = J {\bf s}_1 \cdot {\bf s}_2 + J_D (2 s^z_1 s^z_2 -s^x_1 s^x_2 - s_1^ys_2^y) + D_z (s^x_1s^y_2-s_1^ys_2^x)
    \label{eq:Hspin2}
\end{equation}
where the three contributions account for an isotropic, axial anisotropic and anti-symmetric exchange terms. Only the $q=0$ components appear in Hamiltonian \eqref{eq:Hspin2} because we started from an axially symmetric Hamiltonian \eqref{eq:HamEl}. The values of the couplings are
\begin{subequations}
    \begin{align}
    J &= \frac{2t^2-2\lambda^2/3}{\Delta'}  \\
    J_D &= \frac{4\lambda^2}{3\Delta'}  \\
    D_z &= \frac{4\lambda t}{\Delta'} .
    \end{align}
\end{subequations}


\subsection{Peierls vibrations}
We now consider as a perturbation on $H_0$ coupling of the fermionic system with Peierls vibrations modulating both hopping and spin-orbit interactions, i.e. an Hamiltonian term of the form
\begin{equation}
H_{1P} = \sum_\nu (a_\nu+a^\dagger_\nu)[(t_{1\nu}+i\lambda_{1\nu}) c^\dagger_{\rm j\uparrow} c_{\rm D\uparrow} 
     + (t_{1\nu}-i\lambda_{1\nu}) c^\dagger_{\rm j\downarrow} c_{\rm D\downarrow} ] + {\rm h.c.}, 
    \label{eq:Peierls}
\end{equation}
where we are considering several vibrational modes $\nu$ of energy $\hbar \omega_\nu$ coupled to the same fermionic terms. Note that a diagonal term in the number of bosons $\sum_\nu \hbar\omega_\nu a^\dagger_\nu a_\nu$ must be added to $H_0$. 

We then consider the effect of $H_{1P}$ by second-order perturbation theory analogously to the previous section ($t_1,\lambda_1 \ll \Delta-U$) and we derive an effective spin-spin interaction mediated by Peierls modes. The result an effective Hamiltonian of the same form of Eq. \eqref{eq:Hspin2}, with couplings:
\begin{subequations}
    \begin{align}
    J^P &= 2 \sum_\nu \left(t_{1\nu}^2-\frac{\lambda_{1\nu}^2}{3} \right) f(n_\nu) \approx 2\sum_\nu\left(t_{1\nu}^2-\frac{\lambda_{1\nu}^2}{3} \right)\frac{2n_\nu+1}{\Delta'} \\
    J_D^P &= \sum_\nu\frac{4\lambda_{1\nu}^2}{3} f(n_\nu) \approx  \sum_\nu\frac{4\lambda_{1\nu}^2}{3\Delta'} \; (2n_\nu+1) \\
    D_z^P &= -4\sum_\nu\lambda_{1\nu} t_{1\nu} f(n_\nu) \approx -\sum_\nu \frac{4\lambda_{1\nu} t_{1\nu}}{\Delta'}\; (2n_\nu+1).
    \end{align}
    \label{eq:JPeierls}
\end{subequations}
with 
\begin{equation}
    f(n_\nu)= \frac{n_\nu+1}{U_D-\Delta+\hbar\omega_\nu} + \frac{n_\nu+1}{U_j+\Delta+\hbar\omega_\nu} + \frac{n}{U_D-\Delta-\hbar\omega_\nu} + \frac{n_\nu}{U_j+\Delta-\hbar\omega_\nu} \approx \frac{2n_\nu+1}{\Delta'} 
\end{equation}

and $n_\nu$ is the number of bosons of energy $\hbar\omega_\nu$.
Note that since $|\Delta'|\gg \hbar\omega_\nu$ the result does not depend significantly on $\hbar\omega_\nu$. \\

Remarkably, the perturbative expansion converges even for $t_{1\nu},\lambda_{1\nu}>\hbar \omega_\nu$, as long as 
\begin{equation}
\frac{t_{1\nu}}{\Delta'} \frac{t_{1\nu}}{2\hbar\omega_\nu} n_\nu \ll 1.
\end{equation}
Since $t_{1\nu}/\Delta' \sim 10^{-2}-10^{-3}$, we we can safely investigate regimes with $t_{1\nu},\lambda_{1\nu} \sim 10\hbar\omega$ using the effective Hamiltonian derived above. Indeed, all odd orders are zero and the subsequent (fourth order) correction is $\sim t_{1\nu}^2 n_\nu/\Delta'\hbar\omega_\nu$ smaller than the second order one.
This conclusion is supported by numerical simulations in Fig. 1 of the main text, showing very good agreement between simulations performed with $H$ and $H_{\rm eff}$.

Note that in principle also mixed terms in $H_1$ and $H_{1P}$ could give a second-order correction to the spectrum of $H_0$. However, these terms mix states with different number of bosons with off-diagonal terms of the order of $J^P$, $J^P_D$, $D_z^P$. Since these terms are $\ll \hbar \omega_\nu$, they do not impact significantly the dynamics (their effect being of the order of the difference between the dashed and solid lines in Fig. 1 c,d).

\subsection{Summary of contributions to $H_{\rm eff}$}
In summary, we have three contributions to the exchange: direct Coulomb ($J_{CE}$), electronic super-exchange and vibrationally mediated. The three contribution to $J$ are typically all ferromagnetic (FM) due to the negative sign of $\Delta'$ (the vibrationally mediated term is FM for $t_1 \ge \lambda_1/\sqrt{3}$). 

As discussed in the main text, the vibrationally-mediated contributions are the leading ones,  because of the large values that $t_{1\nu}$ and $\lambda_{1\nu}$ can assume in chiral molecules, of the sum on different modes and of the  $(2n_\nu+1)$ factor which becomes relevant especially for modes at low energies (a few meV) compared to $k_BT$.

For computational reasons we can include only a single vibrational mode in our simulations, with the caveat that the assumed value of $\lambda_1$ and $t_1$ are {\it effective couplings} which take into account contributions  from several modes. 

Therefore, we obtain the following overall expressions for the spin Hamiltonian parameters

\begin{subequations}
    \begin{align}
    J &\approx  \frac{2}{\Delta'} \left[ t^2+(t_1^2-\lambda_1^2/3) (2n+1) \right]  +J_{CE}   \\
    D_z & \approx \frac{4}{\Delta'} \left[\lambda t + \lambda_1 t_1 (2n+1) \right] \\
    J_D & \approx \frac{4\lambda_1^2}{3\Delta'} (2n+1) ,
    \end{align}
    \label{eq:Jtot}
\end{subequations}
where we have dropped the index $\nu$ and considered only a single coupled mode. 

\section{Redfield master equation}

To describe the electron-transfer (ET) dynamics, we consider an interaction between the system and the bath of the form 
\begin{equation}
    H_{SB} =  \sum_r \sum_{\nu = D,A} \kappa_{r,\nu} (X_\nu + X_\nu^\dagger) \; (a_{r,\nu}+a_{r,\nu}^\dagger) 
    \label{eq:Hsb}
\end{equation}
with the operators $X_D = \sum_\sigma c^\dagger_{1\sigma} c_{D \sigma}$ and $X_A = \sum_\sigma c^\dagger_{A\sigma} c_{i=4, \sigma}$ inducing electron 
hopping from the donor excited orbital onto the bridge or from the bridge to the acceptor, respectively. 
In Section~\ref{sec:incoherent_soc} below, we also account for a SOC contribution to the incoherent transfer from donor to bridge (see \eqref{eq:incoherent_soc_samebath} for the respective system-bath coupling operators).

Here $a_{r,\nu}$ is the bosonic annihilation operator for the $r$-th mode of the bath coupled with a strength $\kappa_{r,\nu}$ to either $X_D$ or $X_A$. $X_\nu$ are rank-0 fermionic operators which do not affect the spin of the transferred electron. For simplicity, we do not include further coupling terms between the system and the bath.  \\
We consider temperatures much smaller than the energy gaps driving ET and we describe the time evolution of the system density matrix $\rho$ by the Redfield equation \cite{Tupkary2022}:
\begin{equation} 
\label{eq:Redfield}
 \hbar \frac{d \rho}{d \tau} = -i [H,\rho] 
+  \sum_{\xi=D,A} \Gamma_\xi \left( Y_\xi \rho X_\xi^\dagger  - X_\xi^\dagger Y_\xi \rho  + {\rm h.c.} \right).
\end{equation}
The first term of Eq.~\eqref{eq:Redfield} describes the coherent evolution induced by the Hamiltonian $H$, while $Y_\xi = \sum_{\mu,\nu} \ket{\psi_\mu} \bra{\psi_\nu} \bra{\psi_\mu} X_\xi \ket{\psi_\nu} D_{\mu,\nu}$ and $D_{\mu,\nu}$ are proportional to the bath spectral function and to the Bose-Einstein factor at the energy gap $E_\nu-E_\mu$. In the low-temperature and wide-band limits considered hereafter $D_{\mu,\nu} = \Theta(E_\nu-E_\mu)$. Finally, $\Gamma_\xi$ are the system-bath coupling strengths (or friction coefficients). Unless otherwise noted, we set $\Gamma_D=\Gamma_A=\Gamma$.

\subsection{Spin-orbit assisted incoherent transfer}\label{sec:incoherent_soc}
\begin{figure}[ht]
    \centering
    \includegraphics[width=0.32\linewidth]{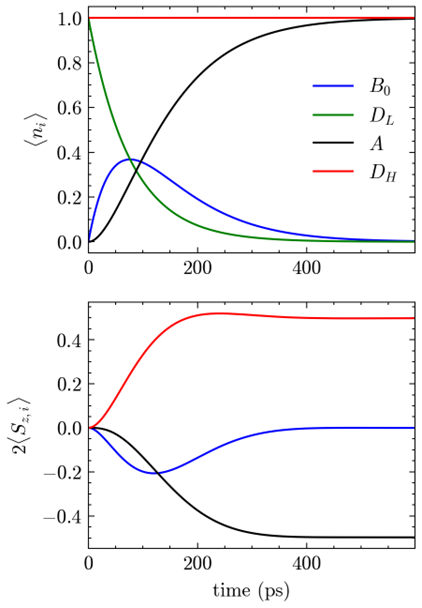}
    \includegraphics[width=0.32\linewidth]{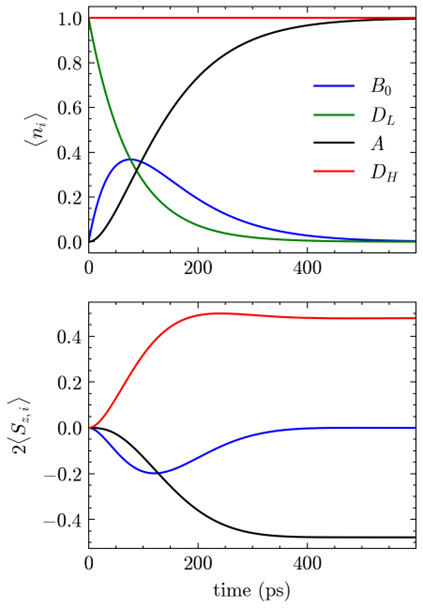}
    \includegraphics[width=0.32\linewidth]{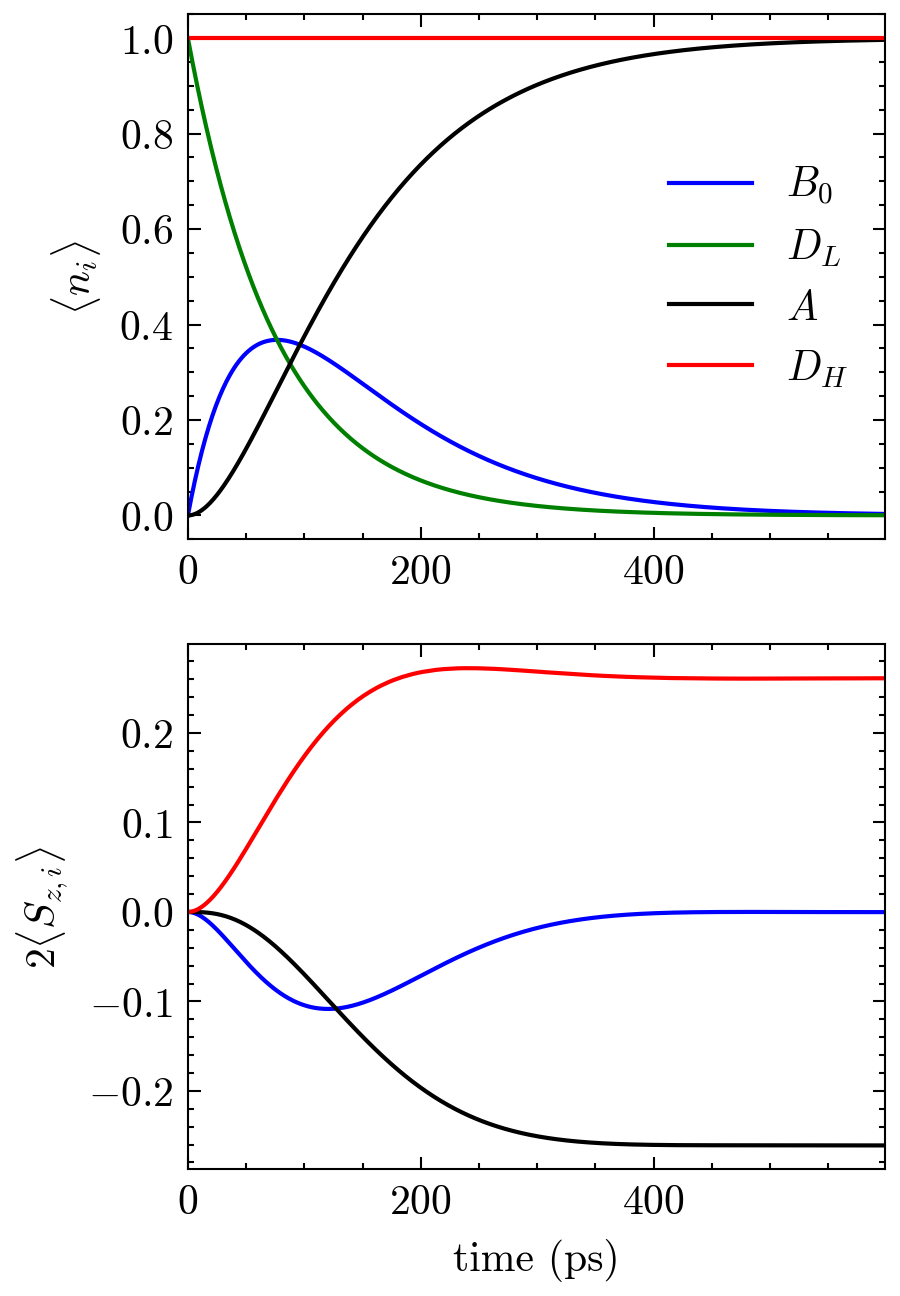}
    \caption{Population and spin polarization evolution accounting for spin-orbit assisted incoherent transfer from donor to bridge, according to approach 1. Parameters are set as describe in Section~\ref{sec:incoherent_soc}. The transfer rates are set as $\Gamma_{D,tot}=\Gamma_A=\SI{4.3e-3}{meV}$, and different columns refer to different values of $\theta$. Left: $\theta=0$. Center: $\theta=\arctan(0.1)$. Right: $\theta=\arctan(0.5)$.}
    \label{fig:soc_incoherent_samebath}
\end{figure}

\begin{figure}[ht]
    \centering
    \includegraphics[width=0.32\linewidth]{dinamica_standard.png}
    \includegraphics[width=0.32\linewidth]{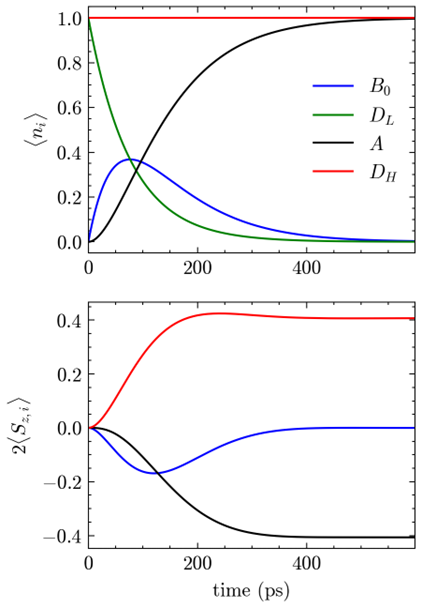}
    \includegraphics[width=0.32\linewidth]{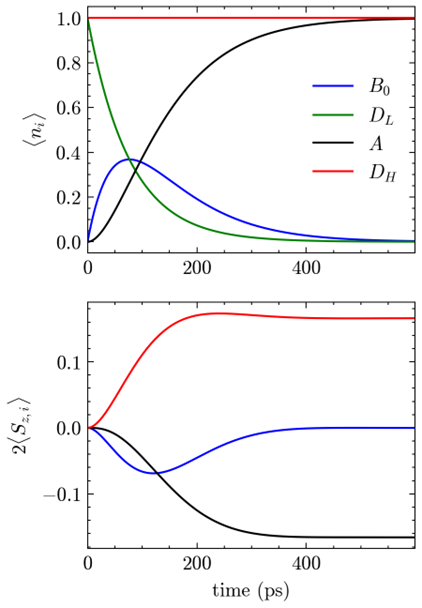}
    \caption{Population and spin polarization evolution accounting for spin-orbit assisted incoherent transfer from donor to bridge, according to approach 2. Parameters are set as described in Section~\ref{sec:incoherent_soc}. The transfer rates are set as $\Gamma_{D,tot}=\Gamma_A=\SI{4.3e-3}{meV}$, and different columns refer to different $\Gamma_{\tilde{D}}/\Gamma_{D}$ ratios.  Left: $\Gamma_{D,tot}=\Gamma_{D}$. Center: $\Gamma_{\tilde{D}}/\Gamma_{D}=1/10$. Right: $\Gamma_{\tilde{D}}/\Gamma_{D}=1/2$.}
    \label{fig:soc_incoherent_diffbath}
\end{figure}

Here we address the effect of SOC in the incoherent transfer from donor to bridge. Numerical simulations (Fig.~\ref{fig:soc_incoherent_samebath} and Fig.~\ref{fig:soc_incoherent_diffbath}) are performed with the set of parameters reported in Table~\ref{tab:microscopic-parameters}.

Spin-orbit assisted incoherent transfer can be introduced in two ways
\begin{enumerate}
    \item Hopping and SOC are both coupled to the same bath. In this case, in order to keep the overall transfer rate equal, the system-bath coupling operator is given by
    \begin{equation}\label{eq:incoherent_soc_samebath}
        \hat{X}_{D} = \cos{\theta}\sum_\sigma c_{1\sigma}^\dag c_{D\sigma} + i\sin{\theta}\sum_{\sigma,\sigma'}c_{1\sigma}^\dag\sigma^z_{\sigma\sigma'} c_{D\sigma'}
    \end{equation}
    \item Hopping and SOC are coupled to different baths. In this case, the system bath coupling Hamiltonian \eqref{eq:Hsb} becomes
    \begin{equation}\label{eq:Hsb_soc}
            H_{SB} =  \sum_r \sum_{\nu = D,\Tilde{D},A} \kappa_{r,\nu} (X_\nu + X_\nu^\dagger) \; (a_{r,\nu}+a_{r,\nu}^\dagger) 
    \end{equation}
    with 
    \begin{equation}
        \hat{X}_{\tilde{D}} = i\sum_{\sigma,\sigma'}c_{1\sigma}^\dag\sigma^z_{\sigma\sigma'} c_{D\sigma'}
    \end{equation}
    The master equation \ref{eq:Redfield} becomes 
    \begin{equation} 
    \label{eq:Redfield2}
    \hbar \frac{d \rho}{d \tau} = -i [H,\rho] 
    +  \sum_{\xi=D,\tilde{D},A} \Gamma_\xi \left( Y_\xi \rho X_\xi^\dagger  - X_\xi^\dagger Y_\xi \rho  + {\rm h.c.} \right).
    \end{equation}

    to keep the overall transfer rate equal we define the total transfer rate from donor to acceptor as 
    \begin{equation}
        \Gamma_{D,tot} = \Gamma_{D} + \Gamma_{\tilde{D}}
    \end{equation}
\end{enumerate}

The inclusion of a small incoherent SOC contribution for the donor-to-bridge transfer following approach (1) has negligible effects on ET dynamics and spin polarization (see left and central panels of Fig.~\ref{fig:soc_incoherent_samebath}). For large (and arguably unrealistic) SOC, the spin polarization is reduced as displayed in the right panel of Fig.~\ref{fig:soc_incoherent_samebath}, where a 2:1 ratio between hopping and SOC is imposed.

Following approach (2) we obtain qualitatively similar results. When the incoherent transfer from donor to bridge is partially due to SOC, the spin polarization accumulated on the acceptor is reduced. This effect is small, albeit non-negligible, when $\Gamma_{\tilde{D}}/\Gamma_{D}=1/10$) and more pronounced when the SOC contribution is large, as displayed in Fig.~\ref{fig:soc_incoherent_diffbath}. 



\clearpage
\newpage

\begin{table}[htbp]
\centering
\caption{Microscopic parameters and the resulting effective spin-coupling parameters for Fig.~\ref{fig:soc_incoherent_samebath} and Fig.~\ref{fig:soc_incoherent_diffbath}. }
\label{tab:microscopic-parameters}

\begin{minipage}[t]{0.48\columnwidth}
\centering
\vspace{0pt}
\begin{tabular*}{\linewidth}{@{\extracolsep{\fill}}lc@{}}
\toprule
\multicolumn{2}{c}{\textbf{Microscopic parameters (meV)}} \\
\midrule
$t$            & 1.0               \\
$t_1$          & 1.8               \\
$\lambda$      & 0.1               \\
$\lambda_1$    & 2.1               \\
$U$            & 3500              \\
$\Delta$       & 5000              \\
$\hbar\omega$  & 2.0               \\
$J_{CE}$       & 0.001             \\
\midrule
$n_{ph}$       & 0                 \\
\bottomrule
\end{tabular*}
\end{minipage}
\hfill
\begin{minipage}[t]{0.48\columnwidth}
\centering
\vspace{0pt}
\begin{tabular*}{\linewidth}{@{\extracolsep{\fill}}lr@{}}
\toprule
\multicolumn{2}{c}{\textbf{Effective spin-spin parameters (meV)}} \\
\midrule
$J$     & $-4.04 \cdot 10^{-3}$ \\
$J_D$   & $-3.24 \cdot 10^{-3}$ \\
$D_z$   & $-8.53 \cdot 10^{-3}$ \\
\bottomrule
\end{tabular*}
\end{minipage}

\end{table}
\clearpage
\newpage

\section{Dependence on the modulated parameters and ET rate}

\begin{figure}[h!]
    \centering
    \includegraphics[width=\linewidth]{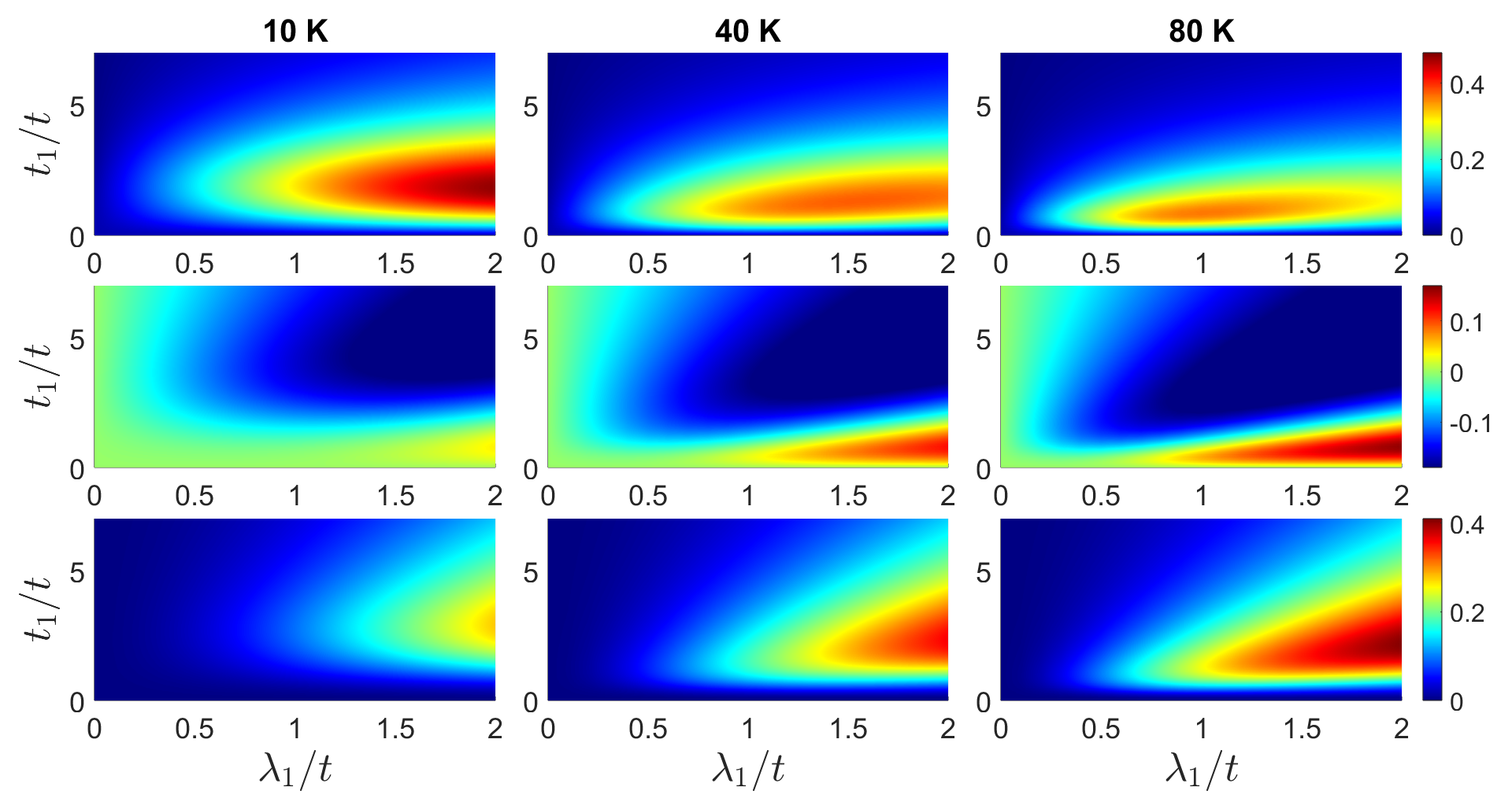}
    \caption{Top line: spin polarization; middle: imaginary singlet-triplet coherence; bottom: triplet component. Parameters as in Fig. 2 of the main text, extended range of $t_1$.}
    \label{fig:colormap2}
\end{figure}

We show below a multi-panel colormap analogous to Fig. 2 of the main text, but assuming a faster ET dynamics.
\begin{figure}[h!]
    \centering
    \includegraphics[width=\linewidth]{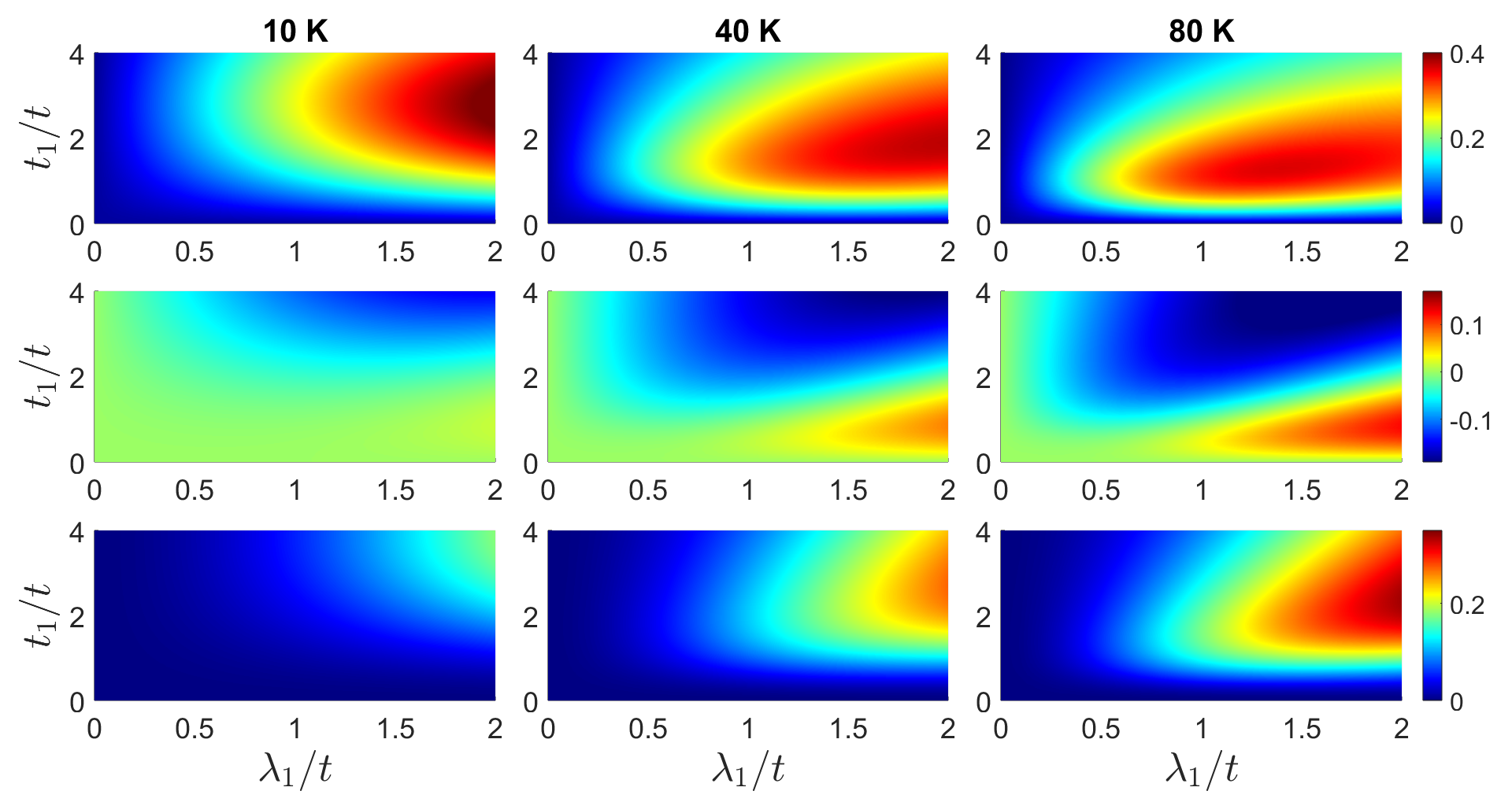}
    \caption{Top line: spin polarization; middle: imaginary singlet-triplet coherence; bottom: triplet component. Parameters as in Fig. 2 of the main text, but with faster ET ($\Gamma=1\times 10^{-2}$ meV).}
    \label{fig:colormap2}
\end{figure}

\newpage
\section{Field-dependence}
\begin{figure}[h!]
    \centering
    \includegraphics[width=0.95\linewidth]{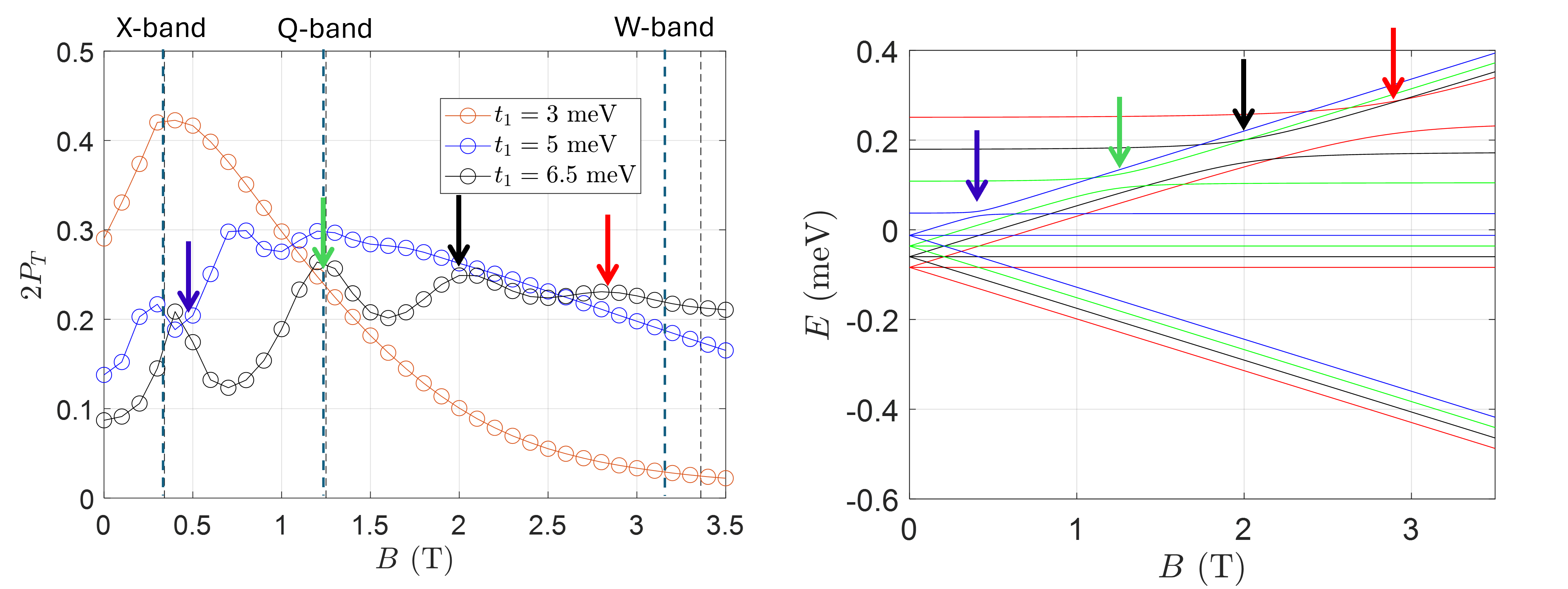}
    \caption{Left: field-dependence of the triplet component at 80 K, $\theta=90^\circ$ between chiral anisotropy axis and external field, with parameters as in Fig. 2 of the main text, $\lambda_1=1$ meV and different $t_1$ as indicated in the legend. 
    Right: level diagram for $t_1=6.5$ meV, with different colors referred to different  $n=0,1,2,3$ and avoided level crossing marked by arrows corresponding to each peak in the triplet component. At larger $n$ the coupling increases, yielding a corresponding broadening of the avoided crossing and of the peak in $P_T$.}
    \label{fig:placeholder}
\end{figure}
\newpage

\section{Maximum achievable spin polarization} \label{50perclimitandbeyond}

\subsection{Origin of the 50\% limit in simplest models} \label{limit}
We consider a minimal kinetic model involving four states: $c^\dagger_{B\uparrow}\ket{\emptyset}$, $c^\dagger_{B\downarrow}\ket{\emptyset}$, $c^\dagger_{A\uparrow}\ket{\emptyset}$, $c^\dagger_{A\downarrow}\ket{\emptyset}$, where $B,A$ represent bridge and acceptor orbitals. The four states correspond to an electron on B and A with spin $\uparrow$ or $\downarrow$. In the following we adopt the notation $B_\sigma=\langle c^\dagger_{B\sigma} c_{B\sigma}\rangle$, with $\sigma=\uparrow,\downarrow$.
In the absence of coupling to the acceptor, we consider a spin polarization on the bridge site evolving periodically as:
\begin{equation}\label{eq:sinewave}
s_{zB}(t) = B_\uparrow(t) - B_\downarrow(t) = a \sin(\omega t)
\end{equation}
where $0 \le a \le 1$ represents the amplitude of the oscillating polarization and $\omega$ its frequency. 
The choice of a sinusoidal form for the spin polarization on the bridge site is dictated by the fact that (i) it satisfies the physical constraint that the initial spin polarization is null, i.e., $B_\uparrow(0) = B_\downarrow(0) = 1/2$, (ii) in a coherent regime where two states are coupled, the polarization must evolve as a monochromatic oscillation, with  $\omega$ proportional to the energy gap of the spin eigenstates, and (iii) when $0 \le a \le 1$ the sine function ensures that the polarization never exceeds the total population. \\
We now include an irreversible population transfer from B to A at a rate $\Gamma$. The time evolution of the populations is governed by the following system of kinetic equations:
\begin{equation}
\begin{cases}
\dot A_\uparrow = \Gamma B_\uparrow
\\
\dot A_\downarrow = \Gamma B_\downarrow 
\\
\dot B_\uparrow = - \Gamma B_\uparrow
\\
\dot B_\downarrow = -\Gamma B_\downarrow 
\end{cases}
\end{equation}
Assuming that at $t=0$ the entire population is on the bridge and since the transfer is spin independent, the population on the bridge decays as $B_{\sigma}(t) = B_\sigma(0) e^{-\Gamma t}$. Consequently, the time-dependent spin polarization on the bridge site is given by:
\begin{equation}
s_{zB}(t) = B_\uparrow(t) - B_\downarrow(t) = a e^{-\Gamma t} \sin(\omega t)
\end{equation}
The rate of change for the spin polarization on the acceptor site $A$ is then:
\begin{equation}
\dot{s}_{zA}(t)= \dot A_\uparrow(t) - \dot A_\downarrow(t) = \Gamma \left[B_\uparrow(t) - B_\downarrow(t) \right] = \Gamma  a e^{-\Gamma t} \sin(\omega t).
\end{equation}
By integrating this expression over time, we obtain the time dependence of the spin polarization at site $A$:
\begin{equation}
s_{zA}(t) = A_\uparrow(t) - A_\downarrow(t) = \Gamma a \int_0^t e^{-\Gamma\tau} \sin(\omega\tau) \, d\tau
\end{equation}
Solving the integral yields the analytical expression for the acceptor spin polarization:
\begin{equation}
s_{zA}(t) = \frac{\Gamma a}{\Gamma^2 + \omega^2} \left[ \omega - e^{-\Gamma t} \bigl( \omega \cos(\omega t) + \Gamma \sin(\omega t) \bigr) \right]
\end{equation}
In the long-time limit ($t \to \infty$), the steady-state spin polarization on the acceptor reaches:
\begin{equation}\label{eq:polarization_infinite_time}
s_{zA}(\infty) = \frac{\Gamma a \omega}{\Gamma^2 + \omega^2}
\end{equation}
To determine the maximum achievable polarization, we analyze the function $f(\Gamma, \omega) = \frac{\Gamma \omega}{\Gamma^2 + \omega^2}$. By setting the derivative with respect to $\omega$ to zero:
\begin{equation}
\frac{df}{d\omega} = \frac{\Gamma(\Gamma^2 + \omega^2) - 2\Gamma\omega^2}{(\Gamma^2 + \omega^2)^2} = \frac{\Gamma(\Gamma^2 - \omega^2)}{(\Gamma^2 + \omega^2)^2} = 0
\end{equation}
we find that the maximum occurs when $\Gamma = \omega$ (assuming $\Gamma > 0$ for a non-trivial transfer). Under this condition, the maximum polarization is:
\begin{equation}
s_{zA}(\infty, \Gamma=\omega) = \frac{a \cdot \Gamma^2}{2\Gamma^2} = \frac{a}{2} \le \frac{1}{2}
\end{equation}
where $s_{zA}(\infty, \Gamma=\omega)$ is limited by the amplitude parameter $a \le 1$.

\subsection{Beyond the 50\% limit: formal analysis}
To overcome this 50\% limit, the system should either deviate from the initial assumption of considering a sinusoidal form for the spin polarization on the bridge site or dropping the assumption of monoexponential population transfer. This latter scenario can be physically obtained with non-Markovian baths. 

Hereafter we prefer to keep the simplest assumption of a Markovian bath leading to an exponential incoherent transfer and focus on the generation of poly-chromatic oscillations of $s_{zB}$. 
From an analytical perspective, overcoming the 50\% limit requires a spin polarization form that rises more steeply than the sine wave and remains near its maximum value for longer times. On the other side, the initial constraints (i) and (iii) detailed in Supplementary Section \ref{limit} must be preserved. A way to satisfy these conditions is to assume a square-wave like profile, which can be represented analytically through a Fourier series of odd harmonics:
\begin{equation}
s_{zB}(t) = e^{-\Gamma t} \,\frac{4a}{\pi}\sum_{n=1,3,5,...}^{\infty} \frac{1}{n} \sin(n\omega t)
\label{series}
\end{equation}
Following the same reasoning of the previous section, the polarization at site A becomes:
\begin{equation}
    s_{zA}(t) = \frac{4\Gamma a}{\pi} \sum_{n=1,3,5,...}^{\infty} \frac{1}{n}\int_0^t e^{-\Gamma \tau}\sin(n\omega \tau)\, d\tau
\end{equation}
Since we are interested in the polarization accumulated over long times, and the integrand decays exponentially, we can extend the upper limit of integration to $+\infty$. The integral then separates into $n$ independent terms, each of the standard Laplace form: $I_n =  \int_0^{\infty} e^{-\Gamma \tau}\sin(n\omega\tau)$ which has as solution $I_n = \frac{n\omega}{\Gamma^2+(n\omega)^2}$.
The resulting polarization becomes:
\begin{equation}
    s_{zA}(t) = \frac{4\Gamma a}{\pi} \sum_{n=1,3,5,...}^{\infty} \frac{1}{n} \frac{n\omega}{\Gamma ^2 + (n\omega)^2} = \frac{4\Gamma a\omega}{\pi} \sum_{n=1,3,5,...}^{\infty}  \frac{1}{\Gamma ^2 + (n\omega)^2}
\end{equation}
To study the convergence of the series, we can use the standard result $\sum_{n=0}^{\infty}\frac{1}{(2n+1)^2+x^2}=\frac{\pi}{4x}\tanh{(\frac{\pi x}{2})}$ with $x = \frac{\Gamma}{\omega}$:
\begin{equation}
    s_{zA}(\infty) = \frac{4\Gamma a}{\pi \omega} \sum_{n=0}^{\infty} \frac{1}{(2n+1)^2 + (\Gamma /\omega)^2} = \frac{4\Gamma a}{\pi \omega} \frac{\pi \omega}{4\Gamma}\tanh{\Bigl(\frac{\pi \Gamma}{2\omega}\Bigr)}= a \tanh{\Bigl(\frac{\pi \Gamma}{2\omega}\Bigr)}
\end{equation}
In this case, for $\omega=\Gamma$ the value of  $s_{zA}(\infty)$ is over $0.9a$, by far exceeding the 50\% limit.
A more physical scenario consists of considering only the first harmonics of the series in Eq.~\ref{series}, corresponding to a system in which the donor interacts with a few bridge sites oscillating, e.g., at frequencies $\omega, 3\omega, 5\omega $. For the truncated series, we drop the $\frac{4}{\pi}$ normalization factor. 
Keeping the normalization factor would cause the truncated profile to overshoot $a$, producing unphysical values of the spin polarization on the bridge sites.
As a consequence, the spin polarization we find at longer times for this three-frequencies example represents a lower bound on what the properly normalized truncated series would yield.
Yet, although this truncation is far from the ideal square-wave limit, it already gives a substantial improvement over the single-frequency case.
We obtain the following spin polarization at longer times on the acceptor:
\begin{equation}
    s_{zA}(\infty) = a \left[ \frac{\Gamma \omega}{\Gamma ^2 + \omega^2} + \frac{\Gamma \omega}{\Gamma ^2 + 9\omega^2} + \frac{\Gamma \omega}{\Gamma ^2 + 25\omega^2} \right]
\end{equation}
which is maximized by taking $\Gamma / \omega=1.31$, resulting in $s_{zA}(\infty)_{\Gamma=1.31\omega} = 0.65 \, a$.
This derivation demonstrates that any physical mechanism 
yielding a polychromatic oscillation on the last site of the bridge (with a steep increase followed by a flat region)  can be exploited to overcome  the 50\% limit. 

\subsection{Physical mechanisms to overcome the 50\% limit: coherent hopping}
\begin{figure}
    \centering
    \includegraphics[width=1.\linewidth]{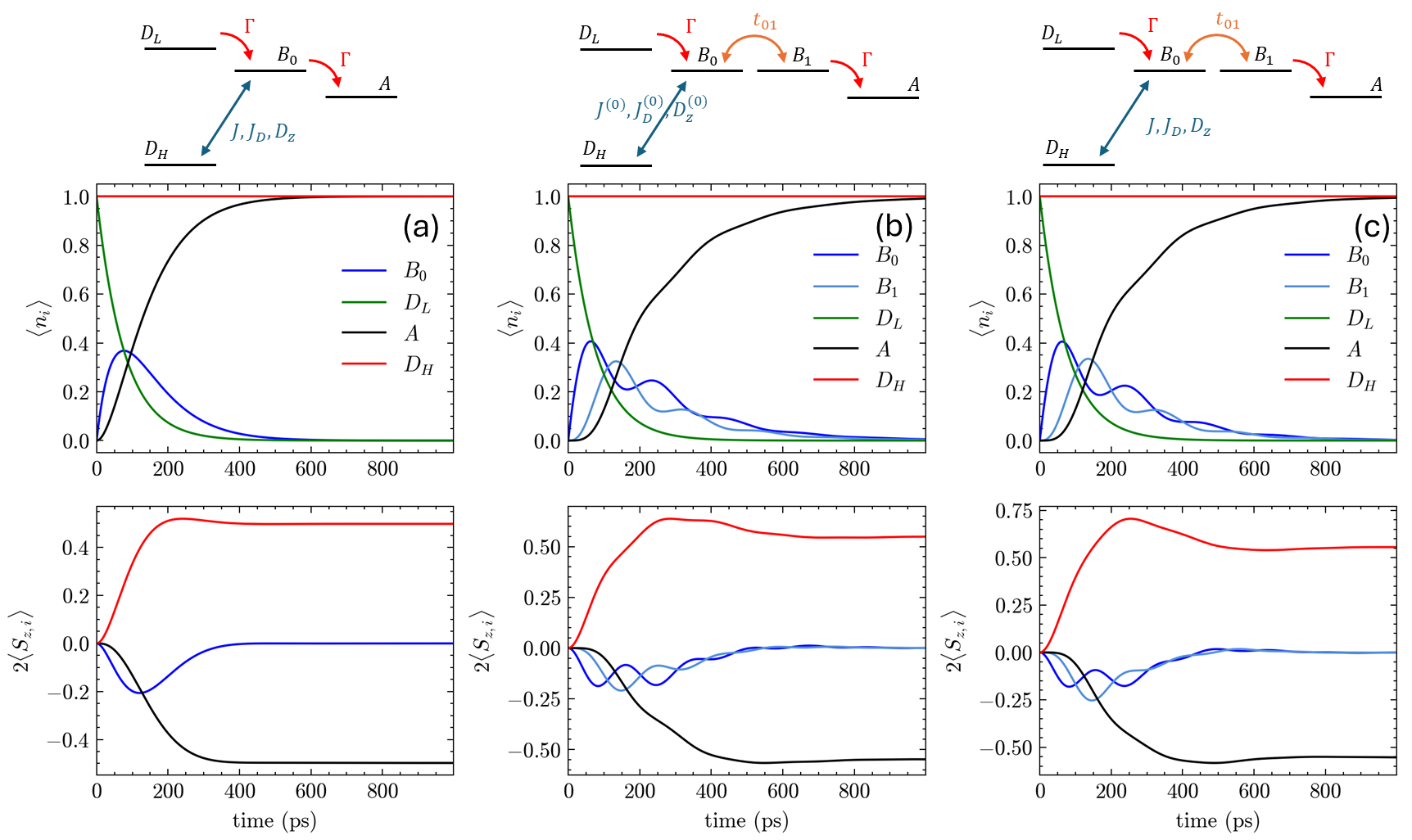}
    \caption{Sketch of the models, population and spin polarization evolution for 
    (a) a one-bridge-site system,
    (b) a two-bridge-site system without spin-spin coupling on the second site,  
    (c) a two-site-bridge system with spin-spin coupling on the second site.
    The model parameters are reported in Table~\ref{tab:dinamiche_over50coherent}. $\Gamma=\SI{4.3e-3}{meV}$ was set for all simulations.}
    \label{fig:dinamiche_over50coherent}
\end{figure}

Perhaps the easiest way to generate multi-frequency coherent oscillations in the spin polarization of the bridge sites is to extend the model described in the main text to account for more than one site on the bridge. As a proof of concept, we compare the spin dynamics for a one-bridge-site model (Fig.~\ref{fig:dinamiche_over50coherent}a) with those obtained for models consisting of two bridge-sites  (Fig.~\ref{fig:dinamiche_over50coherent}b and Fig.~\ref{fig:dinamiche_over50coherent}c). The microscopic parameters (Table~\ref{tab:microscopic-parameters} and Table~\ref{tab:dinamiche_over50coherent}) are chosen so that, with the appropriate transfer rate $\Gamma=\SI{4.3e-3}{meV}$, $\approx 50\%$ spin polarization is accumulated on A for a system with one site on the bridge.

In the case of two or more bridge sites, the one-electron Hamiltonian for the bridge is
\begin{equation}
    H_{bridge} = \sum_{i=0}^N\sum_\sigma \varepsilon_i c_{i,\sigma}^\dag c_{i,\sigma} + \sum_{i,j\neq i}\sum_\sigma t_{ij} c_{i,\sigma}^\dag c_{j,\sigma}+h.c.
\end{equation}
Where $N$ is the number of sites on the bridge, $\varepsilon_i$ is the on-site energy, and $t_{ij}$ is the hopping integral between sites $i$ and $j$.
Specifically, in the simulations from Fig.~\ref{fig:dinamiche_over50coherent}b and c, $t_{01}=\SI{0.1}{meV}$ and $\varepsilon_0=\varepsilon_1$.

\begin{figure}
    \centering
    \includegraphics[width=0.7\linewidth]{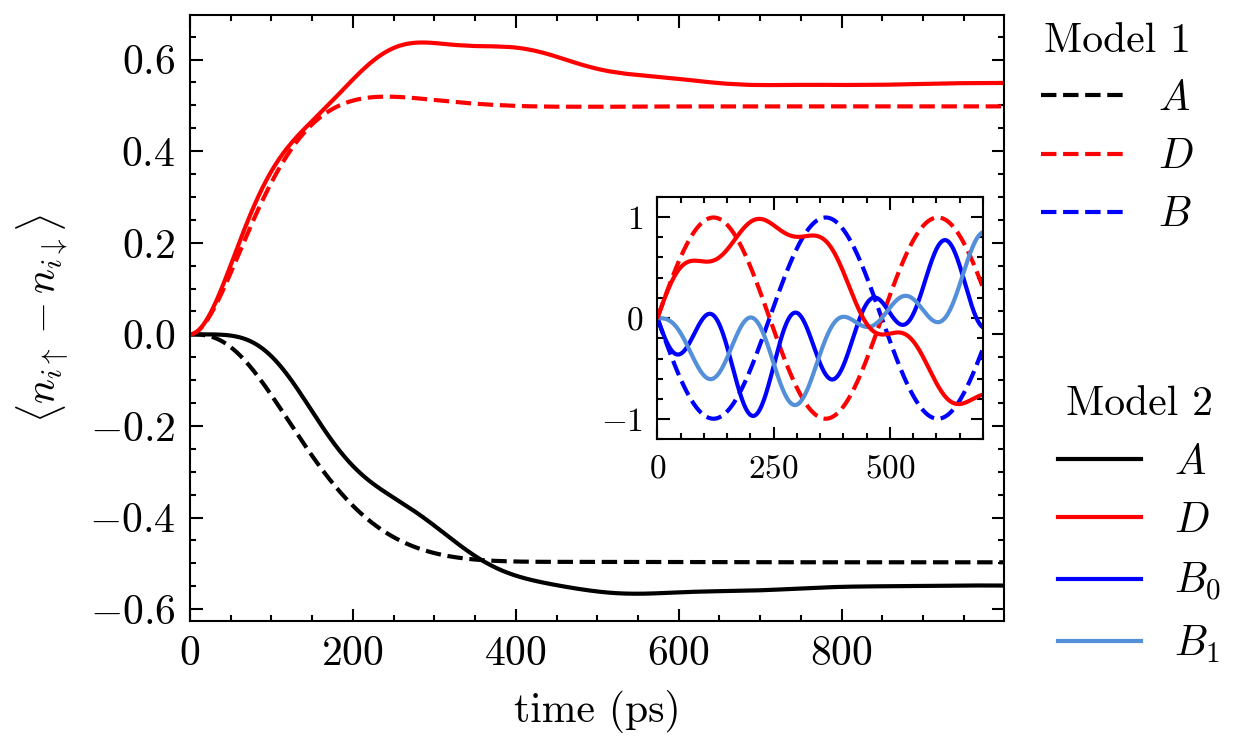}
    \caption{Main plot: simulated time evolution of the spin polarization on D and A for a system with one site on the bridge (dashed, parameters from model a in Table~\ref{tab:dinamiche_over50coherent}) and for a system with two sites on the bridge (continuous, parameters from model b in Table~\ref{tab:dinamiche_over50coherent}). Inset: coherent oscillations of the spin polarization on the donor bridge site(s). The population and local spin polarization evolution for each site are displayed in Fig.~\ref{fig:dinamiche_over50coherent}a and Fig.~\ref{fig:dinamiche_over50coherent}b.}
    \label{fig:over50-main1}
\end{figure}

The addition of a second bridge site, uncoupled to the donor spin, allows one to overcome the 50\% limit on spin polarization on the acceptor. This is evident in Fig.~\ref{fig:over50-main1} and from the comparison between Fig.~\ref{fig:dinamiche_over50coherent}a and Fig.~\ref{fig:dinamiche_over50coherent}b, where the static spin polarization on the acceptor goes from $\approx 50\%$ to $\approx 55\%$. A similar enhancement is observed when a finite spin–spin coupling between $B_1$ and the donor is included (Fig.~\ref{fig:dinamiche_over50coherent}c).

\begin{table}[]
\centering
\caption{Parameters for the simulations in Fig.~\ref{fig:dinamiche_over50coherent} and Fig.~\ref{fig:over50-main1}. $J$, $J_D$, and $D_z$ are taken from Table~\ref{tab:microscopic-parameters}.}
\label{tab:dinamiche_over50coherent}
\begin{tabular}{@{}lccc@{}}
\toprule
            & \textbf{Model a} & \textbf{Model b} & \textbf{Model c} \\ \midrule
$J^{(0)}$   & $J$   & $J$   & $J$   \\
$J_D^{(0)}$ & $J_D$ & $J_D$ & $J_D$ \\
$D_z^{(0)}$ & $D_z$ & $D_z$ & $D_z$ \\
$J^{(1)}$   & -     & 0     & $J/4$ \\
$J_D^{(1)}$ & -     & 0     & $J_D/4$ \\
$D_z^{(1)}$ & -     & 0     & $D_z/3$ \\
$t_{01}$ (meV) & -     & 0.01  & 0.01 \\
\bottomrule
\end{tabular}%
\end{table}

\subsection{Physical mechanisms to overcome the 50\% limit: incoherent transfer}
Spin polarization over 50\% can also be achieved exploiting multistep incoherent transfer among bridge sites spin-coupled to the donor. 
We consider a travelling electron moving through a chain of $n$ bridge sites ($B_i$) towards an acceptor site ($A$):
\[
B_1 \xrightarrow{\Gamma_1} B_2 \xrightarrow{\Gamma_2} \cdots \xrightarrow{\Gamma_n} A
\]
Specifically, we assume here incoherent  transfer, with $\Gamma_i$ being the rate for transfer from site $B_i$ to site $B_{i+1}$. 
While on site $B_i$, the electron is spin-coupled to a second electron on the donor, due to the same mechanism described in the main text, and there is no spin-coupling between electrons on the donor and on the acceptor. Since we are describing photoinduced electron transfer, we assume that the system starts in a singlet state. 
For the sake of simplicity, we assume in this derivation that the leading term in the spin-couplings is the DMI, that induces oscillations of the spin-polarization between -1 and +1, with angular frequency $\omega_i$, as assumed in \eqref{eq:sinewave}, and verified numerically (see inset of Fig.~\ref{fig:dinamiche_over50coherent} for the 1-bridge-site model).

An electron dwelling for a time $t$ on a single site accumulates spin-polarization according to \eqref{eq:sinewave}. In the case of the multi-site model described above, the phases accumulated on each site $i$ (given by $\omega_i t_i$) are additive. For dwell times $(t_0, \dots, t_n)$, the total polarization is:
\begin{equation}
P(t_0, \dots, t_n) = \sin\left(\sum_{i=1}^n \omega_i t_i\right)
\end{equation}

As the process is incoherent and Markovian, the dwell time $t_i$ on site $B_i$ follows an exponential distribution, described by the probability density:
\begin{equation}
p_i(t_i) = \Gamma_i e^{-\Gamma_i t_i}
\end{equation}

Assuming non-reversible incoherent transfer, the joint probability density is:
\begin{equation}
p(t_0, \dots, t_n) = \prod_{i=1}^n \Gamma_i e^{-\Gamma_i t_i}
\end{equation}

The final polarization $P_A$ is the average of $P(t_0, \dots, t_n)$ over all possible dwell times:
\begin{equation}
P_A = \int_0^\infty \cdots \int_0^\infty \left[ \prod_{i=1}^n \Gamma_i e^{-\Gamma_i t_i} \right] \sin\left(\sum_{i=1}^n \omega_i t_i\right) dt_0 \cdots dt_n
\end{equation}

Using the identity $\sin x = \text{Im}(e^{ix})$, and combining the exponentials, we can factorize the integral
\begin{equation}
    P_A = \text{Im} \prod_{i=1}^n \left[ \int_0^\infty \Gamma_i e^{-(\Gamma_i - i\omega_i)t_i} dt_i \right]
\end{equation}

The one-site integral evaluates to:
\begin{equation}
\int_0^\infty \Gamma_i e^{-(\Gamma_i - i\omega_i)t} dt = \frac{\Gamma_i}{\Gamma_i - i\omega_i}
\end{equation}
The compact exact formula for the polarization is therefore:
\begin{equation}\label{eq:polarization_sine_incoherent}
P_A = \text{Im} \prod_{i=1}^n \frac{\Gamma_i}{\Gamma_i - i\omega_i}
\end{equation}
Notice that in the case of a single site \eqref{eq:polarization_sine_incoherent} collapses to \eqref{eq:polarization_infinite_time}, thus recovering the 50\% limit to spin polarization for systems with one bridge site. 

To make the algebra easier in the case of more than one site, we adopt a polar form representation.
Let $\theta_i = \arctan\left(\frac{\omega_i}{\Gamma_i}\right)$. We can rewrite the factor:
\begin{equation}
\frac{\Gamma_i}{\Gamma_i - i\omega_i} = \frac{\Gamma_i(\Gamma_i + i\omega_i)}{\Gamma_i^2 + \omega_i^2} = \cos^2\theta_i + i\sin\theta_i\cos\theta_i = \cos\theta_i e^{i\theta_i}
\end{equation}
Substituting into \eqref{eq:polarization_sine_incoherent} we obtain
\begin{equation}
P_A = \sin\left(\sum_{i=1}^n \theta_i\right) \prod_{i=1}^n \cos\theta_i, \quad \theta_i = \arctan\left(\frac{\omega_i}{\Gamma_i}\right)
\end{equation}

Now that we have an analytical form of the polarization we perform an optimization to maximize $P_A$. Specifically, we take the logarithmic derivative of $f(\theta_1, \dots, \theta_n) = \sin(\Theta) \prod \cos\theta_i$ (where $\Theta = \sum \theta_i$):
\[
\frac{\partial}{\partial \theta_j} \ln f = \cot\Theta - \tan\theta_j = 0
\]
This implies all angles are equal: $\theta_1 = \theta_2 = \dots = \theta_n = \theta$. In other words, the ratios between the frequency of oscillation of spin polarization on site $B_i$, $\omega_i$ and the rate $\Gamma_i$ for transfer from site $B_i$ to site $B_{i+1}$ must be the same for all sites to maximize spin polarization. 

The maximum polarization that can be obtained from a system with $n$ bridge sites is obtained when $\tan\theta = \cot(n\theta)$ is true, which yields:
\begin{equation}
\theta = \frac{\pi}{2(n+1)}
\end{equation}

For a set of oscillation frequencies $\omega_i$, imposed by the spin-couplings, the maximum polarization is achieved when the transfer rates are 
\begin{equation}\label{eq:optimal_rates}
\Gamma_i = \omega_i \cot\left(\frac{\pi}{2(n+1)}\right)
\end{equation}

The maximum polarization for $n$ sites is:
\begin{equation}\label{eq:maximum_polarization}
P_{A,\max}^{(n)} = \cos^{n+1}\left(\frac{\pi}{2(n+1)}\right)
\end{equation}
Thus in the large-$n$ limit, unitary spin polarization can be achieved.

\begin{figure}
    \centering
    \includegraphics[width=1\linewidth]{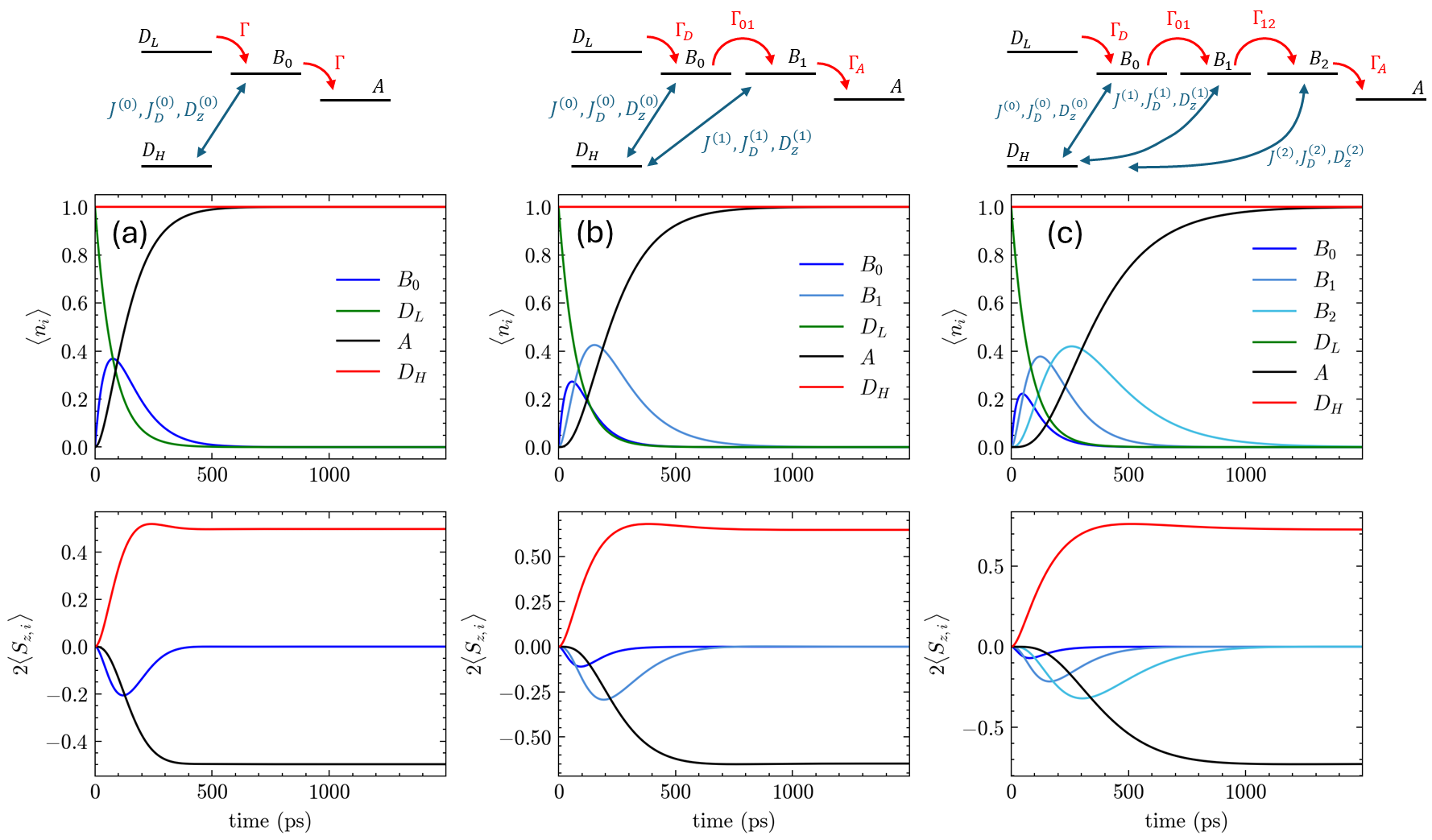}
    \caption{Sketch of the model, population and spin polarization evolution for systems with (a) one, (b) two and (c) three sites on the bridge, spin-spin couplings on each site and incoherent transfer rates are reported in Table~\ref{tab:dinamiche_over50incoherent}.}
    \label{fig:dinamiche_over50incoherent}
\end{figure}

Numerical simulations for systems with one, two, and three sites on the bridge are displayed in Fig.~\ref{fig:over50-main2} and Fig.~\ref{fig:dinamiche_over50incoherent}. The spin-coupling parameters are chosen so that the DMI is the leading term in the spin Hamiltonians  (approximately twice the isotropic exchange) and the incoherent transfer rates are set to respect \eqref{eq:optimal_rates} (the spin-coupling parameters and rates are reported in Table~\ref{tab:dinamiche_over50incoherent}). Specifically, we obtain $\approx 50\%$, $\approx 65\%$, and $\approx 73\%$ spin polarization for systems with one, two, or three sites on the bridge respectively, in agreement with \eqref{eq:maximum_polarization}. 

\begin{table}[]
\centering
\caption{Parameters for the simulations in Fig.~\ref{fig:dinamiche_over50incoherent} and Fig.~\ref{fig:over50-main2}. $J$, $J_D$, and $D_z$ are taken from Table~\ref{tab:microscopic-parameters}. The rates are reported in meV units.}
\label{tab:dinamiche_over50incoherent}
\begin{tabular}{@{}lccc@{}}
\toprule
            & \textbf{One-site} & \textbf{Two-site} & \textbf{Three-site} \\ \midrule
$J^{(0)}$   & $J$   & $J$   & $J$   \\
$J_D^{(0)}$ & $J_D$ & $J_D$ & $J_D$ \\
$D_z^{(0)}$ & $D_z$ & $D_z$ & $D_z$ \\
$J^{(1)}$   & -     & $J/4$ & $J/4$ \\
$J_D^{(1)}$ & -     & $J_D/4$ & $J_D/4$ \\
$D_z^{(1)}$ & -     & $D_z/3$ & $D_z/3$ \\
$J^{(2)}$   & -     & -     & $J/8$ \\
$J_D^{(2)}$ & -     & -     & $J_D/8$ \\
$D_z^{(2)}$ & -     & -     & $D_z/6$ \\
$\Gamma_{D}$ & \num{4.3e-3} & \num{4.3e-3} &  \num{4.3e-3}\\
$\Gamma_{A}$ & \num{4.3e-3} & \num{2.48e-3} & \num{1.73e-3} \\
$\Gamma_{01}$ & - & \num{7.45e-3} & \num{1.04e-2}\\
$\Gamma_{12}$ & - & - & \num{3.46e-3}\\
\bottomrule
\end{tabular}%
\end{table}

\vfill
\newpage

\section{Ab-initio calculations} \label{ab-initio}
Parameters $U, \Delta, t, \text{and}\, \lambda$ employed in the model presented in the main text were derived from ab-initio calculation on the PXX-NMI\textsubscript{2}-NDI molecule \cite{Eckvahl2023}. Specifically, the on-site repulsion $U=\SI{3.5}{eV}$ was adopted from our previous study on the same system \cite{phan_huu_ab_2025}, where it was determined via CAS-sr-DFT calculations.
Also the $\Delta=\SI{5}{eV}$ parameter is extracted from the same work. This value corresponds to the energy difference of the frontier molecular orbitals between the LUMO of the first site of the bridge (the NMI fragment closer to the PXX) and the HOMO localized on the donor at the LC-$\omega$PBE\textbackslash6-31G(d) level of theory, with a range separation parameter $\omega=0.196a_0^{-1}$. 
The hopping term $t$ was obtained by performing DFT calculations on the PXX-NMI\textsubscript{2}-NDI molecule at the LC-$\omega$PBE\textbackslash6-31G(d) and exploiting the Frontier Molecular Orbital (FMO) approach within the QChem software \cite{qchem}, to obtain the electronic coupling between the highest occupied orbital localized on the PXX donor and the lowest unoccupied on the first site of the NMI\textsubscript{2} bridge. In the FMO approach, the electronic coupling is defined as the off-diagonal matrix element of the Kohn-Sham operator between the a donor and an acceptor system. The molecule under study was divided into the donor part, comprising the PXX and the phenyl ring connecting the PXX and the NMI\textsubscript{2}, and the acceptor one, including the rest of the molecule (NMI\textsubscript{2}-NDI). The hopping term was, hence, given by:
\begin{equation}
    t = \bra{\phi_D^{HOMO}}\hat{f}\ket{\phi_A^{NMI}} = \SI{1}{meV}
\end{equation}
with $\hat{f}$ being the Kohn–Sham operator of the donor-acceptor system, $\phi_D^{HOMO}$ is the HOMO localized on the donor and $\phi_A^{NMI}$ the LUMO localized on the first site of the bridge.
Finally, the spin-orbit coupling parameter $\lambda=\SI{0.1}{meV}$ was determined from the Cartesian components ($\lambda_x,\,\lambda_y,\,\lambda_z$) obtained via DFT calculations in ORCA \cite{orca} at the B3LYP\textbackslash def2-TZVP level, using the relation $\lambda = \sqrt{\lambda_x^2 + \lambda_y^2 + \lambda_z^2}$.


\begin{thebibliography}{52}%
\makeatletter
\providecommand \@ifxundefined [1]{%
 \@ifx{#1\undefined}
}%
\providecommand \@ifnum [1]{%
 \ifnum #1\expandafter \@firstoftwo
 \else \expandafter \@secondoftwo
 \fi
}%
\providecommand \@ifx [1]{%
 \ifx #1\expandafter \@firstoftwo
 \else \expandafter \@secondoftwo
 \fi
}%
\providecommand \natexlab [1]{#1}%
\providecommand \enquote  [1]{``#1''}%
\providecommand \bibnamefont  [1]{#1}%
\providecommand \bibfnamefont [1]{#1}%
\providecommand \citenamefont [1]{#1}%
\providecommand \href@noop [0]{\@secondoftwo}%
\providecommand \href [0]{\begingroup \@sanitize@url \@href}%
\providecommand \@href[1]{\@@startlink{#1}\@@href}%
\providecommand \@@href[1]{\endgroup#1\@@endlink}%
\providecommand \@sanitize@url [0]{\catcode `\\12\catcode `\$12\catcode `\&12\catcode `\#12\catcode `\^12\catcode `\_12\catcode `\%12\relax}%
\providecommand \@@startlink[1]{}%
\providecommand \@@endlink[0]{}%
\providecommand \url  [0]{\begingroup\@sanitize@url \@url }%
\providecommand \@url [1]{\endgroup\@href {#1}{\urlprefix }}%
\providecommand \urlprefix  [0]{URL }%
\providecommand \Eprint [0]{\href }%
\providecommand \doibase [0]{https://doi.org/}%
\providecommand \selectlanguage [0]{\@gobble}%
\providecommand \bibinfo  [0]{\@secondoftwo}%
\providecommand \bibfield  [0]{\@secondoftwo}%
\providecommand \translation [1]{[#1]}%
\providecommand \BibitemOpen [0]{}%
\providecommand \bibitemStop [0]{}%
\providecommand \bibitemNoStop [0]{.\EOS\space}%
\providecommand \EOS [0]{\spacefactor3000\relax}%
\providecommand \BibitemShut  [1]{\csname bibitem#1\endcsname}%
\let\auto@bib@innerbib\@empty
\bibitem [{\citenamefont {Naaman}\ \emph {et~al.}(2019)\citenamefont {Naaman}, \citenamefont {Paltiel},\ and\ \citenamefont {Waldeck}}]{Naaman2019}%
  \BibitemOpen
  \bibfield  {author} {\bibinfo {author} {\bibfnamefont {R.}~\bibnamefont {Naaman}}, \bibinfo {author} {\bibfnamefont {Y.}~\bibnamefont {Paltiel}},\ and\ \bibinfo {author} {\bibfnamefont {D.}~\bibnamefont {Waldeck}},\ }\bibfield  {title} {\bibinfo {title} {Chiral molecules and the electron spin},\ }\href {https://doi.org/10.1038/s41570-019-0087-1} {\bibfield  {journal} {\bibinfo  {journal} {Nat. Rev. Chem.}\ }\textbf {\bibinfo {volume} {3}},\ \bibinfo {pages} {250} (\bibinfo {year} {2019})}\BibitemShut {NoStop}%
\bibitem [{\citenamefont {Naaman}\ \emph {et~al.}(2020)\citenamefont {Naaman}, \citenamefont {Paltiel},\ and\ \citenamefont {Waldeck}}]{Naaman2020b}%
  \BibitemOpen
  \bibfield  {author} {\bibinfo {author} {\bibfnamefont {R.}~\bibnamefont {Naaman}}, \bibinfo {author} {\bibfnamefont {Y.}~\bibnamefont {Paltiel}},\ and\ \bibinfo {author} {\bibfnamefont {D.~H.}\ \bibnamefont {Waldeck}},\ }\bibfield  {title} {\bibinfo {title} {Chiral molecules and the spin selectivity effect},\ }\href {https://doi.org/10.1021/acs.jpclett.0c00474} {\bibfield  {journal} {\bibinfo  {journal} {J. Phys. Chem. Lett.}\ }\textbf {\bibinfo {volume} {11}},\ \bibinfo {pages} {3660} (\bibinfo {year} {2020})}\BibitemShut {NoStop}%
\bibitem [{\citenamefont {Bloom}\ \emph {et~al.}(2024)\citenamefont {Bloom}, \citenamefont {Paltiel}, \citenamefont {Naaman},\ and\ \citenamefont {Waldeck}}]{Bloom2024}%
  \BibitemOpen
  \bibfield  {author} {\bibinfo {author} {\bibfnamefont {B.~P.}\ \bibnamefont {Bloom}}, \bibinfo {author} {\bibfnamefont {Y.}~\bibnamefont {Paltiel}}, \bibinfo {author} {\bibfnamefont {R.}~\bibnamefont {Naaman}},\ and\ \bibinfo {author} {\bibfnamefont {D.~H.}\ \bibnamefont {Waldeck}},\ }\bibfield  {title} {\bibinfo {title} {Chiral induced spin selectivity},\ }\href {https://doi.org/10.1021/acs.chemrev.3c00661} {\bibfield  {journal} {\bibinfo  {journal} {Chem. Rev.}\ }\textbf {\bibinfo {volume} {124}},\ \bibinfo {pages} {1950} (\bibinfo {year} {2024})}\BibitemShut {NoStop}%
\bibitem [{\citenamefont {Kumar}\ and\ \citenamefont {Gupta}(2026)}]{Kumar2026}%
  \BibitemOpen
  \bibfield  {author} {\bibinfo {author} {\bibfnamefont {A.}~\bibnamefont {Kumar}}\ and\ \bibinfo {author} {\bibfnamefont {A.}~\bibnamefont {Gupta}},\ }\bibfield  {title} {\bibinfo {title} {Chirality-induced spin selectivity: an interdisciplinary perspective from chemical physics to biology},\ }\href {https://doi.org/10.1039/D5CP04185F} {\bibfield  {journal} {\bibinfo  {journal} {Phys. Chem. Chem. Phys.}\ }\textbf {\bibinfo {volume} {28}},\ \bibinfo {pages} {3812} (\bibinfo {year} {2026})}\BibitemShut {NoStop}%
\bibitem [{\citenamefont {Waldeck}\ \emph {et~al.}(2026)\citenamefont {Waldeck}, \citenamefont {Naaman},\ and\ \citenamefont {Subotnik}}]{Waldeck2026}%
  \BibitemOpen
  \bibfield  {author} {\bibinfo {author} {\bibfnamefont {D.}~\bibnamefont {Waldeck}}, \bibinfo {author} {\bibfnamefont {R.}~\bibnamefont {Naaman}},\ and\ \bibinfo {author} {\bibfnamefont {J.}~\bibnamefont {Subotnik}},\ }\bibfield  {title} {\bibinfo {title} {Detecting chirality-induced spin selectivity in chromophore-linked dna hairpins using photogenerated radical pairs},\ }\bibfield  {journal} {\bibinfo  {journal} {Physics Today}\ }\textbf {\bibinfo {volume} {03}},\ \href {https://doi.org/10.1063/pt.5cb73ded2d} {10.1063/pt.5cb73ded2d} (\bibinfo {year} {2026})\BibitemShut {NoStop}%
\bibitem [{\citenamefont {Eckvahl}\ \emph {et~al.}(2023)\citenamefont {Eckvahl}, \citenamefont {Tcyrulnikov}, \citenamefont {Chiesa}, \citenamefont {Bradley}, \citenamefont {Young}, \citenamefont {Carretta}, \citenamefont {Krzyaniak},\ and\ \citenamefont {Wasielewski}}]{Eckvahl2023}%
  \BibitemOpen
  \bibfield  {author} {\bibinfo {author} {\bibfnamefont {H.~J.}\ \bibnamefont {Eckvahl}}, \bibinfo {author} {\bibfnamefont {N.~A.}\ \bibnamefont {Tcyrulnikov}}, \bibinfo {author} {\bibfnamefont {A.}~\bibnamefont {Chiesa}}, \bibinfo {author} {\bibfnamefont {J.~M.}\ \bibnamefont {Bradley}}, \bibinfo {author} {\bibfnamefont {R.~M.}\ \bibnamefont {Young}}, \bibinfo {author} {\bibfnamefont {S.}~\bibnamefont {Carretta}}, \bibinfo {author} {\bibfnamefont {M.~D.}\ \bibnamefont {Krzyaniak}},\ and\ \bibinfo {author} {\bibfnamefont {M.~R.}\ \bibnamefont {Wasielewski}},\ }\bibfield  {title} {\bibinfo {title} {Direct observation of chirality-induced spin selectivity in electron donor–acceptor molecules},\ }\href {https://doi.org/10.1126/science.adj5328} {\bibfield  {journal} {\bibinfo  {journal} {Science}\ }\textbf {\bibinfo {volume} {382}},\ \bibinfo {pages} {197} (\bibinfo {year} {2023})}\BibitemShut {NoStop}%
\bibitem [{\citenamefont {Eckvahl}\ \emph {et~al.}(2024)\citenamefont {Eckvahl}, \citenamefont {Copley}, \citenamefont {Young}, \citenamefont {Krzyaniak},\ and\ \citenamefont {Wasielewski}}]{Eckvahl2024}%
  \BibitemOpen
  \bibfield  {author} {\bibinfo {author} {\bibfnamefont {H.~J.}\ \bibnamefont {Eckvahl}}, \bibinfo {author} {\bibfnamefont {G.}~\bibnamefont {Copley}}, \bibinfo {author} {\bibfnamefont {R.~M.}\ \bibnamefont {Young}}, \bibinfo {author} {\bibfnamefont {M.~D.}\ \bibnamefont {Krzyaniak}},\ and\ \bibinfo {author} {\bibfnamefont {M.~R.}\ \bibnamefont {Wasielewski}},\ }\bibfield  {title} {\bibinfo {title} {Detecting chirality-induced spin selectivity in randomly oriented radical pairs photogenerated by hole transfer},\ }\href {https://doi.org/10.1021/jacs.4c08706} {\bibfield  {journal} {\bibinfo  {journal} {J. Am. Chem. Soc.}\ }\textbf {\bibinfo {volume} {146}},\ \bibinfo {pages} {24125} (\bibinfo {year} {2024})}\BibitemShut {NoStop}%
\bibitem [{\citenamefont {Latawiec}\ \emph {et~al.}(2025)\citenamefont {Latawiec}, \citenamefont {Chiesa}, \citenamefont {Qiu}, \citenamefont {Tcyrulnikov}, \citenamefont {Young}, \citenamefont {Carretta}, \citenamefont {Krzyaniak},\ and\ \citenamefont {Wasielewski}}]{Latawiec2025}%
  \BibitemOpen
  \bibfield  {author} {\bibinfo {author} {\bibfnamefont {E.~I.}\ \bibnamefont {Latawiec}}, \bibinfo {author} {\bibfnamefont {A.}~\bibnamefont {Chiesa}}, \bibinfo {author} {\bibfnamefont {Y.}~\bibnamefont {Qiu}}, \bibinfo {author} {\bibfnamefont {N.~A.}\ \bibnamefont {Tcyrulnikov}}, \bibinfo {author} {\bibfnamefont {R.~M.}\ \bibnamefont {Young}}, \bibinfo {author} {\bibfnamefont {S.}~\bibnamefont {Carretta}}, \bibinfo {author} {\bibfnamefont {M.~D.}\ \bibnamefont {Krzyaniak}},\ and\ \bibinfo {author} {\bibfnamefont {M.~R.}\ \bibnamefont {Wasielewski}},\ }\bibfield  {title} {\bibinfo {title} {Detecting chirality-induced spin selectivity in chromophore-linked dna hairpins using photogenerated radical pairs},\ }\href {https://doi.org/10.1073/pnas.2515120122} {\bibfield  {journal} {\bibinfo  {journal} {Proc. Natl. Acad. Sci. U.S.A.}\ }\textbf {\bibinfo {volume} {122}},\ \bibinfo {pages} {e2515120122} (\bibinfo {year} {2025})}\BibitemShut {NoStop}%
\bibitem [{\citenamefont {Chiesa}\ \emph {et~al.}(2021)\citenamefont {Chiesa}, \citenamefont {Chizzini}, \citenamefont {Garlatti}, \citenamefont {Salvadori}, \citenamefont {Tacchino}, \citenamefont {Santini}, \citenamefont {Tavernelli}, \citenamefont {Bittl}, \citenamefont {Chiesa}, \citenamefont {Sessoli},\ and\ \citenamefont {Carretta}}]{Chiesa2021}%
  \BibitemOpen
  \bibfield  {author} {\bibinfo {author} {\bibfnamefont {A.}~\bibnamefont {Chiesa}}, \bibinfo {author} {\bibfnamefont {M.}~\bibnamefont {Chizzini}}, \bibinfo {author} {\bibfnamefont {E.}~\bibnamefont {Garlatti}}, \bibinfo {author} {\bibfnamefont {E.}~\bibnamefont {Salvadori}}, \bibinfo {author} {\bibfnamefont {F.}~\bibnamefont {Tacchino}}, \bibinfo {author} {\bibfnamefont {P.}~\bibnamefont {Santini}}, \bibinfo {author} {\bibfnamefont {I.}~\bibnamefont {Tavernelli}}, \bibinfo {author} {\bibfnamefont {R.}~\bibnamefont {Bittl}}, \bibinfo {author} {\bibfnamefont {M.}~\bibnamefont {Chiesa}}, \bibinfo {author} {\bibfnamefont {R.}~\bibnamefont {Sessoli}},\ and\ \bibinfo {author} {\bibfnamefont {S.}~\bibnamefont {Carretta}},\ }\bibfield  {title} {\bibinfo {title} {Assessing the nature of chiral-induced spin selectivity by magnetic resonance},\ }\href {https://doi.org/10.1021/acs.jpclett.1c01447} {\bibfield  {journal} {\bibinfo  {journal} {J. Phys. Chem. Lett.}\ }\textbf {\bibinfo {volume} {12}},\ \bibinfo {pages}
  {6341} (\bibinfo {year} {2021})}\BibitemShut {NoStop}%
\bibitem [{\citenamefont {Chiesa}\ \emph {et~al.}(2025)\citenamefont {Chiesa}, \citenamefont {Privitera}, \citenamefont {Garlatti}, \citenamefont {Allodi}, \citenamefont {Bittl}, \citenamefont {Wasielewski}, \citenamefont {Sessoli},\ and\ \citenamefont {Carretta}}]{Chiesa2025}%
  \BibitemOpen
  \bibfield  {author} {\bibinfo {author} {\bibfnamefont {A.}~\bibnamefont {Chiesa}}, \bibinfo {author} {\bibfnamefont {A.}~\bibnamefont {Privitera}}, \bibinfo {author} {\bibfnamefont {E.}~\bibnamefont {Garlatti}}, \bibinfo {author} {\bibfnamefont {G.}~\bibnamefont {Allodi}}, \bibinfo {author} {\bibfnamefont {R.}~\bibnamefont {Bittl}}, \bibinfo {author} {\bibfnamefont {M.~R.}\ \bibnamefont {Wasielewski}}, \bibinfo {author} {\bibfnamefont {R.}~\bibnamefont {Sessoli}},\ and\ \bibinfo {author} {\bibfnamefont {S.}~\bibnamefont {Carretta}},\ }\bibfield  {title} {\bibinfo {title} {Chirality-induced spin selectivity at the molecular level: A different perspective to understand and exploit the phenomenon},\ }\href {https://doi.org/10.1021/acs.jpclett.5c00755} {\bibfield  {journal} {\bibinfo  {journal} {J. Phys. Chem. Lett.}\ }\textbf {\bibinfo {volume} {16}},\ \bibinfo {pages} {5358} (\bibinfo {year} {2025})}\BibitemShut {NoStop}%
\bibitem [{\citenamefont {Evers}\ \emph {et~al.}(2022)\citenamefont {Evers}, \citenamefont {Aharony}, \citenamefont {Bar-Gill}, \citenamefont {Entin-Wohlman}, \citenamefont {Hedegård}, \citenamefont {Hod}, \citenamefont {Jelinek}, \citenamefont {Kamieniarz}, \citenamefont {Lemeshko}, \citenamefont {Michaeli}, \citenamefont {Mujica}, \citenamefont {Naaman}, \citenamefont {Paltiel}, \citenamefont {Refaely-Abramson}, \citenamefont {Tal}, \citenamefont {Thijssen}, \citenamefont {Thoss}, \citenamefont {van Ruitenbeek}, \citenamefont {Venkataraman}, \citenamefont {Waldeck}, \citenamefont {Yan},\ and\ \citenamefont {Kronik}}]{evers2022}%
  \BibitemOpen
  \bibfield  {author} {\bibinfo {author} {\bibfnamefont {F.}~\bibnamefont {Evers}}, \bibinfo {author} {\bibfnamefont {A.}~\bibnamefont {Aharony}}, \bibinfo {author} {\bibfnamefont {N.}~\bibnamefont {Bar-Gill}}, \bibinfo {author} {\bibfnamefont {O.}~\bibnamefont {Entin-Wohlman}}, \bibinfo {author} {\bibfnamefont {P.}~\bibnamefont {Hedegård}}, \bibinfo {author} {\bibfnamefont {O.}~\bibnamefont {Hod}}, \bibinfo {author} {\bibfnamefont {P.}~\bibnamefont {Jelinek}}, \bibinfo {author} {\bibfnamefont {G.}~\bibnamefont {Kamieniarz}}, \bibinfo {author} {\bibfnamefont {M.}~\bibnamefont {Lemeshko}}, \bibinfo {author} {\bibfnamefont {K.}~\bibnamefont {Michaeli}}, \bibinfo {author} {\bibfnamefont {V.}~\bibnamefont {Mujica}}, \bibinfo {author} {\bibfnamefont {R.}~\bibnamefont {Naaman}}, \bibinfo {author} {\bibfnamefont {Y.}~\bibnamefont {Paltiel}}, \bibinfo {author} {\bibfnamefont {S.}~\bibnamefont {Refaely-Abramson}}, \bibinfo {author} {\bibfnamefont {O.}~\bibnamefont {Tal}}, \bibinfo {author} {\bibfnamefont
  {J.}~\bibnamefont {Thijssen}}, \bibinfo {author} {\bibfnamefont {M.}~\bibnamefont {Thoss}}, \bibinfo {author} {\bibfnamefont {J.~M.}\ \bibnamefont {van Ruitenbeek}}, \bibinfo {author} {\bibfnamefont {L.}~\bibnamefont {Venkataraman}}, \bibinfo {author} {\bibfnamefont {D.~H.}\ \bibnamefont {Waldeck}}, \bibinfo {author} {\bibfnamefont {B.}~\bibnamefont {Yan}},\ and\ \bibinfo {author} {\bibfnamefont {L.}~\bibnamefont {Kronik}},\ }\bibfield  {title} {\bibinfo {title} {Theory of chirality induced spin selectivity: Progress and challenges},\ }\href {https://doi.org/https://doi.org/10.1002/adma.202106629} {\bibfield  {journal} {\bibinfo  {journal} {Adv. Mater.}\ }\textbf {\bibinfo {volume} {34}},\ \bibinfo {pages} {2106629} (\bibinfo {year} {2022})}\BibitemShut {NoStop}%
\bibitem [{\citenamefont {Alwan}\ \emph {et~al.}(2024)\citenamefont {Alwan}, \citenamefont {Sharoni},\ and\ \citenamefont {Dubi}}]{Alwan2024}%
  \BibitemOpen
  \bibfield  {author} {\bibinfo {author} {\bibfnamefont {S.}~\bibnamefont {Alwan}}, \bibinfo {author} {\bibfnamefont {A.}~\bibnamefont {Sharoni}},\ and\ \bibinfo {author} {\bibfnamefont {Y.}~\bibnamefont {Dubi}},\ }\bibfield  {title} {\bibinfo {title} {Role of electrode polarization in the electron transport chirality-induced spin-selectivity effect},\ }\href {https://doi.org/10.1021/acs.jpcc.3c08223} {\bibfield  {journal} {\bibinfo  {journal} {J. Phys. Chem. C}\ }\textbf {\bibinfo {volume} {128}},\ \bibinfo {pages} {6438} (\bibinfo {year} {2024})}\BibitemShut {NoStop}%
\bibitem [{\citenamefont {Fransson}(2025{\natexlab{a}})}]{Fransson2025b}%
  \BibitemOpen
  \bibfield  {author} {\bibinfo {author} {\bibfnamefont {J.}~\bibnamefont {Fransson}},\ }\bibfield  {title} {\bibinfo {title} {What does it take for an organic closed shell molecule to become magnetic?},\ }\href {https://doi.org/10.1021/acs.jpclett.5c03056} {\bibfield  {journal} {\bibinfo  {journal} {J. Phys. Chem. Lett.}\ }\textbf {\bibinfo {volume} {16}},\ \bibinfo {pages} {12202} (\bibinfo {year} {2025}{\natexlab{a}})}\BibitemShut {NoStop}%
\bibitem [{\citenamefont {Zhang}\ \emph {et~al.}(2025)\citenamefont {Zhang}, \citenamefont {Mao}, \citenamefont {Guo},\ and\ \citenamefont {Sun}}]{PhysRevB.111.205417}%
  \BibitemOpen
  \bibfield  {author} {\bibinfo {author} {\bibfnamefont {T.-Y.}\ \bibnamefont {Zhang}}, \bibinfo {author} {\bibfnamefont {Y.}~\bibnamefont {Mao}}, \bibinfo {author} {\bibfnamefont {A.-M.}\ \bibnamefont {Guo}},\ and\ \bibinfo {author} {\bibfnamefont {Q.-F.}\ \bibnamefont {Sun}},\ }\bibfield  {title} {\bibinfo {title} {Dynamical theory of chiral-induced spin selectivity in electron donor--chiral molecule--acceptor systems},\ }\href {https://doi.org/10.1103/PhysRevB.111.205417} {\bibfield  {journal} {\bibinfo  {journal} {Phys. Rev. B}\ }\textbf {\bibinfo {volume} {111}},\ \bibinfo {pages} {205417} (\bibinfo {year} {2025})}\BibitemShut {NoStop}%
\bibitem [{\citenamefont {Matityahu}\ \emph {et~al.}(2016)\citenamefont {Matityahu}, \citenamefont {Utsumi}, \citenamefont {Aharony}, \citenamefont {Entin-Wohlman},\ and\ \citenamefont {Balseiro}}]{Matityahu2016PRB}%
  \BibitemOpen
  \bibfield  {author} {\bibinfo {author} {\bibfnamefont {S.}~\bibnamefont {Matityahu}}, \bibinfo {author} {\bibfnamefont {Y.}~\bibnamefont {Utsumi}}, \bibinfo {author} {\bibfnamefont {A.}~\bibnamefont {Aharony}}, \bibinfo {author} {\bibfnamefont {O.}~\bibnamefont {Entin-Wohlman}},\ and\ \bibinfo {author} {\bibfnamefont {C.~A.}\ \bibnamefont {Balseiro}},\ }\bibfield  {title} {\bibinfo {title} {Spin-dependent transport through a chiral molecule in the presence of spin-orbit interaction and nonunitary effects},\ }\href {https://doi.org/10.1103/PhysRevB.93.075407} {\bibfield  {journal} {\bibinfo  {journal} {Phys. Rev. B}\ }\textbf {\bibinfo {volume} {93}},\ \bibinfo {pages} {075407} (\bibinfo {year} {2016})}\BibitemShut {NoStop}%
\bibitem [{\citenamefont {Fransson}(2025{\natexlab{b}})}]{Fransson2025}%
  \BibitemOpen
  \bibfield  {author} {\bibinfo {author} {\bibfnamefont {J.}~\bibnamefont {Fransson}},\ }\bibfield  {title} {\bibinfo {title} {Chiral induced spin polarized electron current: Origin of the chiral induced spin selectivity effect},\ }\href {https://doi.org/10.1021/acs.jpclett.5c00104} {\bibfield  {journal} {\bibinfo  {journal} {J. Phys. Chem. Lett.}\ }\textbf {\bibinfo {volume} {16}},\ \bibinfo {pages} {4346} (\bibinfo {year} {2025}{\natexlab{b}})}\BibitemShut {NoStop}%
\bibitem [{\citenamefont {Fay}(2021)}]{Fay2021}%
  \BibitemOpen
  \bibfield  {author} {\bibinfo {author} {\bibfnamefont {T.~P.}\ \bibnamefont {Fay}},\ }\bibfield  {title} {\bibinfo {title} {Chirality-induced spin coherence in electron transfer reactions},\ }\href {https://doi.org/10.1021/acs.jpclett.1c00009} {\bibfield  {journal} {\bibinfo  {journal} {J. Phys. Chem. Lett.}\ }\textbf {\bibinfo {volume} {12}},\ \bibinfo {pages} {1407} (\bibinfo {year} {2021})}\BibitemShut {NoStop}%
\bibitem [{\citenamefont {Fay}\ and\ \citenamefont {Limmer}(2021)}]{Fay2021b}%
  \BibitemOpen
  \bibfield  {author} {\bibinfo {author} {\bibfnamefont {T.~P.}\ \bibnamefont {Fay}}\ and\ \bibinfo {author} {\bibfnamefont {D.~T.}\ \bibnamefont {Limmer}},\ }\bibfield  {title} {\bibinfo {title} {Origin of chirality induced spin selectivity in photoinduced electron transfer},\ }\href {https://doi.org/10.1021/acs.nanolett.1c02370} {\bibfield  {journal} {\bibinfo  {journal} {Nano Lett.}\ }\textbf {\bibinfo {volume} {21}},\ \bibinfo {pages} {6696} (\bibinfo {year} {2021})}\BibitemShut {NoStop}%
\bibitem [{\citenamefont {Fransson}(2019)}]{fransson_chirality-induced_2019}%
  \BibitemOpen
  \bibfield  {author} {\bibinfo {author} {\bibfnamefont {J.}~\bibnamefont {Fransson}},\ }\bibfield  {title} {\bibinfo {title} {Chirality-{Induced} {Spin} {Selectivity}: {The} {Role} of {Electron} {Correlations}},\ }\href {https://doi.org/10.1021/acs.jpclett.9b02929} {\bibfield  {journal} {\bibinfo  {journal} {J. Phys. Chem. Lett.}\ }\textbf {\bibinfo {volume} {10}},\ \bibinfo {pages} {7126} (\bibinfo {year} {2019})}\BibitemShut {NoStop}%
\bibitem [{\citenamefont {Chiesa}\ \emph {et~al.}(2024)\citenamefont {Chiesa}, \citenamefont {Garlatti}, \citenamefont {Mezzadri}, \citenamefont {Celada}, \citenamefont {Sessoli}, \citenamefont {Wasielewski}, \citenamefont {Bittl}, \citenamefont {Santini},\ and\ \citenamefont {Carretta}}]{chiesa_many-body_2024}%
  \BibitemOpen
  \bibfield  {author} {\bibinfo {author} {\bibfnamefont {A.}~\bibnamefont {Chiesa}}, \bibinfo {author} {\bibfnamefont {E.}~\bibnamefont {Garlatti}}, \bibinfo {author} {\bibfnamefont {M.}~\bibnamefont {Mezzadri}}, \bibinfo {author} {\bibfnamefont {L.}~\bibnamefont {Celada}}, \bibinfo {author} {\bibfnamefont {R.}~\bibnamefont {Sessoli}}, \bibinfo {author} {\bibfnamefont {M.~R.}\ \bibnamefont {Wasielewski}}, \bibinfo {author} {\bibfnamefont {R.}~\bibnamefont {Bittl}}, \bibinfo {author} {\bibfnamefont {P.}~\bibnamefont {Santini}},\ and\ \bibinfo {author} {\bibfnamefont {S.}~\bibnamefont {Carretta}},\ }\bibfield  {title} {\bibinfo {title} {Many-{Body} {Models} for {Chirality}-{Induced} {Spin} {Selectivity} in {Electron} {Transfer}},\ }\href {https://doi.org/10.1021/acs.nanolett.4c02912} {\bibfield  {journal} {\bibinfo  {journal} {Nano Lett.}\ }\textbf {\bibinfo {volume} {24}},\ \bibinfo {pages} {12133} (\bibinfo {year} {2024})}\BibitemShut {NoStop}%
\bibitem [{\citenamefont {Fransson}(2020)}]{Fransson2020}%
  \BibitemOpen
  \bibfield  {author} {\bibinfo {author} {\bibfnamefont {J.}~\bibnamefont {Fransson}},\ }\bibfield  {title} {\bibinfo {title} {Vibrational origin of exchange splitting and ''chiral-induced spin selectivity},\ }\href {https://doi.org/10.1103/PhysRevB.102.235416} {\bibfield  {journal} {\bibinfo  {journal} {Phys. Rev. B}\ }\textbf {\bibinfo {volume} {102}},\ \bibinfo {pages} {235416} (\bibinfo {year} {2020})}\BibitemShut {NoStop}%
\bibitem [{\citenamefont {Fransson}(2021)}]{Fransson2021}%
  \BibitemOpen
  \bibfield  {author} {\bibinfo {author} {\bibfnamefont {J.}~\bibnamefont {Fransson}},\ }\bibfield  {title} {\bibinfo {title} {Charge redistribution and spin polarization driven by correlation induced electron exchange in chiral molecules},\ }\href {https://doi.org/10.1021/acs.nanolett.1c00183} {\bibfield  {journal} {\bibinfo  {journal} {Nano Lett.}\ }\textbf {\bibinfo {volume} {21}},\ \bibinfo {pages} {3026} (\bibinfo {year} {2021})}\BibitemShut {NoStop}%
\bibitem [{\citenamefont {Das}\ \emph {et~al.}(2022)\citenamefont {Das}, \citenamefont {Tassinari}, \citenamefont {Naaman},\ and\ \citenamefont {Fransson}}]{das_temperature-dependent_2022}%
  \BibitemOpen
  \bibfield  {author} {\bibinfo {author} {\bibfnamefont {T.~K.}\ \bibnamefont {Das}}, \bibinfo {author} {\bibfnamefont {F.}~\bibnamefont {Tassinari}}, \bibinfo {author} {\bibfnamefont {R.}~\bibnamefont {Naaman}},\ and\ \bibinfo {author} {\bibfnamefont {J.}~\bibnamefont {Fransson}},\ }\bibfield  {title} {\bibinfo {title} {Temperature-{Dependent} {Chiral}-{Induced} {Spin} {Selectivity} {Effect}: {Experiments} and {Theory}},\ }\href {https://doi.org/10.1021/acs.jpcc.1c10550} {\bibfield  {journal} {\bibinfo  {journal} {J. Phys. Chem. C}\ }\textbf {\bibinfo {volume} {126}},\ \bibinfo {pages} {3257} (\bibinfo {year} {2022})}\BibitemShut {NoStop}%
\bibitem [{\citenamefont {Chandran}\ \emph {et~al.}(2022)\citenamefont {Chandran}, \citenamefont {Wu}, \citenamefont {Teh}, \citenamefont {Waldeck},\ and\ \citenamefont {Subotnik}}]{Subotnik2022}%
  \BibitemOpen
  \bibfield  {author} {\bibinfo {author} {\bibfnamefont {S.~S.}\ \bibnamefont {Chandran}}, \bibinfo {author} {\bibfnamefont {Y.}~\bibnamefont {Wu}}, \bibinfo {author} {\bibfnamefont {H.-H.}\ \bibnamefont {Teh}}, \bibinfo {author} {\bibfnamefont {D.~H.}\ \bibnamefont {Waldeck}},\ and\ \bibinfo {author} {\bibfnamefont {J.~E.}\ \bibnamefont {Subotnik}},\ }\bibfield  {title} {\bibinfo {title} {{Electron transfer and spin–orbit coupling: Can nuclear motion lead to spin selective rates?}},\ }\href {https://doi.org/10.1063/5.0086554} {\bibfield  {journal} {\bibinfo  {journal} {J. Chem. Phys.}\ }\textbf {\bibinfo {volume} {156}},\ \bibinfo {pages} {174113} (\bibinfo {year} {2022})}\BibitemShut {NoStop}%
\bibitem [{\citenamefont {Peralta}\ \emph {et~al.}(2023)\citenamefont {Peralta}, \citenamefont {Feijoo}, \citenamefont {Varela}, \citenamefont {Gutierrez}, \citenamefont {Cuniberti}, \citenamefont {Mujica},\ and\ \citenamefont {Medina}}]{Cuniberti2023}%
  \BibitemOpen
  \bibfield  {author} {\bibinfo {author} {\bibfnamefont {M.}~\bibnamefont {Peralta}}, \bibinfo {author} {\bibfnamefont {S.}~\bibnamefont {Feijoo}}, \bibinfo {author} {\bibfnamefont {S.}~\bibnamefont {Varela}}, \bibinfo {author} {\bibfnamefont {R.}~\bibnamefont {Gutierrez}}, \bibinfo {author} {\bibfnamefont {G.}~\bibnamefont {Cuniberti}}, \bibinfo {author} {\bibfnamefont {V.}~\bibnamefont {Mujica}},\ and\ \bibinfo {author} {\bibfnamefont {E.}~\bibnamefont {Medina}},\ }\bibfield  {title} {\bibinfo {title} {Spin-phonon coupling in a double-stranded model of dna},\ }\bibfield  {journal} {\bibinfo  {journal} {J. Chem. Phys.}\ }\textbf {\bibinfo {volume} {159}},\ \href {https://doi.org/10.1063/5.0156347} {10.1063/5.0156347} (\bibinfo {year} {2023})\BibitemShut {NoStop}%
\bibitem [{\citenamefont {Rudge}\ \emph {et~al.}(2025)\citenamefont {Rudge}, \citenamefont {Kaspar}, \citenamefont {Smorka}, \citenamefont {Preston}, \citenamefont {Subotnik},\ and\ \citenamefont {Thoss}}]{Rudge2025}%
  \BibitemOpen
  \bibfield  {author} {\bibinfo {author} {\bibfnamefont {S.~L.}\ \bibnamefont {Rudge}}, \bibinfo {author} {\bibfnamefont {C.}~\bibnamefont {Kaspar}}, \bibinfo {author} {\bibfnamefont {R.}~\bibnamefont {Smorka}}, \bibinfo {author} {\bibfnamefont {R.~J.}\ \bibnamefont {Preston}}, \bibinfo {author} {\bibfnamefont {J.}~\bibnamefont {Subotnik}},\ and\ \bibinfo {author} {\bibfnamefont {M.}~\bibnamefont {Thoss}},\ }\bibfield  {title} {\bibinfo {title} {The role of quantum vibronic effects in the spin polarization of charge transport through molecular junctions},\ }\href {https://doi.org/10.1063/5.0270610} {\bibfield  {journal} {\bibinfo  {journal} {J. Chem. Phys.}\ }\textbf {\bibinfo {volume} {162}},\ \bibinfo {pages} {244710} (\bibinfo {year} {2025})}\BibitemShut {NoStop}%
\bibitem [{\citenamefont {Savi}\ \emph {et~al.}(2025{\natexlab{a}})\citenamefont {Savi}, \citenamefont {Celada}, \citenamefont {Phan~Huu}, \citenamefont {Chiesa}, \citenamefont {Carretta},\ and\ \citenamefont {Painelli}}]{NostroJPCL}%
  \BibitemOpen
  \bibfield  {author} {\bibinfo {author} {\bibfnamefont {L.}~\bibnamefont {Savi}}, \bibinfo {author} {\bibfnamefont {L.}~\bibnamefont {Celada}}, \bibinfo {author} {\bibfnamefont {D.~A.}\ \bibnamefont {Phan~Huu}}, \bibinfo {author} {\bibfnamefont {A.}~\bibnamefont {Chiesa}}, \bibinfo {author} {\bibfnamefont {S.}~\bibnamefont {Carretta}},\ and\ \bibinfo {author} {\bibfnamefont {A.}~\bibnamefont {Painelli}},\ }\bibfield  {title} {\bibinfo {title} {Chirality-induced spin selectivity: A minimal model},\ }\href {https://doi.org/10.1021/acs.jpclett.5c01813} {\bibfield  {journal} {\bibinfo  {journal} {J. Phys. Chem. Lett.}\ }\textbf {\bibinfo {volume} {16}},\ \bibinfo {pages} {9107} (\bibinfo {year} {2025}{\natexlab{a}})},\ \bibinfo {note} {pMID: 40857217}\BibitemShut {NoStop}%
\bibitem [{\citenamefont {Fransson}(2026)}]{Fransson2026}%
  \BibitemOpen
  \bibfield  {author} {\bibinfo {author} {\bibfnamefont {J.}~\bibnamefont {Fransson}},\ }\bibfield  {title} {\bibinfo {title} {Breaking of time-reversal symmetry and onsager reciprocity in chiral molecule interfaced with an environment},\ }\href {https://doi.org/10.1063/5.0312951} {\bibfield  {journal} {\bibinfo  {journal} {APL Comput. Phys.}\ }\textbf {\bibinfo {volume} {2}},\ \bibinfo {pages} {016111} (\bibinfo {year} {2026})}\BibitemShut {NoStop}%
\bibitem [{\citenamefont {Phan~Huu}\ \emph {et~al.}(2025)\citenamefont {Phan~Huu}, \citenamefont {Cantarella}, \citenamefont {Bonfà}, \citenamefont {Savi}, \citenamefont {Chiesa}, \citenamefont {Painelli},\ and\ \citenamefont {Carretta}}]{phan_huu_ab_2025}%
  \BibitemOpen
  \bibfield  {author} {\bibinfo {author} {\bibfnamefont {D.~K.~A.}\ \bibnamefont {Phan~Huu}}, \bibinfo {author} {\bibfnamefont {A.}~\bibnamefont {Cantarella}}, \bibinfo {author} {\bibfnamefont {P.}~\bibnamefont {Bonfà}}, \bibinfo {author} {\bibfnamefont {L.}~\bibnamefont {Savi}}, \bibinfo {author} {\bibfnamefont {A.}~\bibnamefont {Chiesa}}, \bibinfo {author} {\bibfnamefont {A.}~\bibnamefont {Painelli}},\ and\ \bibinfo {author} {\bibfnamefont {S.}~\bibnamefont {Carretta}},\ }\bibfield  {title} {\bibinfo {title} {Ab initio parametrization of a generalized {Hubbard} model in a molecule displaying chirality-induced spin selectivity},\ }\href {https://doi.org/10.1038/s43246-025-00821-3} {\bibfield  {journal} {\bibinfo  {journal} {Commun. Mater.}\ }\textbf {\bibinfo {volume} {6}},\ \bibinfo {pages} {1} (\bibinfo {year} {2025})}\BibitemShut {NoStop}%
\bibitem [{\citenamefont {Feng}\ \emph {et~al.}(2025)\citenamefont {Feng}, \citenamefont {Abraham}, \citenamefont {Subotnik},\ and\ \citenamefont {Nitzan}}]{NitzanandSubo2025}%
  \BibitemOpen
  \bibfield  {author} {\bibinfo {author} {\bibfnamefont {J.}~\bibnamefont {Feng}}, \bibinfo {author} {\bibfnamefont {E.}~\bibnamefont {Abraham}}, \bibinfo {author} {\bibfnamefont {J.}~\bibnamefont {Subotnik}},\ and\ \bibinfo {author} {\bibfnamefont {A.}~\bibnamefont {Nitzan}},\ }\bibfield  {title} {\bibinfo {title} {Chiral vibrational modes in small molecules},\ }\href {https://doi.org/10.1063/5.0271584} {\bibfield  {journal} {\bibinfo  {journal} {J. Chem. Phys.}\ }\textbf {\bibinfo {volume} {163}},\ \bibinfo {pages} {014106} (\bibinfo {year} {2025})}\BibitemShut {NoStop}%
\bibitem [{\citenamefont {Bissesar}\ \emph {et~al.}(2022)\citenamefont {Bissesar}, \citenamefont {van Raamsdonk}, \citenamefont {Gibbons},\ and\ \citenamefont {Williams}}]{Bissesar2022}%
  \BibitemOpen
  \bibfield  {author} {\bibinfo {author} {\bibfnamefont {S.}~\bibnamefont {Bissesar}}, \bibinfo {author} {\bibfnamefont {D.~M.~E.}\ \bibnamefont {van Raamsdonk}}, \bibinfo {author} {\bibfnamefont {D.~J.}\ \bibnamefont {Gibbons}},\ and\ \bibinfo {author} {\bibfnamefont {R.~M.}\ \bibnamefont {Williams}},\ }\bibfield  {title} {\bibinfo {title} {Spin orbit coupling in orthogonal charge transfer states: {(TD-)DFT} of pyrene-dimethylaniline},\ }\href {https://doi.org/10.3390/molecules27030891} {\bibfield  {journal} {\bibinfo  {journal} {Molecules}\ }\textbf {\bibinfo {volume} {27}},\ \bibinfo {pages} {891} (\bibinfo {year} {2022})}\BibitemShut {NoStop}%
\bibitem [{\citenamefont {Dhali}\ \emph {et~al.}(2021)\citenamefont {Dhali}, \citenamefont {Phan~Huu}, \citenamefont {Bertocchi}, \citenamefont {Sissa}, \citenamefont {Terenziani},\ and\ \citenamefont {Painelli}}]{dhali2021understanding}%
  \BibitemOpen
  \bibfield  {author} {\bibinfo {author} {\bibfnamefont {R.}~\bibnamefont {Dhali}}, \bibinfo {author} {\bibfnamefont {D.~K.~A.}\ \bibnamefont {Phan~Huu}}, \bibinfo {author} {\bibfnamefont {F.}~\bibnamefont {Bertocchi}}, \bibinfo {author} {\bibfnamefont {C.}~\bibnamefont {Sissa}}, \bibinfo {author} {\bibfnamefont {F.}~\bibnamefont {Terenziani}},\ and\ \bibinfo {author} {\bibfnamefont {A.}~\bibnamefont {Painelli}},\ }\bibfield  {title} {\bibinfo {title} {Understanding tadf: a joint experimental and theoretical study of dmac-trz},\ }\href {http://dx.doi.org/10.1039/D0CP05982J} {\bibfield  {journal} {\bibinfo  {journal} {Phys. Chem. Chem. Phys.}\ }\textbf {\bibinfo {volume} {23}},\ \bibinfo {pages} {378} (\bibinfo {year} {2021})}\BibitemShut {NoStop}%
\bibitem [{\citenamefont {Poh}\ \emph {et~al.}(2024)\citenamefont {Poh}, \citenamefont {Morozov}, \citenamefont {Kazmierczak}, \citenamefont {Hadt}, \citenamefont {Groenhof},\ and\ \citenamefont {Yuen-Zhou}}]{Poh2024}%
  \BibitemOpen
  \bibfield  {author} {\bibinfo {author} {\bibfnamefont {Y.~R.}\ \bibnamefont {Poh}}, \bibinfo {author} {\bibfnamefont {D.}~\bibnamefont {Morozov}}, \bibinfo {author} {\bibfnamefont {N.~P.}\ \bibnamefont {Kazmierczak}}, \bibinfo {author} {\bibfnamefont {R.~G.}\ \bibnamefont {Hadt}}, \bibinfo {author} {\bibfnamefont {G.}~\bibnamefont {Groenhof}},\ and\ \bibinfo {author} {\bibfnamefont {J.}~\bibnamefont {Yuen-Zhou}},\ }\bibfield  {title} {\bibinfo {title} {Alternant hydrocarbon diradicals as optically addressable molecular qubits},\ }\href {https://doi.org/10.1021/jacs.4c04360} {\bibfield  {journal} {\bibinfo  {journal} {J. Am. Chem. Soc.}\ }\textbf {\bibinfo {volume} {146}},\ \bibinfo {pages} {15549} (\bibinfo {year} {2024})}\BibitemShut {NoStop}%
\bibitem [{\citenamefont {Savi}\ \emph {et~al.}(2025{\natexlab{b}})\citenamefont {Savi}, \citenamefont {Barreca}, \citenamefont {Bedogni},\ and\ \citenamefont {Di~Maiolo}}]{savi2025organic}%
  \BibitemOpen
  \bibfield  {author} {\bibinfo {author} {\bibfnamefont {L.}~\bibnamefont {Savi}}, \bibinfo {author} {\bibfnamefont {M.~T.}\ \bibnamefont {Barreca}}, \bibinfo {author} {\bibfnamefont {M.}~\bibnamefont {Bedogni}},\ and\ \bibinfo {author} {\bibfnamefont {F.}~\bibnamefont {Di~Maiolo}},\ }\bibfield  {title} {\bibinfo {title} {Organic diradicals bridged by inverted singlet--triplet units for optical--spin interfaces},\ }\href@noop {} {\bibfield  {journal} {\bibinfo  {journal} {Journal of Chemical Theory and Computation}\ } (\bibinfo {year} {2025}{\natexlab{b}})}\BibitemShut {NoStop}%
\bibitem [{Note1()}]{Note1}%
  \BibitemOpen
  \bibinfo {note} {The electronic system could also be coupled to Holstein modes modulating on-site energies, but these cannot mediate a spin-spin interaction and hence are not considered here}\BibitemShut {NoStop}%
\bibitem [{\citenamefont {Penfold}\ \emph {et~al.}(2018)\citenamefont {Penfold}, \citenamefont {Gindensperger}, \citenamefont {Daniel},\ and\ \citenamefont {Marian}}]{Marian2018}%
  \BibitemOpen
  \bibfield  {author} {\bibinfo {author} {\bibfnamefont {T.~J.}\ \bibnamefont {Penfold}}, \bibinfo {author} {\bibfnamefont {E.}~\bibnamefont {Gindensperger}}, \bibinfo {author} {\bibfnamefont {C.}~\bibnamefont {Daniel}},\ and\ \bibinfo {author} {\bibfnamefont {C.~M.}\ \bibnamefont {Marian}},\ }\bibfield  {title} {\bibinfo {title} {Spin-vibronic mechanism for intersystem crossing},\ }\href {https://doi.org/10.1021/acs.chemrev.7b00617} {\bibfield  {journal} {\bibinfo  {journal} {Chem. rev.}\ }\textbf {\bibinfo {volume} {118}},\ \bibinfo {pages} {6975} (\bibinfo {year} {2018})}\BibitemShut {NoStop}%
\bibitem [{\citenamefont {Manian}\ \emph {et~al.}(2025)\citenamefont {Manian}, \citenamefont {Paz},\ and\ \citenamefont {Yu}}]{manian2025vibronic}%
  \BibitemOpen
  \bibfield  {author} {\bibinfo {author} {\bibfnamefont {A.}~\bibnamefont {Manian}}, \bibinfo {author} {\bibfnamefont {H.}~\bibnamefont {Paz}},\ and\ \bibinfo {author} {\bibfnamefont {H.}~\bibnamefont {Yu}},\ }\bibfield  {title} {\bibinfo {title} {Vibronic contributions to hyperfine-mediated spin kinetics},\ }\href@noop {} {\bibfield  {journal} {\bibinfo  {journal} {ChemRxiv}\ } (\bibinfo {year} {2025})}\BibitemShut {NoStop}%
\bibitem [{\citenamefont {Phan~Huu}\ \emph {et~al.}(2022)\citenamefont {Phan~Huu}, \citenamefont {Saseendran}, \citenamefont {Dhali}, \citenamefont {Franca}, \citenamefont {Stavrou}, \citenamefont {Monkman},\ and\ \citenamefont {Painelli}}]{phan2022thermally}%
  \BibitemOpen
  \bibfield  {author} {\bibinfo {author} {\bibfnamefont {D.~K.~A.}\ \bibnamefont {Phan~Huu}}, \bibinfo {author} {\bibfnamefont {S.}~\bibnamefont {Saseendran}}, \bibinfo {author} {\bibfnamefont {R.}~\bibnamefont {Dhali}}, \bibinfo {author} {\bibfnamefont {L.~G.}\ \bibnamefont {Franca}}, \bibinfo {author} {\bibfnamefont {K.}~\bibnamefont {Stavrou}}, \bibinfo {author} {\bibfnamefont {A.}~\bibnamefont {Monkman}},\ and\ \bibinfo {author} {\bibfnamefont {A.}~\bibnamefont {Painelli}},\ }\bibfield  {title} {\bibinfo {title} {Thermally activated delayed fluorescence: polarity, rigidity, and disorder in condensed phases},\ }\href {https://doi.org/10.1021/jacs.2c05537} {\bibfield  {journal} {\bibinfo  {journal} {J. Am. Chem. Soc.}\ }\textbf {\bibinfo {volume} {144}},\ \bibinfo {pages} {15211} (\bibinfo {year} {2022})}\BibitemShut {NoStop}%
\bibitem [{Note2()}]{Note2}%
  \BibitemOpen
  \bibinfo {note} {In principle, any system where inversion symmetry is broken could display the DMI considered in the present model.}\BibitemShut {Stop}%
\bibitem [{\citenamefont {You}\ \emph {et~al.}(2015)\citenamefont {You}, \citenamefont {Mewes}, \citenamefont {Dreuw},\ and\ \citenamefont {Herbert}}]{FMO}%
  \BibitemOpen
  \bibfield  {author} {\bibinfo {author} {\bibfnamefont {Z.-Q.}\ \bibnamefont {You}}, \bibinfo {author} {\bibfnamefont {J.-M.}\ \bibnamefont {Mewes}}, \bibinfo {author} {\bibfnamefont {A.}~\bibnamefont {Dreuw}},\ and\ \bibinfo {author} {\bibfnamefont {J.~M.}\ \bibnamefont {Herbert}},\ }\bibfield  {title} {\bibinfo {title} {Comparison of the marcus and pekar partitions in the context of non-equilibrium, polarizable-continuum solvation models},\ }\href {https://doi.org/10.1063/1.4936357} {\bibfield  {journal} {\bibinfo  {journal} {The Journal of Chemical Physics}\ }\textbf {\bibinfo {volume} {143}},\ \bibinfo {pages} {204104} (\bibinfo {year} {2015})}\BibitemShut {NoStop}%
\bibitem [{\citenamefont {Neese}\ \emph {et~al.}(2020)\citenamefont {Neese}, \citenamefont {Wennmohs}, \citenamefont {Becker},\ and\ \citenamefont {Riplinger}}]{orca}%
  \BibitemOpen
  \bibfield  {author} {\bibinfo {author} {\bibfnamefont {F.}~\bibnamefont {Neese}}, \bibinfo {author} {\bibfnamefont {F.}~\bibnamefont {Wennmohs}}, \bibinfo {author} {\bibfnamefont {U.}~\bibnamefont {Becker}},\ and\ \bibinfo {author} {\bibfnamefont {C.}~\bibnamefont {Riplinger}},\ }\bibfield  {title} {\bibinfo {title} {The orca quantum chemistry program package},\ }\href {https://doi.org/10.1063/5.0004608} {\bibfield  {journal} {\bibinfo  {journal} {J. Chem. Phys.}\ }\textbf {\bibinfo {volume} {152}},\ \bibinfo {pages} {224108} (\bibinfo {year} {2020})}\BibitemShut {NoStop}%
\bibitem [{\citenamefont {Epifanovsky}\ \emph {et~al.}(2021)\citenamefont {Epifanovsky}, \citenamefont {Gilbert}, \citenamefont {Feng}, \citenamefont {Lee}, \citenamefont {Mao}, \citenamefont {Mardirossian}, \citenamefont {Pokhilko}, \citenamefont {White}, \citenamefont {Coons}, \citenamefont {Dempwolff}, \citenamefont {Gan}, \citenamefont {Hait}, \citenamefont {Horn}, \citenamefont {Jacobson}, \citenamefont {Kaliman}, \citenamefont {Kussmann}, \citenamefont {Lange}, \citenamefont {Lao}, \citenamefont {Levine}, \citenamefont {Liu}, \citenamefont {McKenzie}, \citenamefont {Morrison}, \citenamefont {Nanda}, \citenamefont {Plasser}, \citenamefont {Rehn}, \citenamefont {Vidal}, \citenamefont {You}, \citenamefont {Zhu}, \citenamefont {Alam}, \citenamefont {Albrecht}, \citenamefont {Aldossary}, \citenamefont {Alguire}, \citenamefont {Andersen}, \citenamefont {Athavale}, \citenamefont {Barton}, \citenamefont {Begam}, \citenamefont {Behn}, \citenamefont {Bellonzi}, \citenamefont {Bernard}, \citenamefont {Berquist},
  \citenamefont {Burton}, \citenamefont {Carreras}, \citenamefont {Carter-Fenk}, \citenamefont {Chakraborty}, \citenamefont {Chien}, \citenamefont {Closser}, \citenamefont {Cofer-Shabica}, \citenamefont {Dasgupta}, \citenamefont {de~Wergifosse}, \citenamefont {Deng}, \citenamefont {Diedenhofen}, \citenamefont {Do}, \citenamefont {Ehlert}, \citenamefont {Fang}, \citenamefont {Fatehi}, \citenamefont {Feng}, \citenamefont {Friedhoff}, \citenamefont {Gayvert}, \citenamefont {Ge}, \citenamefont {Gidofalvi}, \citenamefont {Goldey}, \citenamefont {Gomes}, \citenamefont {González-Espinoza}, \citenamefont {Gulania}, \citenamefont {Gunina}, \citenamefont {Hanson-Heine}, \citenamefont {Harbach}, \citenamefont {Hauser}, \citenamefont {Herbst}, \citenamefont {Hernández~Vera}, \citenamefont {Hodecker}, \citenamefont {Holden}, \citenamefont {Houck}, \citenamefont {Huang}, \citenamefont {Hui}, \citenamefont {Huynh}, \citenamefont {Ivanov}, \citenamefont {Jász}, \citenamefont {Ji}, \citenamefont {Jiang}, \citenamefont
  {Kaduk}, \citenamefont {Kähler}, \citenamefont {Khistyaev}, \citenamefont {Kim}, \citenamefont {Kis}, \citenamefont {Klunzinger}, \citenamefont {Koczor-Benda}, \citenamefont {Koh}, \citenamefont {Kosenkov}, \citenamefont {Koulias}, \citenamefont {Kowalczyk}, \citenamefont {Krauter}, \citenamefont {Kue}, \citenamefont {Kunitsa}, \citenamefont {Kus}, \citenamefont {Ladjánszki}, \citenamefont {Landau}, \citenamefont {Lawler}, \citenamefont {Lefrancois}, \citenamefont {Lehtola}, \citenamefont {Li}, \citenamefont {Li}, \citenamefont {Liang}, \citenamefont {Liebenthal}, \citenamefont {Lin}, \citenamefont {Lin}, \citenamefont {Liu}, \citenamefont {Liu}, \citenamefont {Loipersberger}, \citenamefont {Luenser}, \citenamefont {Manjanath}, \citenamefont {Manohar}, \citenamefont {Mansoor}, \citenamefont {Manzer}, \citenamefont {Mao}, \citenamefont {Marenich}, \citenamefont {Markovich}, \citenamefont {Mason}, \citenamefont {Maurer}, \citenamefont {McLaughlin}, \citenamefont {Menger}, \citenamefont {Mewes},
  \citenamefont {Mewes}, \citenamefont {Morgante}, \citenamefont {Mullinax}, \citenamefont {Oosterbaan}, \citenamefont {Paran}, \citenamefont {Paul}, \citenamefont {Paul}, \citenamefont {Pavošević}, \citenamefont {Pei}, \citenamefont {Prager}, \citenamefont {Proynov}, \citenamefont {Rák}, \citenamefont {Ramos-Cordoba}, \citenamefont {Rana}, \citenamefont {Rask}, \citenamefont {Rettig}, \citenamefont {Richard}, \citenamefont {Rob}, \citenamefont {Rossomme}, \citenamefont {Scheele}, \citenamefont {Scheurer}, \citenamefont {Schneider}, \citenamefont {Sergueev}, \citenamefont {Sharada}, \citenamefont {Skomorowski}, \citenamefont {Small}, \citenamefont {Stein}, \citenamefont {Su}, \citenamefont {Sundstrom}, \citenamefont {Tao}, \citenamefont {Thirman}, \citenamefont {Tornai}, \citenamefont {Tsuchimochi}, \citenamefont {Tubman}, \citenamefont {Veccham}, \citenamefont {Vydrov}, \citenamefont {Wenzel}, \citenamefont {Witte}, \citenamefont {Yamada}, \citenamefont {Yao}, \citenamefont {Yeganeh}, \citenamefont
  {Yost}, \citenamefont {Zech}, \citenamefont {Zhang}, \citenamefont {Zhang}, \citenamefont {Zhang}, \citenamefont {Zuev}, \citenamefont {Aspuru-Guzik}, \citenamefont {Bell}, \citenamefont {Besley}, \citenamefont {Bravaya}, \citenamefont {Brooks}, \citenamefont {Casanova}, \citenamefont {Chai}, \citenamefont {Coriani}, \citenamefont {Cramer}, \citenamefont {Cserey}, \citenamefont {DePrince}, \citenamefont {DiStasio}, \citenamefont {Dreuw}, \citenamefont {Dunietz}, \citenamefont {Furlani}, \citenamefont {Goddard}, \citenamefont {Hammes-Schiffer}, \citenamefont {Head-Gordon}, \citenamefont {Hehre}, \citenamefont {Hsu}, \citenamefont {Jagau}, \citenamefont {Jung}, \citenamefont {Klamt}, \citenamefont {Kong}, \citenamefont {Lambrecht}, \citenamefont {Liang}, \citenamefont {Mayhall}, \citenamefont {McCurdy}, \citenamefont {Neaton}, \citenamefont {Ochsenfeld}, \citenamefont {Parkhill}, \citenamefont {Peverati}, \citenamefont {Rassolov}, \citenamefont {Shao}, \citenamefont {Slipchenko}, \citenamefont {Stauch},
  \citenamefont {Steele}, \citenamefont {Subotnik}, \citenamefont {Thom}, \citenamefont {Tkatchenko}, \citenamefont {Truhlar}, \citenamefont {Van~Voorhis}, \citenamefont {Wesolowski}, \citenamefont {Whaley}, \citenamefont {Woodcock}, \citenamefont {Zimmerman}, \citenamefont {Faraji}, \citenamefont {Gill}, \citenamefont {Head-Gordon}, \citenamefont {Herbert},\ and\ \citenamefont {Krylov}}]{qchem}%
  \BibitemOpen
  \bibfield  {author} {\bibinfo {author} {\bibfnamefont {E.}~\bibnamefont {Epifanovsky}}, \bibinfo {author} {\bibfnamefont {A.~T.~B.}\ \bibnamefont {Gilbert}}, \bibinfo {author} {\bibfnamefont {X.}~\bibnamefont {Feng}}, \bibinfo {author} {\bibfnamefont {J.}~\bibnamefont {Lee}}, \bibinfo {author} {\bibfnamefont {Y.}~\bibnamefont {Mao}}, \bibinfo {author} {\bibfnamefont {N.}~\bibnamefont {Mardirossian}}, \bibinfo {author} {\bibfnamefont {P.}~\bibnamefont {Pokhilko}}, \bibinfo {author} {\bibfnamefont {A.~F.}\ \bibnamefont {White}}, \bibinfo {author} {\bibfnamefont {M.~P.}\ \bibnamefont {Coons}}, \bibinfo {author} {\bibfnamefont {A.~L.}\ \bibnamefont {Dempwolff}}, \bibinfo {author} {\bibfnamefont {Z.}~\bibnamefont {Gan}}, \bibinfo {author} {\bibfnamefont {D.}~\bibnamefont {Hait}}, \bibinfo {author} {\bibfnamefont {P.~R.}\ \bibnamefont {Horn}}, \bibinfo {author} {\bibfnamefont {L.~D.}\ \bibnamefont {Jacobson}}, \bibinfo {author} {\bibfnamefont {I.}~\bibnamefont {Kaliman}}, \bibinfo {author} {\bibfnamefont
  {J.}~\bibnamefont {Kussmann}}, \bibinfo {author} {\bibfnamefont {A.~W.}\ \bibnamefont {Lange}}, \bibinfo {author} {\bibfnamefont {K.~U.}\ \bibnamefont {Lao}}, \bibinfo {author} {\bibfnamefont {D.~S.}\ \bibnamefont {Levine}}, \bibinfo {author} {\bibfnamefont {J.}~\bibnamefont {Liu}}, \bibinfo {author} {\bibfnamefont {S.~C.}\ \bibnamefont {McKenzie}}, \bibinfo {author} {\bibfnamefont {A.~F.}\ \bibnamefont {Morrison}}, \bibinfo {author} {\bibfnamefont {K.~D.}\ \bibnamefont {Nanda}}, \bibinfo {author} {\bibfnamefont {F.}~\bibnamefont {Plasser}}, \bibinfo {author} {\bibfnamefont {D.~R.}\ \bibnamefont {Rehn}}, \bibinfo {author} {\bibfnamefont {M.~L.}\ \bibnamefont {Vidal}}, \bibinfo {author} {\bibfnamefont {Z.-Q.}\ \bibnamefont {You}}, \bibinfo {author} {\bibfnamefont {Y.}~\bibnamefont {Zhu}}, \bibinfo {author} {\bibfnamefont {B.}~\bibnamefont {Alam}}, \bibinfo {author} {\bibfnamefont {B.~J.}\ \bibnamefont {Albrecht}}, \bibinfo {author} {\bibfnamefont {A.}~\bibnamefont {Aldossary}}, \bibinfo {author}
  {\bibfnamefont {E.}~\bibnamefont {Alguire}}, \bibinfo {author} {\bibfnamefont {J.~H.}\ \bibnamefont {Andersen}}, \bibinfo {author} {\bibfnamefont {V.}~\bibnamefont {Athavale}}, \bibinfo {author} {\bibfnamefont {D.}~\bibnamefont {Barton}}, \bibinfo {author} {\bibfnamefont {K.}~\bibnamefont {Begam}}, \bibinfo {author} {\bibfnamefont {A.}~\bibnamefont {Behn}}, \bibinfo {author} {\bibfnamefont {N.}~\bibnamefont {Bellonzi}}, \bibinfo {author} {\bibfnamefont {Y.~A.}\ \bibnamefont {Bernard}}, \bibinfo {author} {\bibfnamefont {E.~J.}\ \bibnamefont {Berquist}}, \bibinfo {author} {\bibfnamefont {H.~G.~A.}\ \bibnamefont {Burton}}, \bibinfo {author} {\bibfnamefont {A.}~\bibnamefont {Carreras}}, \bibinfo {author} {\bibfnamefont {K.}~\bibnamefont {Carter-Fenk}}, \bibinfo {author} {\bibfnamefont {R.}~\bibnamefont {Chakraborty}}, \bibinfo {author} {\bibfnamefont {A.~D.}\ \bibnamefont {Chien}}, \bibinfo {author} {\bibfnamefont {K.~D.}\ \bibnamefont {Closser}}, \bibinfo {author} {\bibfnamefont {V.}~\bibnamefont
  {Cofer-Shabica}}, \bibinfo {author} {\bibfnamefont {S.}~\bibnamefont {Dasgupta}}, \bibinfo {author} {\bibfnamefont {M.}~\bibnamefont {de~Wergifosse}}, \bibinfo {author} {\bibfnamefont {J.}~\bibnamefont {Deng}}, \bibinfo {author} {\bibfnamefont {M.}~\bibnamefont {Diedenhofen}}, \bibinfo {author} {\bibfnamefont {H.}~\bibnamefont {Do}}, \bibinfo {author} {\bibfnamefont {S.}~\bibnamefont {Ehlert}}, \bibinfo {author} {\bibfnamefont {P.-T.}\ \bibnamefont {Fang}}, \bibinfo {author} {\bibfnamefont {S.}~\bibnamefont {Fatehi}}, \bibinfo {author} {\bibfnamefont {Q.}~\bibnamefont {Feng}}, \bibinfo {author} {\bibfnamefont {T.}~\bibnamefont {Friedhoff}}, \bibinfo {author} {\bibfnamefont {J.}~\bibnamefont {Gayvert}}, \bibinfo {author} {\bibfnamefont {Q.}~\bibnamefont {Ge}}, \bibinfo {author} {\bibfnamefont {G.}~\bibnamefont {Gidofalvi}}, \bibinfo {author} {\bibfnamefont {M.}~\bibnamefont {Goldey}}, \bibinfo {author} {\bibfnamefont {J.}~\bibnamefont {Gomes}}, \bibinfo {author} {\bibfnamefont {C.~E.}\ \bibnamefont
  {González-Espinoza}}, \bibinfo {author} {\bibfnamefont {S.}~\bibnamefont {Gulania}}, \bibinfo {author} {\bibfnamefont {A.~O.}\ \bibnamefont {Gunina}}, \bibinfo {author} {\bibfnamefont {M.~W.~D.}\ \bibnamefont {Hanson-Heine}}, \bibinfo {author} {\bibfnamefont {P.~H.~P.}\ \bibnamefont {Harbach}}, \bibinfo {author} {\bibfnamefont {A.}~\bibnamefont {Hauser}}, \bibinfo {author} {\bibfnamefont {M.~F.}\ \bibnamefont {Herbst}}, \bibinfo {author} {\bibfnamefont {M.}~\bibnamefont {Hernández~Vera}}, \bibinfo {author} {\bibfnamefont {M.}~\bibnamefont {Hodecker}}, \bibinfo {author} {\bibfnamefont {Z.~C.}\ \bibnamefont {Holden}}, \bibinfo {author} {\bibfnamefont {S.}~\bibnamefont {Houck}}, \bibinfo {author} {\bibfnamefont {X.}~\bibnamefont {Huang}}, \bibinfo {author} {\bibfnamefont {K.}~\bibnamefont {Hui}}, \bibinfo {author} {\bibfnamefont {B.~C.}\ \bibnamefont {Huynh}}, \bibinfo {author} {\bibfnamefont {M.}~\bibnamefont {Ivanov}}, \bibinfo {author} {\bibfnamefont { .}~\bibnamefont {Jász}}, \bibinfo {author}
  {\bibfnamefont {H.}~\bibnamefont {Ji}}, \bibinfo {author} {\bibfnamefont {H.}~\bibnamefont {Jiang}}, \bibinfo {author} {\bibfnamefont {B.}~\bibnamefont {Kaduk}}, \bibinfo {author} {\bibfnamefont {S.}~\bibnamefont {Kähler}}, \bibinfo {author} {\bibfnamefont {K.}~\bibnamefont {Khistyaev}}, \bibinfo {author} {\bibfnamefont {J.}~\bibnamefont {Kim}}, \bibinfo {author} {\bibfnamefont {G.}~\bibnamefont {Kis}}, \bibinfo {author} {\bibfnamefont {P.}~\bibnamefont {Klunzinger}}, \bibinfo {author} {\bibfnamefont {Z.}~\bibnamefont {Koczor-Benda}}, \bibinfo {author} {\bibfnamefont {J.~H.}\ \bibnamefont {Koh}}, \bibinfo {author} {\bibfnamefont {D.}~\bibnamefont {Kosenkov}}, \bibinfo {author} {\bibfnamefont {L.}~\bibnamefont {Koulias}}, \bibinfo {author} {\bibfnamefont {T.}~\bibnamefont {Kowalczyk}}, \bibinfo {author} {\bibfnamefont {C.~M.}\ \bibnamefont {Krauter}}, \bibinfo {author} {\bibfnamefont {K.}~\bibnamefont {Kue}}, \bibinfo {author} {\bibfnamefont {A.}~\bibnamefont {Kunitsa}}, \bibinfo {author} {\bibfnamefont
  {T.}~\bibnamefont {Kus}}, \bibinfo {author} {\bibfnamefont {I.}~\bibnamefont {Ladjánszki}}, \bibinfo {author} {\bibfnamefont {A.}~\bibnamefont {Landau}}, \bibinfo {author} {\bibfnamefont {K.~V.}\ \bibnamefont {Lawler}}, \bibinfo {author} {\bibfnamefont {D.}~\bibnamefont {Lefrancois}}, \bibinfo {author} {\bibfnamefont {S.}~\bibnamefont {Lehtola}}, \bibinfo {author} {\bibfnamefont {R.~R.}\ \bibnamefont {Li}}, \bibinfo {author} {\bibfnamefont {Y.-P.}\ \bibnamefont {Li}}, \bibinfo {author} {\bibfnamefont {J.}~\bibnamefont {Liang}}, \bibinfo {author} {\bibfnamefont {M.}~\bibnamefont {Liebenthal}}, \bibinfo {author} {\bibfnamefont {H.-H.}\ \bibnamefont {Lin}}, \bibinfo {author} {\bibfnamefont {Y.-S.}\ \bibnamefont {Lin}}, \bibinfo {author} {\bibfnamefont {F.}~\bibnamefont {Liu}}, \bibinfo {author} {\bibfnamefont {K.-Y.}\ \bibnamefont {Liu}}, \bibinfo {author} {\bibfnamefont {M.}~\bibnamefont {Loipersberger}}, \bibinfo {author} {\bibfnamefont {A.}~\bibnamefont {Luenser}}, \bibinfo {author} {\bibfnamefont
  {A.}~\bibnamefont {Manjanath}}, \bibinfo {author} {\bibfnamefont {P.}~\bibnamefont {Manohar}}, \bibinfo {author} {\bibfnamefont {E.}~\bibnamefont {Mansoor}}, \bibinfo {author} {\bibfnamefont {S.~F.}\ \bibnamefont {Manzer}}, \bibinfo {author} {\bibfnamefont {S.-P.}\ \bibnamefont {Mao}}, \bibinfo {author} {\bibfnamefont {A.~V.}\ \bibnamefont {Marenich}}, \bibinfo {author} {\bibfnamefont {T.}~\bibnamefont {Markovich}}, \bibinfo {author} {\bibfnamefont {S.}~\bibnamefont {Mason}}, \bibinfo {author} {\bibfnamefont {S.~A.}\ \bibnamefont {Maurer}}, \bibinfo {author} {\bibfnamefont {P.~F.}\ \bibnamefont {McLaughlin}}, \bibinfo {author} {\bibfnamefont {M.~F. S.~J.}\ \bibnamefont {Menger}}, \bibinfo {author} {\bibfnamefont {J.-M.}\ \bibnamefont {Mewes}}, \bibinfo {author} {\bibfnamefont {S.~A.}\ \bibnamefont {Mewes}}, \bibinfo {author} {\bibfnamefont {P.}~\bibnamefont {Morgante}}, \bibinfo {author} {\bibfnamefont {J.~W.}\ \bibnamefont {Mullinax}}, \bibinfo {author} {\bibfnamefont {K.~J.}\ \bibnamefont {Oosterbaan}},
  \bibinfo {author} {\bibfnamefont {G.}~\bibnamefont {Paran}}, \bibinfo {author} {\bibfnamefont {A.~C.}\ \bibnamefont {Paul}}, \bibinfo {author} {\bibfnamefont {S.~K.}\ \bibnamefont {Paul}}, \bibinfo {author} {\bibfnamefont {F.}~\bibnamefont {Pavošević}}, \bibinfo {author} {\bibfnamefont {Z.}~\bibnamefont {Pei}}, \bibinfo {author} {\bibfnamefont {S.}~\bibnamefont {Prager}}, \bibinfo {author} {\bibfnamefont {E.~I.}\ \bibnamefont {Proynov}}, \bibinfo {author} {\bibfnamefont { .}~\bibnamefont {Rák}}, \bibinfo {author} {\bibfnamefont {E.}~\bibnamefont {Ramos-Cordoba}}, \bibinfo {author} {\bibfnamefont {B.}~\bibnamefont {Rana}}, \bibinfo {author} {\bibfnamefont {A.~E.}\ \bibnamefont {Rask}}, \bibinfo {author} {\bibfnamefont {A.}~\bibnamefont {Rettig}}, \bibinfo {author} {\bibfnamefont {R.~M.}\ \bibnamefont {Richard}}, \bibinfo {author} {\bibfnamefont {F.}~\bibnamefont {Rob}}, \bibinfo {author} {\bibfnamefont {E.}~\bibnamefont {Rossomme}}, \bibinfo {author} {\bibfnamefont {T.}~\bibnamefont {Scheele}}, \bibinfo
  {author} {\bibfnamefont {M.}~\bibnamefont {Scheurer}}, \bibinfo {author} {\bibfnamefont {M.}~\bibnamefont {Schneider}}, \bibinfo {author} {\bibfnamefont {N.}~\bibnamefont {Sergueev}}, \bibinfo {author} {\bibfnamefont {S.~M.}\ \bibnamefont {Sharada}}, \bibinfo {author} {\bibfnamefont {W.}~\bibnamefont {Skomorowski}}, \bibinfo {author} {\bibfnamefont {D.~W.}\ \bibnamefont {Small}}, \bibinfo {author} {\bibfnamefont {C.~J.}\ \bibnamefont {Stein}}, \bibinfo {author} {\bibfnamefont {Y.-C.}\ \bibnamefont {Su}}, \bibinfo {author} {\bibfnamefont {E.~J.}\ \bibnamefont {Sundstrom}}, \bibinfo {author} {\bibfnamefont {Z.}~\bibnamefont {Tao}}, \bibinfo {author} {\bibfnamefont {J.}~\bibnamefont {Thirman}}, \bibinfo {author} {\bibfnamefont {G.~J.}\ \bibnamefont {Tornai}}, \bibinfo {author} {\bibfnamefont {T.}~\bibnamefont {Tsuchimochi}}, \bibinfo {author} {\bibfnamefont {N.~M.}\ \bibnamefont {Tubman}}, \bibinfo {author} {\bibfnamefont {S.~P.}\ \bibnamefont {Veccham}}, \bibinfo {author} {\bibfnamefont {O.}~\bibnamefont
  {Vydrov}}, \bibinfo {author} {\bibfnamefont {J.}~\bibnamefont {Wenzel}}, \bibinfo {author} {\bibfnamefont {J.}~\bibnamefont {Witte}}, \bibinfo {author} {\bibfnamefont {A.}~\bibnamefont {Yamada}}, \bibinfo {author} {\bibfnamefont {K.}~\bibnamefont {Yao}}, \bibinfo {author} {\bibfnamefont {S.}~\bibnamefont {Yeganeh}}, \bibinfo {author} {\bibfnamefont {S.~R.}\ \bibnamefont {Yost}}, \bibinfo {author} {\bibfnamefont {A.}~\bibnamefont {Zech}}, \bibinfo {author} {\bibfnamefont {I.~Y.}\ \bibnamefont {Zhang}}, \bibinfo {author} {\bibfnamefont {X.}~\bibnamefont {Zhang}}, \bibinfo {author} {\bibfnamefont {Y.}~\bibnamefont {Zhang}}, \bibinfo {author} {\bibfnamefont {D.}~\bibnamefont {Zuev}}, \bibinfo {author} {\bibfnamefont {A.}~\bibnamefont {Aspuru-Guzik}}, \bibinfo {author} {\bibfnamefont {A.~T.}\ \bibnamefont {Bell}}, \bibinfo {author} {\bibfnamefont {N.~A.}\ \bibnamefont {Besley}}, \bibinfo {author} {\bibfnamefont {K.~B.}\ \bibnamefont {Bravaya}}, \bibinfo {author} {\bibfnamefont {B.~R.}\ \bibnamefont {Brooks}},
  \bibinfo {author} {\bibfnamefont {D.}~\bibnamefont {Casanova}}, \bibinfo {author} {\bibfnamefont {J.-D.}\ \bibnamefont {Chai}}, \bibinfo {author} {\bibfnamefont {S.}~\bibnamefont {Coriani}}, \bibinfo {author} {\bibfnamefont {C.~J.}\ \bibnamefont {Cramer}}, \bibinfo {author} {\bibfnamefont {G.}~\bibnamefont {Cserey}}, \bibinfo {author} {\bibfnamefont {I.}~\bibnamefont {DePrince}, \bibfnamefont {A.~Eugene}}, \bibinfo {author} {\bibfnamefont {J.}~\bibnamefont {DiStasio}, \bibfnamefont {Robert~A.}}, \bibinfo {author} {\bibfnamefont {A.}~\bibnamefont {Dreuw}}, \bibinfo {author} {\bibfnamefont {B.~D.}\ \bibnamefont {Dunietz}}, \bibinfo {author} {\bibfnamefont {T.~R.}\ \bibnamefont {Furlani}}, \bibinfo {author} {\bibfnamefont {I.}~\bibnamefont {Goddard}, \bibfnamefont {William~A.}}, \bibinfo {author} {\bibfnamefont {S.}~\bibnamefont {Hammes-Schiffer}}, \bibinfo {author} {\bibfnamefont {T.}~\bibnamefont {Head-Gordon}}, \bibinfo {author} {\bibfnamefont {W.~J.}\ \bibnamefont {Hehre}}, \bibinfo {author} {\bibfnamefont
  {C.-P.}\ \bibnamefont {Hsu}}, \bibinfo {author} {\bibfnamefont {T.-C.}\ \bibnamefont {Jagau}}, \bibinfo {author} {\bibfnamefont {Y.}~\bibnamefont {Jung}}, \bibinfo {author} {\bibfnamefont {A.}~\bibnamefont {Klamt}}, \bibinfo {author} {\bibfnamefont {J.}~\bibnamefont {Kong}}, \bibinfo {author} {\bibfnamefont {D.~S.}\ \bibnamefont {Lambrecht}}, \bibinfo {author} {\bibfnamefont {W.}~\bibnamefont {Liang}}, \bibinfo {author} {\bibfnamefont {N.~J.}\ \bibnamefont {Mayhall}}, \bibinfo {author} {\bibfnamefont {C.~W.}\ \bibnamefont {McCurdy}}, \bibinfo {author} {\bibfnamefont {J.~B.}\ \bibnamefont {Neaton}}, \bibinfo {author} {\bibfnamefont {C.}~\bibnamefont {Ochsenfeld}}, \bibinfo {author} {\bibfnamefont {J.~A.}\ \bibnamefont {Parkhill}}, \bibinfo {author} {\bibfnamefont {R.}~\bibnamefont {Peverati}}, \bibinfo {author} {\bibfnamefont {V.~A.}\ \bibnamefont {Rassolov}}, \bibinfo {author} {\bibfnamefont {Y.}~\bibnamefont {Shao}}, \bibinfo {author} {\bibfnamefont {L.~V.}\ \bibnamefont {Slipchenko}}, \bibinfo {author}
  {\bibfnamefont {T.}~\bibnamefont {Stauch}}, \bibinfo {author} {\bibfnamefont {R.~P.}\ \bibnamefont {Steele}}, \bibinfo {author} {\bibfnamefont {J.~E.}\ \bibnamefont {Subotnik}}, \bibinfo {author} {\bibfnamefont {A.~J.~W.}\ \bibnamefont {Thom}}, \bibinfo {author} {\bibfnamefont {A.}~\bibnamefont {Tkatchenko}}, \bibinfo {author} {\bibfnamefont {D.~G.}\ \bibnamefont {Truhlar}}, \bibinfo {author} {\bibfnamefont {T.}~\bibnamefont {Van~Voorhis}}, \bibinfo {author} {\bibfnamefont {T.~A.}\ \bibnamefont {Wesolowski}}, \bibinfo {author} {\bibfnamefont {K.~B.}\ \bibnamefont {Whaley}}, \bibinfo {author} {\bibfnamefont {I.}~\bibnamefont {Woodcock}, \bibfnamefont {H.~Lee}}, \bibinfo {author} {\bibfnamefont {P.~M.}\ \bibnamefont {Zimmerman}}, \bibinfo {author} {\bibfnamefont {S.}~\bibnamefont {Faraji}}, \bibinfo {author} {\bibfnamefont {P.~M.~W.}\ \bibnamefont {Gill}}, \bibinfo {author} {\bibfnamefont {M.}~\bibnamefont {Head-Gordon}}, \bibinfo {author} {\bibfnamefont {J.~M.}\ \bibnamefont {Herbert}},\ and\ \bibinfo
  {author} {\bibfnamefont {A.~I.}\ \bibnamefont {Krylov}},\ }\bibfield  {title} {\bibinfo {title} {Software for the frontiers of quantum chemistry: An overview of developments in the q-chem 5 package},\ }\href {https://doi.org/10.1063/5.0055522} {\bibfield  {journal} {\bibinfo  {journal} {J. Chem. Phys.}\ }\textbf {\bibinfo {volume} {155}},\ \bibinfo {pages} {084801} (\bibinfo {year} {2021})}\BibitemShut {NoStop}%
\bibitem [{Note3()}]{Note3}%
  \BibitemOpen
  \bibinfo {note} {Since the orientation of the spin-orbit coupling vector depends on the details of the molecular structure, we consider for $\lambda $ in our minimal axial model the magnitude of the spin-orbit coupling vector, as shown in the Supporting Information.}\BibitemShut {Stop}%
\bibitem [{\citenamefont {Luo}\ and\ \citenamefont {Hore}(2021)}]{Luo2021}%
  \BibitemOpen
  \bibfield  {author} {\bibinfo {author} {\bibfnamefont {J.}~\bibnamefont {Luo}}\ and\ \bibinfo {author} {\bibfnamefont {P.~J.}\ \bibnamefont {Hore}},\ }\bibfield  {title} {\bibinfo {title} {Chiral-induced spin selectivity in the formation and recombination of radical pairs: cryptochrome magnetoreception and epr detection},\ }\href {https://doi.org/10.1088/1367-2630/abed0b} {\bibfield  {journal} {\bibinfo  {journal} {New J. Phys.}\ }\textbf {\bibinfo {volume} {23}},\ \bibinfo {pages} {043032} (\bibinfo {year} {2021})}\BibitemShut {NoStop}%
\bibitem [{\citenamefont {Ren}\ and\ \citenamefont {Hore}(2023)}]{Ren2023}%
  \BibitemOpen
  \bibfield  {author} {\bibinfo {author} {\bibfnamefont {Y.}~\bibnamefont {Ren}}\ and\ \bibinfo {author} {\bibfnamefont {P.~J.}\ \bibnamefont {Hore}},\ }\bibfield  {title} {\bibinfo {title} {Conditions for epr detection of chirality-induced spin selectivity in spin-polarized radical pairs in isotropic solution},\ }\href {https://doi.org/10.1063/5.0171700} {\bibfield  {journal} {\bibinfo  {journal} {J. Chem. Phys.}\ }\textbf {\bibinfo {volume} {159}},\ \bibinfo {pages} {145104} (\bibinfo {year} {2023})}\BibitemShut {NoStop}%
\bibitem [{Note4()}]{Note4}%
  \BibitemOpen
  \bibinfo {note} {In the present minimal axial model we obtain an AC for any $\theta \protect \neq 0$. For non-axial SOC and hence DMI we will get ACs also for $\theta =0$}\BibitemShut {NoStop}%
\bibitem [{Note5()}]{Note5}%
  \BibitemOpen
  \bibinfo {note} {In the liquid crystals alignment we are considering both $\mathinner {|{T_+}\rangle }$ and $\mathinner {|{T_-}\rangle }$ undergo an avoided level crossing for oppositely oriented sets of molecules in the solution.}\BibitemShut {Stop}%
\bibitem [{\citenamefont {Shiddiq}\ \emph {et~al.}(2016)\citenamefont {Shiddiq}, \citenamefont {Komijani}, \citenamefont {Duan}, \citenamefont {Gaita-Ariño}, \citenamefont {Coronado},\ and\ \citenamefont {Hill}}]{Hill2016}%
  \BibitemOpen
  \bibfield  {author} {\bibinfo {author} {\bibfnamefont {M.}~\bibnamefont {Shiddiq}}, \bibinfo {author} {\bibfnamefont {D.}~\bibnamefont {Komijani}}, \bibinfo {author} {\bibfnamefont {Y.}~\bibnamefont {Duan}}, \bibinfo {author} {\bibfnamefont {A.}~\bibnamefont {Gaita-Ariño}}, \bibinfo {author} {\bibfnamefont {E.}~\bibnamefont {Coronado}},\ and\ \bibinfo {author} {\bibfnamefont {S.}~\bibnamefont {Hill}},\ }\bibfield  {title} {\bibinfo {title} {Enhancing coherence in molecular spin qubits via atomic clock transitions},\ }\href {https://doi.org/10.1038/nature16984} {\bibfield  {journal} {\bibinfo  {journal} {Nature}\ }\textbf {\bibinfo {volume} {531}},\ \bibinfo {pages} {348} (\bibinfo {year} {2016})}\BibitemShut {NoStop}%
\bibitem [{\citenamefont {Aiello}\ \emph {et~al.}(2022)\citenamefont {Aiello}, \citenamefont {Abendroth}, \citenamefont {Abbas}, \citenamefont {Afanasev}, \citenamefont {Agarwal}, \citenamefont {Banerjee}, \citenamefont {Beratan}, \citenamefont {Belling}, \citenamefont {Berche}, \citenamefont {Botana}, \citenamefont {Caram}, \citenamefont {Celardo}, \citenamefont {Cuniberti}, \citenamefont {Garcia-Etxarri}, \citenamefont {Dianat}, \citenamefont {Diez-Perez}, \citenamefont {Guo}, \citenamefont {Gutierrez}, \citenamefont {Herrmann}, \citenamefont {Hihath}, \citenamefont {Kale}, \citenamefont {Kurian}, \citenamefont {Lai}, \citenamefont {Liu}, \citenamefont {Lopez}, \citenamefont {Medina}, \citenamefont {Mujica}, \citenamefont {Naaman}, \citenamefont {Noormandipour}, \citenamefont {Palma}, \citenamefont {Paltiel}, \citenamefont {Petuskey}, \citenamefont {Ribeiro-Silva}, \citenamefont {Saenz}, \citenamefont {Santos}, \citenamefont {Solyanik-Gorgone}, \citenamefont {Sorger}, \citenamefont {Stemer}, \citenamefont
  {Ugalde}, \citenamefont {Valdes-Curiel}, \citenamefont {Varela}, \citenamefont {Waldeck}, \citenamefont {Wasielewski}, \citenamefont {Weiss}, \citenamefont {Zacharias},\ and\ \citenamefont {Wang}}]{Aiello2022}%
  \BibitemOpen
  \bibfield  {author} {\bibinfo {author} {\bibfnamefont {C.~D.}\ \bibnamefont {Aiello}}, \bibinfo {author} {\bibfnamefont {J.~M.}\ \bibnamefont {Abendroth}}, \bibinfo {author} {\bibfnamefont {M.}~\bibnamefont {Abbas}}, \bibinfo {author} {\bibfnamefont {A.}~\bibnamefont {Afanasev}}, \bibinfo {author} {\bibfnamefont {S.}~\bibnamefont {Agarwal}}, \bibinfo {author} {\bibfnamefont {A.~S.}\ \bibnamefont {Banerjee}}, \bibinfo {author} {\bibfnamefont {D.~N.}\ \bibnamefont {Beratan}}, \bibinfo {author} {\bibfnamefont {J.~N.}\ \bibnamefont {Belling}}, \bibinfo {author} {\bibfnamefont {B.}~\bibnamefont {Berche}}, \bibinfo {author} {\bibfnamefont {A.}~\bibnamefont {Botana}}, \bibinfo {author} {\bibfnamefont {J.~R.}\ \bibnamefont {Caram}}, \bibinfo {author} {\bibfnamefont {G.~L.}\ \bibnamefont {Celardo}}, \bibinfo {author} {\bibfnamefont {G.}~\bibnamefont {Cuniberti}}, \bibinfo {author} {\bibfnamefont {A.}~\bibnamefont {Garcia-Etxarri}}, \bibinfo {author} {\bibfnamefont {A.}~\bibnamefont {Dianat}}, \bibinfo {author}
  {\bibfnamefont {I.}~\bibnamefont {Diez-Perez}}, \bibinfo {author} {\bibfnamefont {Y.}~\bibnamefont {Guo}}, \bibinfo {author} {\bibfnamefont {R.}~\bibnamefont {Gutierrez}}, \bibinfo {author} {\bibfnamefont {C.}~\bibnamefont {Herrmann}}, \bibinfo {author} {\bibfnamefont {J.}~\bibnamefont {Hihath}}, \bibinfo {author} {\bibfnamefont {S.}~\bibnamefont {Kale}}, \bibinfo {author} {\bibfnamefont {P.}~\bibnamefont {Kurian}}, \bibinfo {author} {\bibfnamefont {Y.-C.}\ \bibnamefont {Lai}}, \bibinfo {author} {\bibfnamefont {T.}~\bibnamefont {Liu}}, \bibinfo {author} {\bibfnamefont {A.}~\bibnamefont {Lopez}}, \bibinfo {author} {\bibfnamefont {E.}~\bibnamefont {Medina}}, \bibinfo {author} {\bibfnamefont {V.}~\bibnamefont {Mujica}}, \bibinfo {author} {\bibfnamefont {R.}~\bibnamefont {Naaman}}, \bibinfo {author} {\bibfnamefont {M.}~\bibnamefont {Noormandipour}}, \bibinfo {author} {\bibfnamefont {J.~L.}\ \bibnamefont {Palma}}, \bibinfo {author} {\bibfnamefont {Y.}~\bibnamefont {Paltiel}}, \bibinfo {author} {\bibfnamefont
  {W.}~\bibnamefont {Petuskey}}, \bibinfo {author} {\bibfnamefont {J.~C.}\ \bibnamefont {Ribeiro-Silva}}, \bibinfo {author} {\bibfnamefont {J.~J.}\ \bibnamefont {Saenz}}, \bibinfo {author} {\bibfnamefont {E.~J.~G.}\ \bibnamefont {Santos}}, \bibinfo {author} {\bibfnamefont {M.}~\bibnamefont {Solyanik-Gorgone}}, \bibinfo {author} {\bibfnamefont {V.~J.}\ \bibnamefont {Sorger}}, \bibinfo {author} {\bibfnamefont {D.~M.}\ \bibnamefont {Stemer}}, \bibinfo {author} {\bibfnamefont {J.~M.}\ \bibnamefont {Ugalde}}, \bibinfo {author} {\bibfnamefont {A.}~\bibnamefont {Valdes-Curiel}}, \bibinfo {author} {\bibfnamefont {S.}~\bibnamefont {Varela}}, \bibinfo {author} {\bibfnamefont {D.~H.}\ \bibnamefont {Waldeck}}, \bibinfo {author} {\bibfnamefont {M.~R.}\ \bibnamefont {Wasielewski}}, \bibinfo {author} {\bibfnamefont {P.~S.}\ \bibnamefont {Weiss}}, \bibinfo {author} {\bibfnamefont {H.}~\bibnamefont {Zacharias}},\ and\ \bibinfo {author} {\bibfnamefont {Q.~H.}\ \bibnamefont {Wang}},\ }\bibfield  {title} {\bibinfo {title} {A
  chirality-based quantum leap},\ }\href {https://doi.org/10.1021/acsnano.1c01347} {\bibfield  {journal} {\bibinfo  {journal} {ACS Nano}\ }\textbf {\bibinfo {volume} {16}},\ \bibinfo {pages} {4989} (\bibinfo {year} {2022})}\BibitemShut {NoStop}%
\bibitem [{\citenamefont {Chiesa}\ \emph {et~al.}(2023)\citenamefont {Chiesa}, \citenamefont {Privitera}, \citenamefont {Macaluso}, \citenamefont {Mannini}, \citenamefont {Bittl}, \citenamefont {Naaman}, \citenamefont {Wasielewski}, \citenamefont {Sessoli},\ and\ \citenamefont {Carretta}}]{Chiesa2023}%
  \BibitemOpen
  \bibfield  {author} {\bibinfo {author} {\bibfnamefont {A.}~\bibnamefont {Chiesa}}, \bibinfo {author} {\bibfnamefont {A.}~\bibnamefont {Privitera}}, \bibinfo {author} {\bibfnamefont {E.}~\bibnamefont {Macaluso}}, \bibinfo {author} {\bibfnamefont {M.}~\bibnamefont {Mannini}}, \bibinfo {author} {\bibfnamefont {R.}~\bibnamefont {Bittl}}, \bibinfo {author} {\bibfnamefont {R.}~\bibnamefont {Naaman}}, \bibinfo {author} {\bibfnamefont {M.~R.}\ \bibnamefont {Wasielewski}}, \bibinfo {author} {\bibfnamefont {R.}~\bibnamefont {Sessoli}},\ and\ \bibinfo {author} {\bibfnamefont {S.}~\bibnamefont {Carretta}},\ }\bibfield  {title} {\bibinfo {title} {Chirality-induced spin selectivity: An enabling technology for quantum applications},\ }\href {https://doi.org/https://doi.org/10.1002/adma.202300472} {\bibfield  {journal} {\bibinfo  {journal} {Adv. Mater.}\ }\textbf {\bibinfo {volume} {35}},\ \bibinfo {pages} {2300472} (\bibinfo {year} {2023})}\BibitemShut {NoStop}%
\bibitem [{Note6()}]{Note6}%
  \BibitemOpen
  \bibinfo {note} {Indeed, only for very small intra-site hopping (on the same order of magnitude of the effective spin-spin couplings) the 50\% limit on spin polarization is surpassed}\BibitemShut {NoStop}%
\bibitem [{\citenamefont {Tupkary}\ \emph {et~al.}(2022)\citenamefont {Tupkary}, \citenamefont {Dhar}, \citenamefont {Kulkarni},\ and\ \citenamefont {Purkayastha}}]{Tupkary2022}%
  \BibitemOpen
  \bibfield  {author} {\bibinfo {author} {\bibfnamefont {D.}~\bibnamefont {Tupkary}}, \bibinfo {author} {\bibfnamefont {A.}~\bibnamefont {Dhar}}, \bibinfo {author} {\bibfnamefont {M.}~\bibnamefont {Kulkarni}},\ and\ \bibinfo {author} {\bibfnamefont {A.}~\bibnamefont {Purkayastha}},\ }\bibfield  {title} {\bibinfo {title} {Fundamental limitations in lindblad descriptions of systems weakly coupled to baths},\ }\href {https://doi.org/10.1103/PhysRevA.105.032208} {\bibfield  {journal} {\bibinfo  {journal} {Phys. Rev. A}\ }\textbf {\bibinfo {volume} {105}},\ \bibinfo {pages} {032208} (\bibinfo {year} {2022})}\BibitemShut {NoStop}%
\end{thebibliography}


\end{document}